\documentclass[a4paper,12pt]{article}
\usepackage{amssymb}
\usepackage[latin1]{inputenc}
\usepackage[british]{babel}
\usepackage{natbib}
\usepackage{graphicx}
\usepackage[pdftex]{lscape}
\usepackage{verbatim}
\usepackage{amsmath}
\usepackage{arydshln}
\usepackage{amsthm}
\usepackage{amssymb}
\usepackage{amsfonts}
\usepackage{setspace}
\usepackage{latexsym}
\usepackage{a4}
\usepackage{bbm}
\usepackage{mathrsfs}
\usepackage{fixmath}
\usepackage{tikz}   
\usepackage{subfigure}
\usepackage{float}
\usepackage{booktabs,caption}
\usepackage[flushleft]{threeparttable}

\newcommand{\rk}{\mathrm{rank}}
\newcommand{\ve}{\mathrm{vec}\,}
\newcommand{\Chol}{\Sigma_{tr}}
\newcommand{\Choli}{\Sigma_{tr}^{-1}}

\newcommand{\Qr}{\ensuremath\mathcal{Q}(\phi\,|F,S)}

\newcommand{\e}{\varepsilon}

\newcommand{\Dtn}{\tilde{D}_n}

\newcommand{\Stri}{\Sigma_{tr}^{-1}}

\ifdefined\O
    \renewcommand{\O}{\ensuremath\mathcal{O}\left(n\right)}
\else
    \newcommand{\O}{\ensuremath\mathcal{O}\left(n\right)}
\fi
\ifdefined\Re
    \renewcommand{\Re}{\ensuremath\mathbb{R}}
\else
    \newcommand{\Re}{\ensuremath\mathbb{R}}
\fi
\ifdefined\A
    \renewcommand{\A}{\ensuremath\mathcal{A}_0(\phi)}
\else
    \newcommand{\A}{\ensuremath\mathcal{A}_0(\phi)}
\fi
\ifdefined\Ar
    \renewcommand{\Ar}{\ensuremath\mathcal{A}_R(\phi)}
\else
    \newcommand{\Ar}{\ensuremath\mathcal{A}_R(\phi)}
\fi
\ifdefined\AR
    \renewcommand{\AR}{\ensuremath\mathcal{A}_R}
\else
    \newcommand{\AR}{\ensuremath\mathcal{A}_R}
\fi
\ifdefined\Qr
    \renewcommand{\Qr}{\ensuremath\mathcal{Q}_R(\phi)}
\else
    \newcommand{\Qr}{\ensuremath\mathcal{Q}_R(\phi)}
\fi
\ifdefined\Aro
    \renewcommand{\Aro}{\ensuremath\mathcal{A}_{R,0}(\phi)}
\else
    \newcommand{\Aro}{\ensuremath\mathcal{A}_{R,0}(\phi)}
\fi

\def\thesection{\Roman{section}}
\makeatletter

\newcommand*\circled[1]{\tikz[baseline=(char.base)]{
            \node[shape=circle,draw,inner sep=2pt] (char) {#1};}}

\newtheorem{prop}{Proposition}
\newtheorem{corol}{Corollary}
\newtheorem{algo}{Algorithm}
\theoremstyle{definition}
\newtheorem{defin}{Definition}
\newtheorem{condition}{Condition}

\date{}

\begin{document}

\title{ \vspace{-2.5cm} \huge Locally- but not Globally-identified SVARs{\Large \thanks{We would like to thank Luca Fanelli, Gabriele Fiorentini, Riccardo Lucchetti, Sophocles Mavroeidis, Geert Mesters, Ulrich M\"{u}ller, Luca Neri, Mikkel Plagborg-M\o ller, Michele Piffer, Francesco Ravazzolo, Morten Ravn, Matthew Read, and Frank Schorfheide for beneficial discussions and comments. We have also benefited from presenting this work at the 12th International Conference on Computational and Financial Econometrics in Pisa (December 2018), Workshop in SVARs in QMUL (June 2019), 30th $(EC)^2$ Conference on Identification in Macroeconomics in Oxford (December 2019), the Econometric Society World Congress 2020 (August 2020), and First Joint Workshop of Applied Macro- and Microeconomics, University of Bozen (December 2021). We thank Thomas Carr for excellent research assistance. 
Financial support from the ESRC through the ESRC Centre for Microdata Methods and Practice (CeMMAP) (grant number RES-589-28-0001) and the European Research Council (Starting grant No. 715940) is gratefully acknowledged. Emanuele Bacchiocchi gratefully acknowledges financial support from Italian Ministry of University and Research (PRIN 2022, Grant 20229PFAX5) and the University of Bologna (RFO grants).}}}

\author{%
\begin{tabular}{ccc}
Emanuele Bacchiocchi\thanks{University of Bologna, Department of Economics. Email: e.bacchiocchi@unibo.it} &  & 
Toru Kitagawa\thanks{Brown University, Department of Economics. Email: toru\_kitagawa@brown.edu} \\ 
University of Bologna &  & Brown University \\ 
\end{tabular}%
}


\date{This draft: 1 April 2025}
\maketitle

\vspace{-0.5cm}
\begin{abstract}
This paper analyzes Structural Vector Autoregressions (SVARs) where identification of structural parameters holds locally but not globally. In this case there exists a set of isolated structural parameter points that are observationally equivalent under the imposed restrictions. Although the data do not inform us which observationally equivalent point should be selected, the common frequentist practice is to obtain one as a maximum likelihood estimate and perform impulse response analysis accordingly. For Bayesians, the lack of global identification translates to non-vanishing sensitivity of the posterior to the prior, and the multi-modal likelihood gives rise to computational challenges as posterior sampling algorithms can fail to explore all the modes. This paper overcomes these challenges by proposing novel estimation and inference procedures. We characterize a class of identifying restrictions and circumstances that deliver local but non-global identification, and the resulting number of observationally equivalent parameter values. We propose algorithms to exhaustively compute all admissible structural parameters given reduced-form parameters and utilize them to sample from the multi-modal posterior. In addition, viewing the set of observationally equivalent parameter points as the identified set, we develop Bayesian and frequentist procedures for inference on the corresponding set of impulse responses. An empirical example illustrates our proposal. 
\end{abstract}


\begin{flushleft}
\end{flushleft}

\newpage


\section{Introduction}
\label{sec:intro}

Macroeconomic policy analysis makes extensive use of impulse response analysis based on Structural Vector Autoregressions (SVARs). Various types of identifying assumptions have been proposed, including equality and sign restrictions, and analytical investigation of whether they point- or set-identify the objects of interest is an active area of research. The seminal work of \cite{RWZ10} (henceforth RWZ) shows a necessary and sufficient condition for zero restrictions to achieve global identification. This class of zero restrictions, however, does not exhaust the universe of zero and non-zero equality restrictions that are relevant in practice. Questions regarding identification, estimation, and inference when identification is not global remain largely open.  

This paper focuses on a class of SVARs where the adopted identification strategies guarantee local identification but do not attain global identification. The set of observationally equivalent structural parameters then consists of multiple isolated points, which implies that the likelihood can have multiple peaks of the same height. Such locally- but non-globally identified SVARs appear in various settings of practical relevance. Examples include non-zero restrictions which set the structural parameters to calibrated values, non-recursive zero restrictions, equality restrictions across shocks and/or equations, heteroskedastic SVARs, and SVARs characterized by structural breaks with across-regime restrictions on the structural coefficients. These classes of models have been largely used in empirical applications and, some of them, like heteroskedastic SVARs, continue to play a relevant role in applied macroeconomics.\footnote{See Section \ref{sec:ex} for a detailed review of empirical applications using locally-identified SVARs.} Although the data do not inform us which observationally equivalent point should be selected, the common frequentist practice is to obtain one as a maximum likelihood estimator and perform impulse response analysis as if it were the only maximizer. This approach is problematic as different maximizers of the likelihood may imply very different impulse responses. In our view, this practice is prevalent due to the lack of an efficient algorithm that can uncover all the local maxima. Standard Bayesian analysis also faces challenges when the likelihood has multiple modes. First, the lack of global identification leads the posterior to remain sensitive to the choice of prior even asymptotically. Second, posterior sampling algorithms may fail to explore all the modes, resulting in an inaccurate approximation of the posterior.  

This paper proposes methods for estimation and inference that overcome these challenges. We first characterize a class of equality and sign restrictions that delivers local but non-global identification. Second, we show a necessary and sufficient condition for local identification that can be easily checked under a general class of equality constraints imposed on the structural parameters or functions of them. Third, we investigate how many observationally equivalent parameter values exist under such identifying restrictions, and propose algorithms to exhaustively compute them given reduced-form parameter values. Specifically, we exploit the orthogonal matrix parametrization of \citet{Uhlig2005} and \citet{RWZ10} and pin down the observationally equivalent parameter points (conditional on the reduced-form parameters) by sequentially exhausting the admissible orthogonal vectors satisfying the imposed restrictions, or in some cases solving a system of polynomial equations. We provide an intuitive geometric exposition that illustrates the mechanism driving the lack of global identification and the number of observationally equivalent parameter values. As a byproduct, we also characterize the set of reduced-form parameter values that yield no admissible structural parameters (i.e, an empty identified set) despite the condition for local identification being met.

Our proposal for computing the identified set contributes to standard Bayesian inference by simplifying and stabilising sampling from the multi-modal posterior. The way we obtain a draw of the structural parameters or an impulse response from the posterior incorporates the following three steps. The first step is to obtain a draw of the reduced-form parameters either by directly sampling it or transforming a draw of the structural parameters into the reduced-form parameters. In the second step, given the draw of the reduced-form parameters, we compute all the observationally equivalent orthogonal matrices using our proposed algorithm. In the third step, we draw one of the observationally equivalent orthogonal matrices according to the probability weights implied by the prior distribution. Combining the draws of the reduced-form parameters and orthogonal matrix provides a draw of the structural parameters and impulse responses. Thus-constructed new draw stochastically moves across the modes supported by the prior. Hence, merging these extra steps into Gibbs  or Metropolis-Hasting algorithm helps us explore all the posterior modes of the structural parameters.

Bayesian inference for non-identified parameters requires specifying a prior over the observationally equivalent parameter values and its influence to posterior remains even asymptotically. This phenomenon persists to the current case of only locally-identified structural parameters. To deal with the case where the user cannot form the prior or wants to draw prior-free frequentist inference, we propose projection-based frequentist inference procedures for the impulse responses. Viewing the set of observationally equivalent parameter points as the identified set (a set-valued map from the reduced-form parameters to the set of observationally equivalent structural parameters), we extend the approach of \citet{NT14}, \citet{KT13}, and \cite{GK18}, designed primarily for models with interval identified sets, to cases where the identified set consists of a finite number of points. Specifically, we consider projecting the posterior credible region for the reduced-form parameters to the impulse responses through the discrete identified set mapping. This approach obtains asymptotically frequentist valid confidence intervals in the presence of local identification. A complication unique to the current case of discrete identified set is how to label these observationally equivalent parameter points in a manner consistent over different values of the reduced-form parameters. We propose two different ways to do so. As shown in \citet{GK18}, our frequentist procedures can be interpreted as (multiple prior) robust Bayesian procedures performing global sensitivity analysis with respect to  a certain class of priors.

To illustrate our proposal, we apply the method to a locally identified non-recursive New-Keynesian monetary policy SVAR. We show that, when a single element is selected from the identified set, the choice of the element can lead to significantly different and arguably contradictory results. We perform Bayesian inference with the prior that equally weights these observationally equivalent contradictory impulse responses, and show that the posterior distribution, approximated by our sampling procedure, well captures these contradictory impulse responses by its multimodality. Our proposals of asymptotically valid frequentist inference also explore all the admissible impulse responses and provide their summary by interval estimates. 

In a more sophisticated empirical application, instead, we show to what extent the issue of local identification can affect the SVARs identified through heteroskedasticity. In this case, the heteroskedasticity allows to obtain a statistical identification of the model. The identification of the shock (or shocks) of interest, instead, is performed \textit{a posteriori}, through a critical inspection of the impulse responses, that should satisfy appropriate features suggested by the economic theory. Inspired by a recent work by \cite{CH2019AEJ}, we try to identify a monetary policy shock based on the two volatility regimes associated to the `great inflation' and `great moderation', largely documented in the literature. We obtain that two observationally equivalent structural shocks have the desirable characteristics of a monetary policy shock. The methodology developed in the present paper allows to treat both of them as perfect candidates and produces statistical inference on the impulse responses when they are jointly considered.   


\subsection{Related literature}
\label{sec:literature}

The theory of identification for linear simultaneous equation models has a long history in econometrics. See \citet{Dhrymes78}, \citet{Fisher66} and \citet{Hausman83}, among others. 
\cite{Rothenberg71ECTA} analyses identification in parametric models. Building on this, \cite{Giannini92} proposes a criterion for verifying local identification for SVAR models. This criterion takes the form of rank conditions for the Hessian matrix of the average likelihood. It is much weaker than the necessary and sufficient condition for global identification shown in \citet{RWZ10}. 
The focus of the present paper is the class of identifying restrictions that satisfies the former but not the latter.
Once local identification is guaranteed, \cite{Giannini92} proposes estimating the parameters of the SVAR by numerically maximizing the likelihood. This approach is also recommended by the textbooks 
\cite{AG}, \cite{Hamilton94}, \cite{LutBook06}, and \cite{KLbook}. For the locally identified models considered in this paper, however, the maximum likelihood estimate is not necessarily unique, 
and a typical numerical maximization routine will select only one point in a non-systematic manner (e.g. depending on a choice of initial value). \cite{SZ99} and \cite{HWZ07} include discussions 
of the existence of multiple likelihood peaks due to local identification. 

Following \citet{Uhlig2005}, \citet{RWZ10}, \cite{ARW18} and \citet{MSG13}, we parameterize an SVAR by its reduced-form VAR-parameters and the orthogonal matrix relating its reduced-form error covariance matrix and structural parameters. Fixing the reduced-form parameters, finding all the observationally equivalent structural parameters reduces to finding all the admissible orthogonal matrices that satisfy the imposed identifying restrictions. Compared to expressing the non-linear equation system by the reduced form and structural parameters, this formulation is advantageous in terms of geometric interpretability and analytical tractability. In addition, it simplifies not only assessing local identification (e.g., \citeauthor{mn07}, \citeyear{mn07}), but also obtaining all the solutions given the reduced-form parameters.

Our paper is related to the growing literature on SVARs that are set-identified through sign and zero restrictions (\citeauthor{Faust98} \citeyear{Faust98}; \citeauthor{CanovadeNicolo02JME} 
\citeyear{CanovadeNicolo02JME}; \citeauthor{Uhlig2005} \citeyear{Uhlig2005}; \citeauthor{MU09} \citeyear{MU09}, \citeauthor{ARW18} \citeyear{ARW18}, \citeauthor{GafarovOlea14} \citeyear{GafarovOlea14}, \citeauthor{GK18} \citeyear{GK18}, \citeauthor{MSG13} \citeyear{MSG13}, among others). The identified set of impulse responses in this class of models is a set with a positive measure if non-empty, whereas the identified set here consists of a finite number of isolated points, each corresponding to a solution of a non-linear system of equations. This difference in the topological features of the identified set distinguishes our inferential procedure from these works.  

Our proposals of asymptotically frequentist-valid inference build on Bayesian approach to inference on identified set as considered in \citet{CCT18}, \citet{GK18}, \citet{KT13}, \citet{LS13}, \citet{MS12}, and \citet{NT14}. To our knowledge, however, none of these proposals have been applied to the case where the identified set consists of isolated points. As discussed in \citet{GK18}, our approach for drawing frequentist inference has a robust Bayes interpretation, where ambiguity within the identified set is introduced through a set of unrevisable priors. In this sense, it can be appealing to Bayesians who cannot form a credible prior for the structural parameters or want to perform global sensitivity analysis. Depending on the application, the class of priors considered in \citet{GK18} could be too large. In such cases, refining the set of priors would be sensible. This can be done, for instance, by applying the approaches considered in \citet{GKV18} and \citet{GKU19}, although we do not present them in this paper.

The results and proposals of this paper, from identification to estimation and inference, can also contribute to the literature that bridges Dynamic Stochastic General Equilibrium (DSGE) and VAR models. The solution of a linearized DSGE model can be summarized by a state-space representation that implies, under appropriate invertibility conditions, an (infinite order) SVAR subject to specific identifying restrictions (see,  \citeauthor{CEV06NBER} \citeyear{CEV06NBER}, \citeauthor{FRSW07AER} \citeyear{FRSW07AER}, and \citeauthor{Ravenna07JME} \citeyear{Ravenna07JME}, for example). As stressed by \citeauthor{Canova05Book} (\citeyear{Canova05Book}, chapter 4) among others, popular identification schemes that lead to global identification, such as the Cholesky decomposition, cannot be justified in a large class of DSGE models. Hence, if the mapping between the DSGE and the SVAR is unique as in \citeauthor{CEV06NBER} (\citeyear{CEV06NBER}, Proposition 1), DSGE-based identifying restrictions can result in local (but not global) identification. This is due to the non-recursive nature of the identification scheme, and the possible multiplicity of solutions characterizing the DSGE model. See \cite{Iskrev10JME}, \cite{KomNg11ECTA} and \cite{QuTka12QE} for DSGE models, and \cite{ASZ19} for local identification in linear rational expectation models. This paper is related to the DSGE literature for two reasons: firstly, DSGE models may imply non-recursive identifying restrictions in an SVAR, resulting thus in local identification of the SVAR, and, secondly, our estimation and inference methods to handle local identification can also be extended to locally identified (linearised) DSGE models.

Although identification analysis of this paper  mainly focuses on Gaussian SVARs, the inferential proposals  we propose can also contribute to the growing literature on 
identification using non-Gaussianity and/or heteroskedasticity. Concerning the former, \cite{LMS17JoE}, \cite{LL2021JBES}, \cite{GMR2017JoE} exploit higher moments and Independent Component Analysis (ICA) to point identify all or a subset of the structural shocks of SVARs. For the latter, \cite{Sen_Fio01}, \cite{Rigobon03}, \cite{Lan_Lut2008}, \cite{BPSS21}, \cite{Lewis21Restud}, \cite{SRP23} among others, propose to use the heteroskedasticity in the data to reach point identification for factor models, simultaneous equations, and SVARs. 
Without an enough number of restrictions that can pin down the labeling of the structural shocks, the identification under these approaches are inherently local but not global due to the observationally equivalent representations through permutations of the structural shocks or structural equations. As pursued in \cite{DW21}, adding sign restrictions motivated via economic models to non-Gaussian SVARs helps reduce the number of admissible impulse responses, while it is not known if the sign restrictions can guarantee global identification in non-Gaussian SVARs. \cite{SRP23}, instead, propose combining heteroskedasticity and external instruments, that, when available, can help point identifying the shocks of interest. Our inferential proposals can be implemented in these models if we can compute the set of admissible structural parameters given the reduced-form parameters (e.g., moments of the data). The empirical application in Section \ref{sec:EmpApp} goes in this direction.

\bigskip

The remainder of the paper is organized as follows. Section \ref{sec:def} introduces notation and  basic definitions of identification in SVAR models. Section \ref{sec:ex} provides a set of examples of locally- but not globally-identified SVARs and motivates our research. Section \ref{sec:identif} is dedicated to the theory of identification in SVARs whose identifying restrictions take the form of equality and sign restrictions. It also presents a set of necessary and sufficient conditions for local identification in SVARs, as well as a discussion on local identification in heteroskedastic SVARs (hereafter, HSVARs). Section \ref{sec:Estimation} presents algorithms for computing observationally equivalent parameter values in SVARs and HSVARs, and Section \ref{sec:InfLocId} proposes inference methods that accommodate frequentist, Bayesian, and robust Bayesian perspectives. Section \ref{sec:EmpApp} presents an empirical example based on HSVAR, and Section \ref{sec:conclusion} concludes. Further results on local identification are reported in Appendices \ref{app:Geom}, \ref{app:FurResId}, \ref{app:Dnt}, \ref{app:AlgorithmNH}; the proofs omitted from the main text are presented in Appendix \ref{app:Proofs}. A second empirical application, mainly dedicated to illustrate the details of the implementation of our methodology is reported in Appendix \ref{sec:EmpAppNK}.

\section{Econometric framework}
\label{sec:def}


Let $y_t$ be a $n\times 1$ vector of variables observed over $t=1\ldots T$. The SVAR model is specified as
\begin{equation}
\label{eq:SVAR}
 A_0 y_t = a + \sum_{j=1}^{p}A_j y_{t-j}+\e_t
\end{equation}
where $\e_t$ is a $n\times 1$ multivariate normal white noise process with null expected value and covariance matrix equal to the identity matrix $I_n$. The quantities $A_0,\,A_1,\ldots,\,A_p$ are $n\times n$ matrices of parameters, and $a$ is a $n\times 1$ vector of constant terms. The set of structural parameters is denoted by $A=(A_0,A_+)\in \mathcal{A} \subset \Re^{(n+m)n}$, with $m\equiv np+1$ and $A_+\equiv (a,\,A_1,\ldots,\,A_p)$ being a $n\times m$ matrix. We also assume that the initial conditions $y_1,\,\ldots,\, y_p$ are given, and $A_0$ to be invertible.

The reduced-form representation of the SVAR, obtained by pre-multiplying by the inverse of $A_0$, is the standard VAR model
\begin{equation}
\label{eq:VAR}
 y_t = b + \sum_{j=1}^{p}B_j y_{t-j}+u_t
\end{equation}
where $B_j=A_0^{-1}A_j$, $j=1,\ldots, p$, $b=A_0^{-1}a$, $u_t=A_0^{-1}\e_t$ and $E(u_t\,u_t^\prime)\equiv\Sigma=A_0^{-1}A_0^{-1\prime}$. The set of reduced-form parameters is $\phi=(B,\,\Sigma)\in \Phi \subset \Re^{n+n^2p}\times \Omega$, where $B=(b,\,B_1,\ldots,\,B_p)$ and $\Omega$ is the space of positive semi-definite matrices.

Assuming further that the VAR in Eq. (\ref{eq:VAR}) is invertible, it has the VMA$(\infty)$ representation:
\begin{equation}
y_t = \mu+\sum_{j=0}^{\infty}C_j (B) u_{t-j}\nonumber\\
                = \mu+\sum_{j=0}^{\infty}C_j (B) A_0^{-1}\e_{t-j}\label{eq:VMA}\nonumber
\end{equation}
where $C_j (B)$ is the \textit{j}-th coefficient matrix of the inverted lag polynomial $\left(I_n-\sum_{j=1}^{p} B_j L^j\right)^{-1}$ and $\mu = (I_n-B_1-\ldots-B_p)^{-1}b$. We define the impulse response matrix at horizon $h$ ($IR^h$)
and the long-run cumulative impulse response matrix ($CIR^\infty$) to be
\begin{eqnarray}
IR^h & = & C_h (B) A_0^{-1},\label{eq:IRh}\\
CIR^\infty & = & \sum_{j=0}^{\infty}IR^j = \left(\sum_{j=0}^{\infty}C_j (B)\right) A_0^{-1},\label{eq:CIR}
\end{eqnarray}

In what follows throughout, we denote the Cholesky decomposition of $\Sigma$ by $\Sigma = \Chol\Chol^\prime$, where $\Chol$ is the unique lower-triangular Cholesky factor with non-negative diagonal elements. The column vectors of $\Chol^{-1}$ and $\Chol^\prime$ are denoted by $\Chol^{-1} \equiv (\tilde{\sigma}_1,\,\tilde{\sigma}_2,\ldots,\,\tilde{\sigma}_n)$ and $\Chol^\prime \equiv (\sigma_1,\,\sigma_2,\ldots,\,\sigma_n)$. The $i$-th entry of $\tilde{\sigma}_j$ and $\sigma_j$ are denoted by $\tilde{\sigma}_{j,i}$ and $\sigma_{j,i}$, respectively.


\subsection{Identification of SVAR models}
\label{sec:identification}

Identification analysis of SVAR models concerns solving $\Sigma = A_0^{-1} A_0^{-1\prime}$ to decompose the reduced form error variance-covariance matrix $\Sigma$ into the matrix of structural coefficients $A_0$. Following \cite{Uhlig2005}, any structural matrix $A_0$ defined by a rotation of the Cholesky factor $A_0=Q^\prime\Choli$ admits the decomposition $\Sigma=A_0^{-1}A_0^{-1\prime}$ and, given the reduced-form parameters $\phi$, the set of related $A_0$ matrices can be represented by $\A\equiv\{A_0=Q^\prime\Choli\,:\,Q\in\O\}$, where $\O$ is the set of $n\times n$ orthogonal matrices. Let $R$ be generic notation denoting identifying restrictions. The identifying restrictions constrain the possible values of $A$ to a subset of $\mathcal{A}$. We denote this subset by $\mathcal{A}_R$, and its projection for $A_0$ by $\mathcal{A}_{R,0}$. Accordingly, let $\Phi_R \subset \Phi$ be the space of reduced-form parameters formed by projecting $A \in \mathcal{A}_R$, and let
\begin{equation}     
        \Aro \equiv \mathcal{A}_0 (\phi) \cap \mathcal{A}_{R,0}, \label{eq:Ar0(phi)}
\end{equation}
which is non-empty for $\phi \in \Phi_{R}$.

We define global and local identification for an SVAR as follows.

\begin{defin}[Global identification]
\label{def:global}
        An SVAR model is globally identified under identifying restrictions $R$ if for almost every $A \in \mathcal{A}_R$ there is no other
        observationally equivalent $A$ in $\mathcal{A}_R$. 
\end{defin}

\begin{defin}[Local identification]
\label{def:local}
        An SVAR model is locally identified under identifying restrictions $R$ if for almost every $A \in \mathcal{A}_R$, 
        there exists an open neighborhood $G$ such that $G \cap \mathcal{A}_R$ contains no other observationally equivalent $A$. 
\end{defin}

Some remarks on these two notions of identification are in order. An equivalent definition of global identification would be that, for almost every $\phi \in \Phi_R$, there exists a unique corresponding structural parameter point. In other words, $\Aro$ is singleton-valued at almost every $\phi \in \Phi_R$. In addition, the case where $\Phi_R = \Phi$, i.e. the imposed identifying assumptions are not observationally restrictive, is what RWZ refer to as \textit{exact identification}. In contrast, the definition of local identification says that, if there are multiple observationally equivalent structural parameter points, they must be far apart. This implies that for almost every $\phi \in \Phi_R$, if $\Aro$ is not singleton, it consists of isolated points. In Proposition 2 below, we characterize a class of locally identified SVARs. For this class of SVARs, the space of reduced-form parameters $\Phi$ can be partitioned into three subsets. 
The first, of positive measure, contains parameters for which the model is locally- but not globally-identified; the second, of positive measure, on which there is no structural parameter satisfying the identifying assumption (i.e., $\Aro$ is empty); and the third, of measure zero, on which the model is globally identified. This feature of locally identified SVARs stands in contrast to exactly identified SVARs and globally and over-identified SVARs, where the mapping from the reduced-form parameter space $\Phi$ to structural parameters that satisfy the identifying restrictions is guaranteed to be either singleton-valued or empty at almost every $\phi \in \Phi$.



\section{Locally-identified SVARs: some examples}
\label{sec:ex}

To motivate our research, in this section we provide some examples of SVAR models that could be locally but not globally identified. \cite{HWZ07} discuss local identification as a normalization problem. As we will see in the next section, in the presence of non-homogeneous equality restrictions and/or across-shock restrictions, proper sign normalization restrictions are not enough to resolve the issue of local identification in SVARs. The examples below illustrate that the local identification issue is of practical relevance.


\subsection{Calibrated identifying restrictions and restrictions across shocks or across equations}
\label{sec:NonHomo}

One strategy employed in the literature is to calibrate some parameters instead of estimating them. Calibration can be viewed as imposing non-homogeneous restrictions and, as we will show in the simple example in Section \ref{sec:geo}, this can lead to local identification. For example, \cite{BlanPerotti02} and \cite{BW86} impose non-zero values for some structural parameters in $A_0$, based on external information. \cite{AH95}, \cite{DK11} and \cite{Kilian10}, instead of imposing fixed values, explore a grid 
of possible values for some structural parameters in order to provide robustness checks for their main model specification. 

Cross-equation restrictions have been investigated in the classical literature of simultaneous equation systems (\citeauthor{Fisher66} \citeyear{Fisher66}, and \citeauthor{Kelly75} \citeyear{Kelly75}).\footnote{In particular, \cite{Kelly75} presents cases in which economic theory might suggest imposing such restrictions. However, constraining parameters across equations is conditional on the kind of normalization considered. In simultaneous equation systems, normalization rules were generally based on imposing a unit coefficient for the variable playing the role of endogenous variable in that specific equation. In the parametrization proposed by RWZ for SVAR models the normalization rule instead consists of imposing unit variance on the uncorrelated structural shocks. In this case, imposing restrictions on elasticities across equations would involve non-linear restrictions on the estimated coefficients. In fact, to obtain the elasticities, we need to normalize the coefficient for the \textit{endogenous} variable in each equation. See \cite{HWZ07} and \cite{WZ03} for specific details on the normalization issue in SVAR models.} Among others, examples in this direction can be found in 
\cite{DM80} and related papers on demand systems, where cross-equation restrictions are imposed in order to test for the Slutsky symmetry assumptions. See \cite{Kilian13} for a simple example on restrictions across equations in a bivariate SVAR. Similar situations arise in SVARs when restrictions are imposed on impulse responses to different structural shocks. As we will see, in both approaches, such kinds of constraints involve restrictions across the columns of the orthogonal matrix $Q$, that are not contemplated by RWZ. As for calibrated parameters, this identification strategy can lead to local identification.


\subsection{Non-recursive SVAR models}
\label{sec:NonRec}

RWZ provide an example of a locally- but not globally-identified SVAR. This example involves non-recursive causal ordering restrictions and has practical importance, as we illustrate below.   

\cite{Cochrane06} considers the following New-Keynesian model for inflation $\pi_t$, output gap $x_t$, and the nominal interest rate $i_t$:
\begin{eqnarray}
\pi_t & = & \beta E_t\:\pi_{t+1}\:+\: \kappa x_t \:+\:u_t^s\nonumber\\
x_t     & = & E_t\:x_{t+1} \:-\:\tau(i_t-E_t\:\pi_{t+1}) \:+\:u_t^d\label{eq:DSGE}\\
i_t     & = & \phi_\pi\,\pi_t \:+\:u_t^{mp}\nonumber
\end{eqnarray}
with $u_t^s$, $u_t^d$ and $u_t^{mp}$ being, respectively, the independent supply, demand, and monetary policy shocks with variances  $\sigma_s^2$, $\sigma_d^2$ and $\sigma_{mp}^2$. \cite{FWZ07DSGE} show that this model can be written as an SVAR of the form
\begin{equation}
A_0\,y_t \:=\:\e_t\nonumber
\end{equation}
where $y_t=(\pi_t,\,x_t,\,i_t)^\prime$ is the vector of observable variables, $\e_t=(\e_t^s,\,\e_t^d,\,\e_t^{mp})$ collects the unit-variance uncorrelated structural shocks and 
\begin{equation}
\label{eq:DSGEsimmat}
    A_0=\left(\begin{array}{ccc}
    a_{11} & a_{12} & 0\\
    0 & a_{22} & a_{23}\\
    a_{31} & 0 & a_{33}
    \end{array}\right).
\end{equation}
Note that there is a well-defined mapping between the parameters in $A_0$ and those in the DSGE representation Eq. (\ref{eq:DSGE}).\footnote{As pointed out by \cite{Canova05Book}, zero restrictions implied by DSGE models do not match the recursive identification schemes common in SVAR analyses.}


Given this identification scheme, \cite{RWZ08} show the existence of two different $A_0$ matrices, both of them admissible according to the zero restrictions in Eq. (\ref{eq:DSGEsimmat}), and thus local identification. 
In Appendix \ref{sec:EmpAppNK}, we use this example to show the practical implementation of all theoretical results and algorithms proposed in the paper. Other examples of non-recursive SVARs, among others, are \cite{sims86}, \cite{Bernanke86}, \cite{BW86} and \cite{SimsZha06AER}.


\subsection{SVAR with breaks}
\label{sec:SVARWB}

\cite{BF15} and \cite{BK_SVARWB} consider SVARs with breaks in the structural error variances and regime-dependent structural coefficients. They consider identifying assumptions that restrict some structural parameters to being invariant across the regimes. 

Suppose that the two regimes are characterized by two different reduced-form error covariance matrices $\Sigma_1$ and $\Sigma_2$, which are related to the regime-dependent structural parameters through
\begin{equation}
\label{eq:HSVAR} 
\Sigma_1=A_{01}^{-1}A_{01}^{-1\prime}\hspace{1cm}\text{and}\hspace{1cm}\Sigma_2=A_{02}^{-1}A_{02}^{-1\prime},
\end{equation}
where $A_{01}$ and $A_{02}$ are the matrices of regime-specific structural parameters. Let $Q_1$ and $Q_2$ be the regime specific orthogonal matrices mapping the reduced-form error variances to the structural coefficients,
\begin{equation}
\label{eq:HSVARalt} 
A_{01}=Q_1^\prime\,\Sigma_{1,tr}^{-1}\hspace{1cm}\text{and}\hspace{1cm}A_{02}=Q_2^\prime\,\Sigma_{2,tr}^{-1}
\end{equation}
with  $Q_i=[q_{1(i)},\,\ldots,q_{n(i)}]$, $\forall\:i,\,\in\,\{\,1,\,2\,\}$. We denote the $j$-th column vector of $\Sigma_{i,tr}'$ by $\sigma_{j(i)}$ for $j=1,2$ and $i=1,2$.

For simplicity, consider a bivariate SVAR with two regimes. Impose the following identifying restrictions:
\begin{equation}
\label{eq:HSVARrest}
\begin{array}{lcl}
(A_{01})_{[1,2]}^{-1}=0 & \text{\hspace{1cm}and\hspace{1cm}} & (A_{01})_{[2,1]}^{-1}=(A_{02})_{[2,1]}^{-1}.
\end{array}
\end{equation}
The first zero restriction makes the SVAR in the first regime a standard triangular one, that according to RWZ is globally identified. Once we uniquely pin down the matrix $A_{01}$, the second restriction reduces to the case of a calibrated identifying restriction for the SVAR in the second regime. Hence, the problem of identification for structural parameters in the second regime exactly corresponds to the example that will be discussed in Section \ref{sec:geo}, in which local identification holds with two distinct admissible solutions.


\subsection{Heteroskedastic SVAR (HSVAR)}
\label{sec:HSVAR}

Since \cite{Rigobon03} studies a simultaneous equation model with breaks in the volatility of structural shocks, HSVAR is a topic of active research. See, e.g., \cite{Lan_Lut2008}, \cite{Sims22}, \cite{BPSS21}, and \cite{Lewis21Restud,Lewis22Restat} for recent contributions. 

As shown in \cite{BBKM} and \cite{LMNS21} among others, we can formalize identification of HSVAR as an eigenvalue decomposition problem, and it clarifies how local identification of $A_0$ occurs. Suppose we observe a break in the covariance matrix of the reduced-form errors from $\Sigma_1$ to $\Sigma_2$ at time $T_B$. 
The key identifying assumption in HSVAR is that the break in the reduced-form error variance is characterized solely by the volatility break of the structural shocks, keeping any of the structural coefficients invariant. 
Normalizing the variance-covariance matrix of the the structural shocks before the break to the identity matrix, let a diagonal matrix $\Lambda$ with strictly positive diagonal elements be the variance-covariance matrix of the structural shocks after the break. This assumption leads to 
\begin{equation}
        \label{eq:CovarVARWB}
        E(u_tu_t^\prime)=\left\{\begin{array}{lcl}
        \Sigma_1 = A_0^{-1} A_0^{-1\prime} & & \text{if }\quad 1\leq t\leq T_{B}\\
        \Sigma_2 = A_0^{-1} \Lambda A_0^{-1\prime}& & \text{if }\quad T_{B}< t \leq T. 
        \end{array}\right.
\end{equation}
Let $\Sigma_{1,tr}$ be the lower triangular Cholesky decomposition of $\Sigma_1$ with positive diagonal elements and, as before, let $Q \in \mathcal{O}(n)$ be an orthonormal matrix. \cite{BBKM} show that the identification issue reduces to
\begin{equation}
\label{eigen decomposition}
\begin{array}{l}
A_0^{-1} = \Sigma_{1,tr} Q\\
\Sigma_{1,tr}^{-1} \Sigma_2 \Sigma_{1,tr}^{-1\prime} = Q \Lambda Q^{\prime},   
\end{array}
\end{equation}
where the second equation in the display is an eigen-decomposition problem of the identified matrix $\Sigma_{1,tr}^{-1} \Sigma_2 \Sigma_{1,tr}^{-1\prime}$, where $\Lambda$ collects the eigenvalues and the columns of $Q$ are the corresponding eigenvectors. 

Note that even with the sign-normalization of the eigenvectors imposed, the eigen-decomposition of a positive definite matrix is not unique since for any permutation matrix $P$, 
$\tilde{Q}=Q P$ and $\tilde{\Lambda}=P\Lambda P^\prime$ is also a solution for the eigen-decomposition problem. There are $n$ distinct decompositions if the eigenvalues in $\Lambda$ are all distinct, and this multiplicity corresponds to local identification of the structural parameters in $A_0$. This local identification issue corresponds to indeterminacy of labeling of the structural shocks, and to achieve global identification one needs to assume the knowledge of the ordering of the magnitudes of the volatility ratios among the structural shocks. 

The empirical macroeconomics literature discusses identification through the concepts of ``statistical identification'' and ``economic identification.'' The former refers to a unique decomposition of the reduced-form errors into structural shocks without pinning down their labels (i.e., economic interpretation), while the latter refers to a stronger concept of identification that determines not only the unique decomposition of the structural shocks but also of their labels. 
The concept of economic identification corresponds to our definition of global identification for $A_0$ and the subsequent economically meaningful impulse response analysis, whereas solely the statistical identification leads to local identification of $A_0$ only. 

As we discuss further details in Section \ref{sec:IdHSVAR}, it is common empirical practice to additionally impose identifying restrictions so as to reduce the number of admissible decompositions or to attain the economic identification. However, as we illustrate in our empirical application in Section \ref{sec:EmpApp}, a set of available restrictions can fail to yield a unique eigen-decomposition, resulting in $A_0$ and the impulse responses only locally identified.    


\subsection{Proxy-SVAR}
\label{sec:proxySVAR}

A set of identifying restrictions similar to the non-recursive zero restrictions discussed above can appear when the identification strategy exploits proxy variables for the structural shocks.

Consider again a three-variable SVAR. Instead of imposing zero restrictions directly on any element of $A_0$, we consider observable variables that proxy some of the underlying structural shocks. The idea of using proxy variables to identify the structural impulse responses has been considered in \citet{SW12} and \citet{MR2013}, amongst others. We restrict our analysis to SVARs and focus on identification of the full system of SVARs rather than subset identification of the impulse responses.\footnote{The proxy-variable identification strategy has been shown to be useful for non-invertible structural MA models. See \cite{SW18} and \cite{PMW19}.}    
To be specific, consider introducing the external variables $m_t=(m_{1t},m_{2t},m_{3t})'$, each of which acts as a proxy for some contemporaneous structural shocks. Following \citet{AF19}, \citet{ARW21}, and \citet{GKR19}, we augment $m_t$ into the original SVAR,
\begin{equation}
\begin{pmatrix} A_0 & O \\ \Gamma_{1} & \Gamma_2 \end{pmatrix} \begin{pmatrix} y_t \\ m_t \end{pmatrix} = \begin{pmatrix} \epsilon_t \\ \nu_t \end{pmatrix}, \mspace{15mu} (\epsilon_t, \nu_t)' \sim \mathcal{N}(0,I_{6 \times 6}), \label{augmented SVAR}
\end{equation} 
where $O$ is $3 \times 3$ matrix of zeros, $\Gamma_1$ and $\Gamma_2$ are $3 \times 3$ coefficient matrices in the augmented equations, and the shocks $\nu_t$ in the second block component of the augmented system are interpreted as measurement errors in the proxy variables. Inverting Eq. (\ref{augmented SVAR}) leads to 
\begin{equation}
m_t = - \Gamma_2^{-1} \Gamma_1 A_0^{-1} \epsilon_t + \Gamma_2^{-1} \nu_t. \label{mt equation}
\end{equation} 
In the Proxy-SVAR approach, the identifying restrictions are zero restrictions on the covariance matrix of $m_t$ and $\epsilon_t$.  Consider imposing the following restrictions:
\begin{equation}
E(m_t \epsilon_t') = \begin{pmatrix} 0 & \rho_{12} & \rho_{13} \\ \rho_{21} & 0 & \rho_{23} \\ \rho_{31} & \rho_{32} & 0 \end{pmatrix} \label{exogeneity}
\end{equation}
where $\rho_{ij}$ is the (unconstrained) covariance of $m_{it}$ and $\epsilon_{jt}$. The zero-covariance restrictions represented in Eq. (\ref{exogeneity}) imply that variable $m_{it}$, $i=1,2,3$, proxies a combination of the structural shocks \textit{excluding} $\epsilon_{it}$. Combining Eq. (\ref{mt equation}) with Eq. (\ref{exogeneity}) and substituting $A_0^{-1} = \Sigma_{tr} Q$, $Q = [q_1, q_2, q_3]$, the exogeneity restrictions of Eq. (\ref{exogeneity}) can be expressed as
\begin{align}
(e_1' \Gamma_2^{-1} \Gamma_1 \Sigma_{tr})q_1 & = 0, \notag \\ 
(e_2' \Gamma_2^{-1} \Gamma_1 \Sigma_{tr})q_2 & = 0, \label{exogeneity2} \\
(e_3' \Gamma_2^{-1} \Gamma_1 \Sigma_{tr})q_3 & = 0. \notag
\end{align}
Since $\Gamma_2^{-1}\Gamma_1$ can be identified by the covariance matrix of the reduced-form VAR errors in the augmented system (\ref{augmented SVAR}), the zero restrictions of Eq. (\ref{exogeneity2}) have the same form as Eq. (\ref{eq:DSGEsimmat}). Hence, Proxy-SVAR identification under the exogeneity restrictions Eq. (\ref{exogeneity}) delivers local but non-global identification of the $A_0$ matrix.\footnote{\label{footnote:proxyIV} \cite{AF19}, in their motivating example, provide a specification of a proxy-SVAR that is locally but not globally identified, being the imposed restrictions non-recursive. See Eq. (17), page 958, in their paper.}


\section{Theory of identification in SVAR models}
\label{sec:identif}

This section is dedicated to the theory of identification in SVAR models. We first consider classical SVARs, where identification is obtained through imposing restrictions on the parameter space. We discuss about the restrictions to impose, the conditions for checking global and local identification, the number of admissible solutions to expect (at most), and finally, through a simple example, we show to what extent the identification issue can be seen as a geometric problem. The last part of the section is devoted to the theory of local identification in heteroskedastic SVAR models (HSVARs). 


\subsection{Normalization, sign, zero and non-zero identifying restrictions}
\label{sec:restriction}


\vspace{0.7cm}
\noindent
\textit{Zero and non-zero equality restrictions}
\vspace{0.4cm}

\noindent
 The standard approach in the literature for obtaining identification of SVARs is to impose equality restrictions either on the structural parameters or particular linear and non-linear functions of them.\footnote{Alternative proposals of identification strategies, discussed in other sections of the paper, include the use of external instruments as in \cite{MR2013} and \cite{SW18}, heteroskedasticity of the structural shocks as in \cite{Rigobon03}, \cite{BF15} and \cite{Bacchiocchi17}, and the presence of non-normality as in \cite{LL10JBES} and \cite{LMS17JoE}.} 

Following RWZ, we represent identifying restrictions as restrictions on the reduced-form parameters $\phi$ and the column vectors $(q_1,\,q_2,\ldots,\,q_n)$ of the orthogonal matrix $Q$:

\begin{eqnarray}
\big((i,j)\text{-th element of }A_0^{-1}\big) = c & \Longleftrightarrow &(e_i^\prime \Chol)q_j=c,\label{eq:restr1}\\
\big((i,j)\text{-th element of }A_0\big) = c & \Longleftrightarrow &(\Choli e_j)^\prime q_i=c,\label{eq:restr2}\\
\big((i,j)\text{-th element of }A_l\big) = c & \Longleftrightarrow &(\Choli B_l e_j)^\prime q_i=c,\label{eq:restr3}\\
\big((i,j)\text{-th element of }CIR^\infty\big) = c & \Longleftrightarrow &\bigg[e_i^\prime \sum\limits_{h=0}^{\infty} C_h(B)\Chol\bigg]q_j=c,\label{eq:restr4}\\
\big(\text{linear restriction between }(i,j)\text{-th}\hspace{0.5cm}&&\nonumber\\
\text{and }(h,k)\text{-th elements of }A_0^{-1}\big) & \Longleftrightarrow &(e_i^\prime \Chol)q_j-d(e_h^\prime \Chol)q_k=c,\label{eq:restr5}\\
\big(\text{linear restriction between }(i,j)\text{-th}\hspace{0.5cm}&&\nonumber\\
\text{and }(h,k)\text{-th elements of }A_0\big) & \Longleftrightarrow &(\Choli e_j )^\prime q_i-d(\Choli e_k )^\prime q_h=c,\label{eq:restr6}
\end{eqnarray}
where $e_i$ is the $i$-th column of the identity matrix $I_n$, and $c$ and $d$ are known  scalars. Great part of the literature has dealt with zero restrictions, that, in our formulation corresponds, in all Eq.s (\ref{eq:restr1})-(\ref{eq:restr6}), to the case of $c=0$. 
For instance, RWZ and \cite{GK18} consider restrictions of the form Eq.s (\ref{eq:restr1})-(\ref{eq:restr4}), with $c=0$. Our representation of the restrictions is more general since we allow for non-zero restrictions, $c\neq 0$. 

Eq. (\ref{eq:restr1}) and Eq. (\ref{eq:restr2}) cover short-run identifying restrictions including the causal ordering restrictions of \citet{sims80} and \citet{Bernanke86}. Eq. (\ref{eq:restr3}) corresponds to restrictions on some of the right-hand side variables in the structural equations. Eq. (\ref{eq:restr4}) corresponds to long-run identifying restriction as considered in \citet{BlanQuah1}. The last two restrictions, in Eq.s (\ref{eq:restr5})-(\ref{eq:restr6}), correspond to constraints across the impulse responses and across the equations, respectively. In both cases, we consider the general case of possible non-zero restrictions, allowing for $c\neq 0$ (e.g., two coefficients in two distinct equations sum to unity).\footnote{Starting from Eq. (\ref{eq:restr5}) it is straightforward to extend the restrictions to $CIR^\infty$ or 
responses at any horizon. Similarly, starting from Eq. (\ref{eq:restr6}), we can also restrict the elements $A_1,\ldots,A_p$ across two or more equations. The general form to be given in Eq. (\ref{eq:GenFormRest}) can accommodate these two extensions.}



We represent these equality restrictions by
\begin{eqnarray}
        \textbf{F}(\phi,Q) & \equiv &
        \left(
        \begin{array}{cccc}
                F_{11}(\phi) & F_{12}(\phi) & \cdots & F_{1n}(\phi)\\
                F_{21}(\phi) & F_{22}(\phi) & \cdots & F_{2n}(\phi)\\
                \vdots & \vdots & \ddots & \vdots\\
                F_{n1}(\phi) & F_{n2}(\phi) & \cdots & F_{nn}(\phi)
        \end{array}
        \right)
        \left(
        \begin{array}{c}
                q_1\\
                q_2\\
                \vdots\\
                q_n
        \end{array}
        \right)-
        \left(
        \begin{array}{c}
                c_1\\
                c_2\\
                \vdots\\
                c_n
        \end{array}
        \right)=\bf{0}\nonumber\\
        &&\nonumber\\
        & \equiv & \textbf{F}(\phi)\text{vec}\, Q-\bf{c}=\bf{0} \label{eq:GenFormRest}
\end{eqnarray}
where $F_{ij}(\phi)$, $1 \leq i , j \leq n$, is a matrix of dimension $f_{i}\times n$,  which depends only on the reduced-form parameters $\phi=(B,\,\Sigma)$. The dimension of $\textbf{F}(\phi)$ is $f\times n^2$, where $f=f_1+\cdots+f_n$ denotes the total number of restrictions imposed. We allow $f_i = 0$ for some $i$, in which case the $i$-th block row in $\textbf{F}(\phi)$ is null. Finally, $\text{vec}\, Q\equiv (q_1^\prime,\,\ldots q_n^\prime)^\prime$ is the vectorization of $Q$, and $ \textbf{c} \equiv (c_1^\prime\,\ldots,c_n^\prime)^\prime$ is a vector of known constants with length $f$, where each $c_i$ is a $f_i \times 1$ vector.

If $F_{ij}(\phi)=0$ for all $i\neq j$, there are no cross equation restrictions or restrictions across the effects of the shocks. If $c_i=0$ for all $i$, then only zero restrictions are imposed. This representation of the identifying restrictions is in line with \cite{Hamilton94}, \cite{LutBook06} and \cite{BL18}, all of which allow non-zero and across-shock restrictions. We provide the following formal definitions.

\begin{defin}[Triangular restrictions]
\label{def:tri}
        Referring to Eq. (\ref{eq:GenFormRest}), the restrictions are said to be \textit{triangular} if $F_{ij}(\phi)=0$ for $j>i$, and $f_i=n-i$, for $i=1,\ldots,n$. 
\end{defin}

\begin{defin}[Homogeneous and non-homogeneous restrictions]
\label{def:homo}
        Referring to Eq. (\ref{eq:GenFormRest}), the restrictions are said to be \textit{homogeneous} if $\bf{c}=\bf{0}$ and \textit{non-homogeneous} if $\bf{c}\neq\bf{0}$.
\end{defin}

\begin{defin}[Recursive restrictions]
\label{def:rec}
        Referring to Eq. (\ref{eq:GenFormRest}), the restrictions are said to be \textit{recursive} if the restrictions are triangular and homogeneous.
\end{defin}

Defined in this way, triangular restrictions pin down a unique ordering of the variables with $\textbf{F}(\phi)$ in Eq. (\ref{eq:GenFormRest}) becoming a lower-triangular block matrix. Otherwise, our framework allows for the ordering of variables to be non-unique. Even with the order of variables fixed, if the restrictions include across-shock restrictions, then Eq. (\ref{eq:GenFormRest}) allows for a multiple block-matrix representation. The general identification results of this section are valid independent of how the variables are ordered or how the imposed restrictions are represented within Eq. (\ref{eq:GenFormRest}), unless the triangular structure is assumed explicitly.
%

$\textbf{F}(\phi)$ can be decomposed into the product of two matrices: a) a selection matrix $Z$, and b) a transformation of the reduced-form parameters $z(\phi)$. In this direction, following \cite{ARW18}, consider a mapping between the standard parametrization of the structural form, as in Eq. (\ref{eq:SVAR}), and the so called orthogonal reduced-form parametrization. Let $g$ be a function with domain the set of all structural parameters $A$ defined as
\begin{equation}
    \label{eq:OrtRedForm}
    g(A_0,A_+) = \bigg(\underbrace{A_0^{-1}A_+}_{B}, \underbrace{A_0^{-1}A_0^{-1^\prime}}_{\Sigma}, \underbrace{\Big[\mathrm{Chol}\big(A_0^{-1}A_0^{-1^\prime}\big)\Big]^\prime A_0^\prime}_{Q}\bigg)
\end{equation}
where $\mathrm{Chol}(\cdot)$ is a function extracting the Cholesky factor of a symmetric and positive definite matrix.\footnote{In this paper we consider the Cholesky factor, but any differentiable decomposition can be used, as in \cite{ARW18}.} This function is invertible and its inverse is given by
\begin{equation}
    \label{eq:OrtRedFormInv}
    g^{-1}(B,\Sigma,Q) = \bigg(\underbrace{Q^\prime \Stri}_{A_0}, \underbrace{Q^\prime\Stri B}_{A_+}\bigg).
\end{equation}
The common practice of imposing linear restrictions on a function of the structural parameters, say $h(A_0,A_+)$, with domain the set of structural parameters and codomain the set of $n\times k$ matrices, for some $k>0$, can be restated as follows
\begin{equation}
    \label{eq:RestrFunction}
    Z_j\bigg[h\Big(g^{-1}(\Sigma,B,Q)\Big)\bigg]^\prime e_j =
    Z_j\bigg[h\Big(g^{-1}(\Sigma,B,I_n)\Big)\bigg]^\prime Q e_j = c_j,     
\end{equation}
with $Z_j$ a simple selection matrix, of dimension $f_j\times k$, selecting the elements of $\big[h(\,\cdot\,)\big]^\prime$ to constrain. For obtaining the result in Eq. (\ref{eq:RestrFunction}) we have used the following condition on the transformation function $h$:

\begin{condition}
    \label{cond:adm}
        The transformation $h(\cdot)$, with the domain $\mathcal{A}$, is admissible if and only if for any $Q\in\O$ and $(A_0,A_+)\in \mathcal{A}$, $h(Q^\prime A_0,Q^\prime A_+)=Q^\prime h(A_0,A_+)$. 
\end{condition}

Moreover, as we allow for restrictions across the different columns of $Q$, we collect the constraints all together and obtain
\begin{equation}
    \label{eq:RestrFunctionAll}
    Z\bigg[I_n\otimes h\Big(g^{-1}(\Sigma,B,I_n)\Big)\bigg]^\prime 
    \left(\begin{array}{c}q_1\\\vdots\\q_n\end{array}\right)  = \textbf{c}
    \hspace{0.5cm}\Longleftrightarrow \hspace{0.5cm} 
    Z\Big( I_n \otimes z( \phi ) \Big)^\prime
    \left(\begin{array}{c}q_1\\\vdots\\q_n\end{array}\right) - \textbf{c} = \textbf{0},
\end{equation}
where it is emphasized to what extent the matrix $F(\phi)$ in Eq. (\ref{eq:GenFormRest}) can be seen as the product of an $f\times nk$ selection matrix $Z$, collecting all the $Z_j$ matrices on the main diagonal and further known numbers (in general $1$ or $-1$) for possible restrictions across shocks, and a transformation of the reduced-form parameters, $z(\phi)=h\Big(g^{-1}(\Sigma,B,I_n)\Big)$. 

Finally, we introduce the following technical condition that will be used in the next sections for proving some of our results.

\begin{condition}
    \label{cond:reg}
    The transformation $h(\cdot)$, with the domain $\mathcal{A}$, is regular if and only if $\mathcal{A}$ is open and $h$ is continuously differentiable with $h^\prime (A_0,A_+)$ of rank $kn$ for all $(A_0,A_+)\in \mathcal{A}$. 
\end{condition}

Since all the restrictions considered above in Eq.s (\ref{eq:restr1})-(\ref{eq:restr6}) concern the elements of $A_0$, $A_0^{-1}$, or $A_+$, or the linear transformations thereof, these two conditions are satisfied as far as $\mathcal{A}$ is defined as an open subset of nonsingular matrices.


\vspace{0.7cm}
\noindent
\textit{Sign Normalization restrictions}
\vspace{0.4cm}

\noindent
Let $D$ be any diagonal matrix with entries of either $+1$ or $-1$. Since $D$ is an orthogonal matrix, if $(A_0,A_+)$ satisfies the restrictions in Eq.s (\ref{eq:restr1})-(\ref{eq:restr6}), then by Condition \ref{cond:adm}, $(DA_0,DA_+)$ will satisfy the restrictions, too. Our identification analysis imposes sign normalization restrictions. As in RWZ, we impose the following general form of sign normalization:\footnote{\cite{WZ03} and \cite{HWZ07} discuss about the importance and impact of normalization on the statistical inference, and provide practical strategies for choosing the adequate one.}

\begin{defin}[Sign normalization]
\label{def:SignNorm}
        A normalization rule can be characterized by a set $N\in \mathcal{A}$ such that for any structural parameter point $(A_0,A_+)\in \mathcal{A}$, there exists a unique $n\times n$ diagonal matrix $D$ with plus or minus one on the main diagonal such that $(DA_0,DA_+)\in N$, or equivalently, $(DQ^\prime\Stri,DQ^\prime\Stri B)\in N$.
\end{defin}


\vspace{0.7cm}
\noindent
\textit{Sign restrictions}
\vspace{0.4cm}

\noindent
In addition to equality restrictions, sign restrictions can be imposed on impulse responses or structural parameters. These sign restrictions can be seen as additional constraints on the columns of the $Q$ matrix. Suppose we impose $s_{h,i} \leq n $ number of sign restrictions on the impulse responses to \textit{i}-th shock at \textit{h}-th horizon. They can be expressed as
\begin{equation}
\label{eq:SignRestr_ih}
S_{h,i}(\phi)q_i\,\geq\,\textbf{0},
\end{equation}
where $S_{h,i}\equiv D_{h,i}\,C_h(B) \Chol$ is a $s_{h,i}\times n$ matrix, $D_{h,i}$ is the $s_{h,i}\times n$ signed selection matrix, which indicates by $1$ ($-1$) the impulse responses whose signs are restricted to being positive (negative), and $C_h(B)$ is from the definition of an impulse response as in Eq. (\ref{eq:IRh}). The inequality in Eq. (\ref{eq:SignRestr_ih}) is component-wise. Sign restrictions on structural parameters are linear inequality constraints on the columns of the matrix $Q$, so can also be accommodated. Stacking all the $S_{h,i}$ matrices involving sign restrictions on $q_i$ at different horizons into a matrix $S_i$, we have
\begin{equation}
\label{eq:SignRestr_i}
S_{i}(\phi)q_i\,\geq\,\textbf{0}.
\end{equation}
We represent the set of all sign restrictions by 
\begin{equation}
\label{eq:SignRest}
\textbf{S}(\phi,Q)\geq \textbf{0}. 
\end{equation}

\vspace{0.7cm}
\noindent
\textit{Admissible structural parameters and identified set}
\vspace{0.4cm}

\noindent
Given identifying restrictions of the form introduced above, we hereafter let $R$ be the collection of restrictions $ \big\{\textbf{F}(\phi,Q)=0, \textbf{S}(\phi,Q) \geq 0, (DQ^\prime\Stri,DQ^\prime\Stri B)\in N\big\}$, or $R = (F,S) $ for short. We call $A=(A_0,A_+)$ \textit{admissible} if it satisfies $R$. The set of all these admissible structural parameters can be represented by
\begin{equation}
\label{eq:Arestr}
        \Ar\equiv\Big\{(A_0,A_+)=(Q^\prime\Choli,Q^\prime\Choli B)\in \mathcal{A} \cap N \,:\,Q\in\O,\,\:\textbf{F}(\phi,Q)=\textbf{0},\,\:\textbf{S}(\phi,Q)\geq\textbf{0}\Big\}\nonumber.
\end{equation}
The projection of $\Ar$ for $A_0$ gives $\mathcal{A}_{R,0}(\phi)$ as defined in Eq. (\ref{eq:Ar0(phi)}).
The identified set for $Q$ is defined as the set of admissible orthogonal matrices given the reduced-form parameters:
\begin{equation}
\label{eq:Qrestr}
\Qr\equiv\Big\{Q\in\O\::\,\:\textbf{F}(\phi,Q)=\textbf{0},\,\:\textbf{S}(\phi,Q)\geq\textbf{0},
\,\:(DQ^\prime\Stri,DQ^\prime\Stri B)\in N\Big\}\nonumber.
\end{equation}

The objects of interest may also include transformations of structural parameters such as impulse response functions. We denote a scalar parameter of interest by $\eta=\eta(\phi,Q)$ and define its identified set as
\begin{equation}
\label{eq:IS}
IS_{\eta}(\phi) \equiv\Big\{\eta(\phi,Q)\::\:Q\in\Qr\Big\}\nonumber.
\end{equation}
When $\eta(\phi,Q)$ is an impulse response, then
\begin{equation}
\label{eq:Trasf}
    \eta(\phi,Q) = IR_{ij}^h=e_i^\prime C_h(B)\Chol Q e_j \equiv c_{ih}^\prime(\phi)\,q_j, \nonumber
\end{equation}
where $IR_{ij}^h$ is the (\textit{i,j})-th element of $IR^h$ and $c_{ih}^\prime(\phi)$ is the \textit{i}-th row of $C_h(B)\Chol$. 


When $A$ is globally identified, $IS_{\eta}(\phi)$ is a singleton for almost every $\phi \in \Phi_R$. If $A$ is only locally identified, $IS_{\eta}(\phi)$ can be a set of multiple isolated points generated by observationally equivalent structural parameters. Local identification can be viewed as a special case of set identification, although it is not covered by standard set identification analysis where the identified set is typically an interval or a set with positive Lebesgue measure.


\subsection{Conditions for local identification}
\label{sec:IdentCond}

This section presents conditions for global and local identification when the identifying restrictions are equality restrictions given in the form Eq. (\ref{eq:GenFormRest}). In the case of local identification, we present an analytical characterization of the number of observationally equivalent structural parameter values. 

We begin with the well known condition for global identification developed in Theorem 7 of RWZ and recently extended in \cite{BKglob20}.\footnote{Theorem 7 of \cite{RWZ10} claims that under a set of regularity conditions, the exact identification of an SVAR holds if and only if $f_i = n - i$ for all $i=1, \dots, n$. \citet{BKglob20} show that relaxing one of their regularity conditions, the condition of $f_i = n - i$ for all $i=1, \dots, n$ is no longer sufficient and it needs to be augmented by rank conditions, which, in the current setting, is equivalent to Eq. (\ref{eq:prop1i=1}) and Eq. (\ref{eq:NonRed}). See \cite{BKglob20} for further details. } This condition for global identification acts as a reference point in our discussion of local identification.

\begin{prop}[Necessary and sufficient condition for global identification, RWZ and \cite{BKglob20}]
\label{prop:GlobIdent}
    Consider an SVAR with identifying restrictions of the form Eq. (\ref{eq:restr1}) - Eq. (\ref{eq:restr6}) collected in $\textbf{F}(\phi,Q)$. Assume the restrictions to be recursive according to Definition \ref{def:rec}, i.e. $F_{ij}(\phi)=0$ for $i\neq j$, and $\textbf{c}=\textbf{0}$.     
    
    \noindent  
    The SVAR is globally identified at $A=\left(A_0,A_+\right) \in \AR$ if and only if the following conditions hold at $\phi$ implied by $A$:
    \begin{enumerate}
    \item It holds
        \begin{equation}
            \rk\big(F_{11}(\phi)^\prime, \tilde{\sigma}_1\big) = n. \label{eq:prop1i=1}
        \end{equation}
    \item Let $q_1$ be a unit length vector satisfying $F_{11}(\phi)\,q_1=0$ and the sign normalization restriction, which is unique under Eq. (\ref{eq:prop1i=1}). 
        For $i=2,\ldots,n$
        \begin{equation}
            \label{eq:NonRed}
            \rk\big(F_{ii}(\phi)^\prime, q_1, \dots, q_{i-1}, \tilde{\sigma}_i\big) = n, 
        \end{equation}  
        holds, where the orthonormal vectors $q_2,\ldots,q_n$ solve                 
        \begin{equation}
            \label{eq:q2qn}
            \big(F_{ii}(\phi)^{\prime}, q_1, \dots, q_{i-1}\big)'q_i= \bf0
        \end{equation}   
        sequentially, and satisfy the sign normalization restrictions.
    \end{enumerate}
\end{prop}
 
This proposition characterizes a boundary separating cases where an SVAR is globally identified and cases where it is not guaranteed to be globally identified. In what follows, we consider departures from this proposition's conditions for global identification, and show implications for local identification and the failure of global identification. 
Specifically, we allow for non-recursive restrictions as stated in Definition \ref{def:rec}, and for non-homogeneous ones as in Definition \ref{def:homo}. In addition, we can also allow across-equations restrictions of the form Eq. (\ref{eq:restr5})- Eq. (\ref{eq:restr6}). 
With this expanded set of identifying restrictions, Proposition \ref{prop:LocIdent} derives a rank condition that is necessary and sufficient for local identification. \cite{LutBook06}, \citeauthor{Hamilton94} (\citeyear{Hamilton94}, page 334) and \cite{BL18} provide similar conditions for local identification in a setting that is less general in terms of the kind of restrictions that can be imposed.\footnote{\label{footnote_CEE} \cite{CEE99} discuss a rank condition for local identification for SVAR models subject to zero restrictions on the elements of $A_0$. Our approach is more general than theirs by allowing for non-homogeneous restrictions and restrictions on objects other than $A_0$.} Their rank condition is 
expressed in terms of the structural parameter matrices $A$, while our Proposition \ref{prop:LocIdent} presents the rank condition in terms of the coefficient matrix of the equality restrictions \textit{$\textbf{F}(\phi)$} and the orthogonal matrix $Q$.

\begin{prop}[Rank condition - necessary and sufficient condition for local identification]
\label{prop:LocIdent}
        Consider an SVAR with equality restrictions of the form Eq. (\ref{eq:restr1}) - Eq. (\ref{eq:restr6}) collected in $\textbf{F}(\phi,Q)$. 
        Let $\tilde{D}  _n$ be the $n^2 \times n(n-1)/2$ full-column rank matrix such that for any $n(n-1)/2$-dimensional vector $v$, $\tilde{D}_n\,v \equiv \text{vec }(H)$ holds, 
        where $H$ is an $n\times n$ skew-symmetric matrix satisfying $H = -H^\prime$ 
        (see Appendix D for the specific construction of $\tilde{D}_n$ for $n=2,3,4$). 

        \noindent (i) The SVAR is locally identified at $A=\left(A_0,A_+\right) \in \AR$ if and only if
        \begin{equation}
        \label{eq:RankCond}
                \rk \:\bigg[\textbf{F}(\phi)\big(I_n\otimes Q\big)\tilde{D}_n\bigg] = n(n-1)/2
        \end{equation}
        holds, where the reduced-form parameters $\phi=(B,\,\Sigma)\in \Phi$ and the orthogonal matrix $Q\in \Qr$ are such that         $(B,\,\Sigma,\,Q)=g(A_0,A_+)=\Big(A_0^{-1}A_+,A_0^{-1}A_0^{-1\prime},Chol(A_0^{-1}A_0^{-1\prime})^\prime A_0^\prime\Big)$. 
        Hence, a necessary condition for the rank condition Eq. (\ref{eq:RankCond}) is $f = \sum_{i=1}^{n} f_i \geq n(n-1)/2$.
 
        \noindent (ii) Let $\mathcal{K}$ be the set of structural parameters in $\AR$ satisfying the rank condition of Eq. (\ref{eq:RankCond}), 
        \begin{equation}
                \mathcal{K} \equiv \left\{ A \in \AR : \rk \:\bigg[ \textbf{F}(\phi)\big(I_n\otimes Q\big)\tilde{D}_n\bigg] = 
                n(n-1)/2 \right\}. \notag
        \end{equation}
        If Conditions \ref{cond:adm} and \ref{cond:reg} hold, then either $\mathcal{K}$ is empty or the complement of $\mathcal{K}$ in $\AR$ is of measure zero in $\mathcal{A}_R$.
\end{prop}

\begin{proof} See Appendix \ref{app:Proofs}. \end{proof}

Statement (i) of this proposition provides a necessary and sufficient condition for local identification at a given $A \in \AR$ in the form of a rank condition for a matrix that is a function of $A$, i.e., $(\phi,Q)$ is a function of $A$. 
Importantly, in contrast to Proposition \ref{prop:GlobIdent}, we do not impose any constraints on $\mathbf{F}$ or $\mathbf{c}$, i.e, they are allowed to be neither triangular nor homogeneous. 
Eq. (\ref{eq:RankCond}) as stated, however, offers limited practical implication since it only confirms the SVAR to be identified at a particular $A \in \AR$ and remains silent about whether such point-identified $A$ is common or rare in $\AR$. 

Statement (ii) of this proposition makes the rank condition Eq. (\ref{eq:RankCond}) useful by showing that it holds either nowhere or almost everywhere in the constrained parameter space $\mathcal{A}_R$. This means that, similar to the proposals following Theorem 3 in RWZ and Theorem 1 in \cite{BL18}, one can assess local identification by randomly generating structural parameters $A \in \mathcal{A}_R$ and checking whether the rank condition holds or not. Specifically, we can consider drawing reduced-form parameters $\phi \in \Phi_R$ from its prior or posterior and solving a constrained non-linear optimization problem of the form\footnote{This minimization problem is constrained by the orthogonality constraints $Q'\,Q = I_n$, which is known as Stiefel manifold following \cite{Stiefel}. \cite{EAT98} develop algorithms for optimization in the Stiefel manifold, while \cite{Manopt} propose a Matlab toolbox for optimization on manifolds including the Stiefel one. A Matlab code for this optimization is available from the authors upon request.}
\begin{equation}
    \label{eq:RandQ}
    \begin{array}{l}
    \text{arg }\underset{Q\in\:\Re^{n^2}}{\text{min}}\:\: \bigg( \textbf{F}(\phi)\text{vec}\, Q-\textbf{c}\bigg)^\prime \bigg( \textbf{F}(\phi)\text{vec}\, Q-\textbf{c}\bigg)\\
    \text{s.t. } \text{ diag }(Q^\prime\Choli)\geq 0, \:\:\textbf{S}(\phi,Q)\geq \textbf{0} \text{ and } Q^\prime Q=I_n.
    \end{array}
\end{equation}
If the value of the optimization is zero, then the obtained $Q$ is an admissible orthogonal matrix at the given $\phi$. If such an admissible $Q$ satisfies the rank condition in Eq. (\ref{eq:RankCond}), then the SVAR is locally identified at $(\phi,Q)$. 
If the rank condition is not met, the SVAR is not locally identified at $(\phi,Q)$. 
Proposition \ref{prop:LocIdent} (ii) says that only one of the two possibilities occurs with positive measure, while the other has zero measure. Hence, by checking the rank condition at a parameter value drawn from a probability distribution supporting $\mathcal{A}_R$ or $\Phi_R$, we can learn whether the rank condition holds nowhere or almost everywhere on the restricted space of structural parameters. 

The next corollary rephrases the rank condition in Proposition \ref{prop:LocIdent} in terms of the restrictions imposed directly on a function of the structural parameters, $h(A_0,A_+)$.

\begin{corol}
    \label{corol:RankCond}
    Under the assumptions in claims (i) and (ii) of Proposition \ref{prop:LocIdent}, the SVAR is locally identified at the parameter point $(A_0,A_+)\in \AR$ if and only if
    \begin{equation}
        \label{eq:RankCondCorol}
        \rk \:\bigg[Z\Big(I_n\otimes h(A_0,A_+)^\prime\Big)\tilde{D}_n\bigg] = n(n-1)/2.
    \end{equation}
\end{corol}

\begin{proof} 
    See Appendix \ref{app:Proofs} and, specifically, the proof of Proposition \ref{prop:LocIdent}.
\end{proof}

The corollary, together with the point (ii) in Proposition \ref{prop:LocIdent}, adds practicality to the check of identification, avoiding the solution of the optimization problem in Eq. (\ref{eq:RandQ}). A natural algorithm could be to a) randomly generate values for $h(A_0,A_+)$ satisfying the restrictions, and then b) check for the rank condition in Eq. (\ref{eq:RankCondCorol}).
If the condition holds, the SVAR is locally identified almost everywhere. On the contrary, we repeat the two steps either until the rank is satisfied, or until a maximum number of random draws is reached. If in all cases the rank is deficient, we have reasonable confidence that the SVAR is not identified. 

If the rank condition of Proposition \ref{prop:LocIdent} or Corollary \ref{corol:RankCond} fails, we generally have set-identification for $A$ and the impulse responses, i.e., there is a  continuum of observationally equivalent structural parameters given $\phi$. Such implied condition for set-identification is more general than the sufficient condition for set identification of SVARs shown in \cite{ARW18} and \cite{GK18}.

In many empirical applications the interest is only on one single shock, or at most on a small subset of them. The condition in Proposition \ref{prop:LocIdent} could be, thus, too stringent. 
The next result allows to check whether a subset of shocks is locally identified. First of all, without loss of generality, let the shock be ordered such that $f_1\geq f_2\geq \ldots\geq f_n$. 
Conditional on this ordering, let the shocks be partitioned into two groups such that: a) the shocks of interest belong to the first group of $s$ shocks, and b) the shocks in the former group do not present restrictions with those of the latter. These features allow to write the restrictions as
\begin{equation}
\label{eq:RestrSubset}
        \left(\begin{array}{cc}\textbf{F}_{11}(\phi) & 0 \\ 0 & \textbf{F}_{22}(\phi)\end{array}\right)
        \left(\begin{array}{c}\textbf{q}_1\\\textbf{q}_2\end{array}\right)-\left(\begin{array}{c}\textbf{c}_1\\\textbf{c}_2\end{array}\right) = \textbf{0}
\end{equation}
where the orthogonal matrix $Q$ has been partitioned as $Q = \big[\:Q_1\:|\:Q_2\:\big]$, with $\textbf{q}_1=\ve\,Q_1$ and $\textbf{q}_2=\ve\,Q_2$. 

\begin{corol}[Rank condition - necessary and sufficient condition for local identification of a subset of shocks]
\label{prop:LocIdentSubset}
        Consider an SVAR with equality restrictions of the form Eq. (\ref{eq:restr1}) - Eq. (\ref{eq:restr6}) collected as in Eq. (\ref{eq:RestrSubset}), 
        where the shocks are ordered such that $f_1\geq f_2\geq \ldots\geq f_n$.
        A necessary and sufficient condition for the first $s$ shocks to be locally identified at the parameter point $A=\left(A_0,A_+\right) \in \mathcal{A}_R$ is that
        \begin{equation}
            \label{eq:RankCondSubset}
            \rk \:\left(\begin{array}{c}\textbf{F}_{11}(\phi)\\ N_{ns}\left(I_n\otimes Q_1^\prime\right)\end{array}\right) = ns
        \end{equation}
        where $N_{ns}\equiv 1/2(I_{ns}+K_{ns})$ is a known matrix with $K_{ns}$ being the commutation matrix.\footnote{A commutation matrix $K$ is defined such that, given a generic matrix $A$, then $K\,\ve\,A = \ve\,A^\prime$. See \cite{mn07} for some properties of the commutation matrix.}
\end{corol}

\begin{proof} See Appendix \ref{app:Proofs}. \end{proof}

The implementation of the rank condition in Corollary \ref{prop:LocIdentSubset} can be easily performed by a slight modification of the procedure used for the general condition in Proposition \ref{prop:LocIdent}. We first derive an admissible matrix $Q_1$ by solving a constrained non-linear optimization problem as in Eq. (\ref{eq:RandQ}), where $Q_1$ substitutes the entire $Q$ and $\textbf{c}_1$ the entire $\textbf{c}$. The second step, thus, consists in checking the rank condition in Eq. (\ref{eq:RankCondSubset}) using the obtained $Q_1$ matrix.

Allowing only triangular identifying restrictions as defined in Definition \ref{def:tri}, the next proposition provides a simple necessary and sufficient condition for the rank condition of Proposition \ref{prop:LocIdent} (i). In contrast to Proposition \ref{prop:GlobIdent}, it obtains a necessary and sufficient condition for local identification with no restrictions imposed on $\mathbf{c}$.

\begin{prop}[Necessary and sufficient condition for local identification in triangular SVARs]
        \label{prop:SuffLocIdent}
        Consider an SVAR with triangular identifying restrictions according to Definition \ref{def:tri}. Let 
        $\tilde{F}_{ii}(\phi) = F_{11}(\phi)$ for $i=1$, and 
        \begin{equation}
            \label{eq:Fiit}
                \tilde{F}_{ii}(\phi)=\big(F_{ii}^{\prime}(\phi),q_1, 
                \dots, q_{i-1} \big)'  
        \end{equation}
        for $i=2,\ldots,n$, where $q_1,\ldots,q_i$ are the first $i$ column vectors of $Q\in\O$ satisfying the equality restrictions $\textbf{F}(\phi)\text{vec}\, Q-\bf{c}=\bf{0}$ given $\phi \in \Phi_R$. The rank condition of Eq. (\ref{eq:RankCond}) holds at $(\phi,Q)$ if and only if $\rk\big(\tilde{F}_{ii}(\phi)\big) = n-1$ holds for all $i=1,\ldots,n$.
\end{prop}

\begin{proof} See Appendix \ref{app:Proofs}. \end{proof}

Since the rank condition of Proposition \ref{prop:LocIdent} (i) is necessary and sufficient for local identification, the condition shown in Proposition \ref{prop:SuffLocIdent} is also necessary and sufficient for local identification for SVARs under triangular identifying restrictions. Moreover, if Conditions \ref{cond:adm} and \ref{cond:reg} hold, the claim of Proposition \ref{prop:LocIdent} (ii) carries over to the setting of Proposition \ref{prop:SuffLocIdent}, so knowing that the condition shown in Proposition \ref{prop:SuffLocIdent} holds at one or a few $\phi \in \Phi_R$ drawn from its prior or posterior allows us to conclude local identification holds almost everywhere in the parameter space. The condition in Proposition \ref{prop:SuffLocIdent} exploits sequential determination of $q_i$, $i=1, \dots, n$, given $\phi$, so checking it does not require nonlinear optimization for $Q$.\footnote{\label{footnote:Qconstruction} To see this, let $i=1$ and consider rank condition $\rk\big(F_{11}(\phi)\big) = n-1$. If this is met, we solve the non-homogeneous system of equations $F_{11}(\phi)q_1 = c_1$ for $q_1$. Solutions of the system can be represented as $q_1=c_1+F_{11,\perp}(\phi)z$ for some $z$ with appropriate dimension, where $F_{11,\perp}(\phi)$ is the null space of $F_{11}(\phi)$. If the rank condition is met, $F_{11,\perp}(\phi)$ is a vector and $z$ is a scalar. Moreover, as $q_1$ must have unit length, then $\big(c_1+F_{11,\perp}(\phi)z\big)^\prime \big(c_1+F_{11,\perp}(\phi)z\big)=1$ forms a simple univariate quadratic equation, whose two solutions give the two admissible $q_1$ vectors. They will be used to check the rank condition for $i=2$. If the rank condition $\rk\big(\tilde{F}_{22}(\phi)\big) = n-1$ holds, then we proceed similarly to the $i=1$ case to obtain at least two admissible values of $q_2$ by solving the non-homogeneous system of equations of the form $F_{12}(\phi)q_1+F_{22}(\phi)q_2=c_2$, where $q_1$ is given from the initial step. We repeat this process for further $i$ until $i=n$.}

The proof of Proposition \ref{prop:SuffLocIdent} leads to the following corollary showing a necessary and sufficient condition for the local identification of impulse responses to a particular shock.

\begin{corol}[Necessary and Sufficient condition for local identification of the \textit{j}-th shock in triangular SVARs]
\label{corol:ParId}
        Under the assumptions of Proposition \ref{prop:SuffLocIdent}, the impulse responses for the j-th structural shock, $1 \leq j \leq n$, are locally identified
        at the parameter point $A=\left(A_0,A_+\right) \in \mathcal{A}_R$ if and only if $\rk(\tilde{F}_{ii}(\phi)) = n-1$ holds for all $i=1,\ldots,j$.
\end{corol}


\subsection{The number of observationally equivalent parameter points}
\label{sec:IdentNumb}

The results presented so far are silent about how many observationally equivalent structural parameter points there are. As the next proposition shows, our constructive identification argument through the orthogonal matrix $Q$ allows us to characterize the number of observationally equivalent parameter points.

\begin{prop}[Number of locally identified points]
\label{prop:LocVsLack}
        Consider an SVAR with equality restrictions of the form Eq. (\ref{eq:restr1})-Eq. (\ref{eq:restr6}) collected in $\textbf{F}(\phi,Q) = \bf{0}$. Given $\phi \in \Phi$ and provided that the rank condition in Eq. (\ref{eq:RankCond}) is met, the number of admissible $Q$ matrices ($Q$ matrices solving $\textbf{F}(\phi,Q)=\bf{0}$) is zero or  finite.
        In particular, if the equality identifying restrictions are triangular, the number of admissible $Q$ matrices is at most $2^n$. If the equality identifying restrictions are non-triangular, the number of admissible $Q$ matrices is at most $2^{n(n+1)/2}$.
\end{prop}

\begin{proof} See Appendix \ref{app:Proofs}. \end{proof}

The proposition provides an upper bound for the number of locally identified observationally equivalent parameter points. It corresponds to the maximal number of modes that the likelihood of the structural parameters can have. The maximum number of observationally equivalent structural parameters is considerably lower when the SVAR is identified through triangular equality restrictions rather than non-triangular restrictions. The intuition for this result is that, if the identification of the columns of $Q$ can be performed recursively, the equations concerning the orthogonality conditions among the columns of $Q$ are linear, rather than quadratic. 

In comparison to the exact (global) identification case of RWZ and Proposition \ref{prop:GlobIdent}, Proposition \ref{prop:LocVsLack} highlights that non-homogenous restrictions ($\bf c \neq0$) lead to the possibility that, given $\phi \in \Phi$, (i) an admissible $Q$ does not exist, or (ii) the admissible $Q$ is no longer unique. Adding sign restrictions to the sign normalization restrictions can reduce the number of admissible $Q$'s, but cannot generally guarantee uniqueness of the admissible $Q$'s. Section \ref{sec:geo} below illustrates the transition from exact global identification to local identification through a simple example. 

The following corollary, instead, focuses on the identification of a subset of all the structural shocks and derives the maximum number of
solutions for this specific case. We consider, thus, the same situation as in Corollary \ref{prop:LocIdentSubset} and assume that the 
rank condition therein is met. 

\begin{corol}[Number of locally identified points for a subset of shocks]
\label{prop:LocVsLackSubset}
        Consider an SVAR with equality restrictions of the form Eq.s (\ref{eq:restr1})-(\ref{eq:restr6}) collected as in Eq. (\ref{eq:RestrSubset}). Let $s$ indicates the set of shocks of interest. Given $\phi \in \Phi$ and provided that the rank condition in Eq. (\ref{eq:RankCondSubset}) is met, the number of admissible $Q_1$ matrices is zero or finite. In particular, the number of admissible $Q_1$ matrices is at most $2^{s(s+1)/2}$.
\end{corol}

\begin{proof} The result is a by-product of Proposition \ref{prop:LocVsLack} and can be easily derived from its proof. 
\end{proof}


\subsection{The geometry of identification}
\label{sec:geo}

We present an intuitive geometric exposition for why the introduction of nonhomogeneous restrictions Eq. (\ref{eq:restr5}) and/or across-shock restrictions Eq. (\ref{eq:restr6}) can lead to local identification. This exposition also provides intuition for the number of local identified parameter points shown in Proposition \ref{prop:LocVsLack}. Appendix A provides the algebraic analysis behind our geometric discussion.

To make exposition as simple as possible, consider a bivariate VAR with a single non-homogeneous identifying restriction imposed on the structural parameters:
\begin{equation}
\label{eq:biVARconstr}
(A_0)_{[1,1]}^{-1}=c \:\Longleftrightarrow\: (e_1^\prime \Chol)\,q_1 = c
\end{equation}
where $c>0$ is a known (positive) scalar and $e_1$ is the first column of $I_2$. Denoting the first column of $\Chol^\prime=\left(\begin{array}{cc} \sigma_{1,1}&\sigma_{2,1}\\0&\sigma_{2,2}\end{array}\right)$ by $\sigma_1 = (\sigma_{1,1},0)'$,  this identifying restriction can be written as $\sigma_1^\prime\,q_1=c$. 
\begin{figure}
\caption{Identification of $q_1$ in the bivariate SVAR with non-zero restriction.}
\label{fig:Baby1}
\begin{center}
\begin{tikzpicture}[scale=2.5]
\draw [->] (0,0) -- (2.4,0);
\draw [->] (1.2,-1.2) -- (1.2,1.2);
\draw [red, ultra thick] (2,-1.1) -- (2,1.1);
\draw [black, ultra thick][->] (1.2,0) -- ++(325:1);
\node [right] at (1.4,-0.45) {$q_1^{(2)}$};
\draw [black, ultra thick][->] (1.2,0) -- ++(35:1);
\node [right] at (1.5,0.5) {$q_1^{(1)}$};
\draw (1.2,0) circle [radius=1];
\node at (2.7,-1) {Restriction};
\node at (2.7,-1.3) { $(A_0)_{[1,1]}^{-1} = c\,\Longleftrightarrow\,(e_1^\prime \Chol)\,q_1 = c$};
\end{tikzpicture}
\end{center}
{\scriptsize{\textit{Notes}: The vertical red line represents the non-zero restriction $(A_0)_{[1,1]}^{-1} = c$. The two black arrows represent the identified vectors $q_1^{(1)}$ and $q_1^{(2)}$.}}
\end{figure}
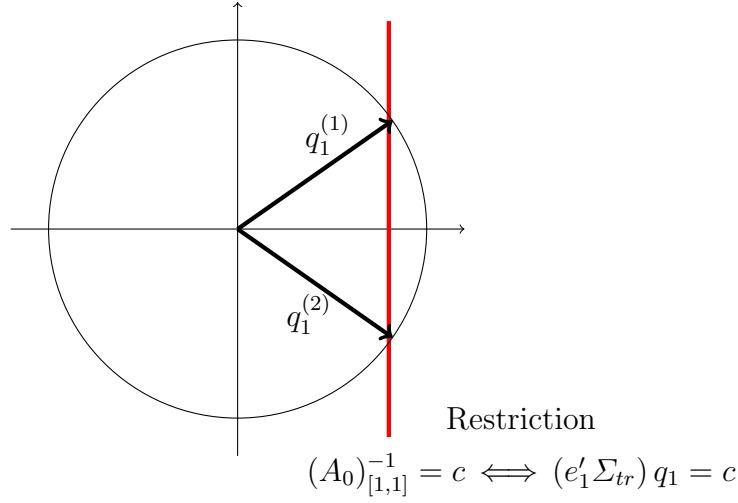

Hence, given $\phi$, $q_1$ must satisfy the two equations,
\begin{equation}
\label{eq:biVARsysq1}
\left\{
\begin{array}{rcl}
\sigma_1^\prime\,q_1 &=& c\\
q_1^\prime\,q_1 &=& 1.
\end{array}
\right.\nonumber
\end{equation}
Figure \ref{fig:Baby1} depicts these two constraints. Letting the x-axis correspond to the vector $\sigma_1$, the set of $q_1$ vectors satisfying the first constraint is a vertical line whose location is determined by $\sigma_1$ and $c$. The second constraint imposes that $q_1$ lies on the unit circle. Points at the intersection of the vertical line and the unit circle, if any exist, are solutions to this system of equations. 

When the imposed restriction is a zero restriction ($c=0$), the vertical line passes through the origin and intersects the circle at two points. The two solutions for $q_1$, $q_1^{(1)}$ and $q_1^{(2)}$,  are symmetric across the origin, and the sign normalization restriction in Definition \ref{def:SignNorm} is guaranteed to rule one of them out (see Appendix A for details). Thus, the first column of $Q$ is globally identified. 

The vertical line in Figure \ref{fig:Baby1} corresponds to a non-zero restriction ($c> 0$). If the vertical line is perfectly tangent to the unit circle, we continue to have global identification. Otherwise, there are two distinct solutions for $q_1$, as shown  in Figure \ref{fig:Baby1}. Compared to the case where $c=0$, a crucial difference is that there are some values of $\phi$ and $c$ where the sign normalization restriction cannot rule out one solution. In this case, they are both admissible and the first column of $Q$ is locally- but not globally-identified.\footnote{For $\phi \notin \Phi_F$, the vertical line does not intersect the unit circle, and no real solution for $q_1$ exists. If $c \neq 0$, the identifying restriction becomes observationally restrictive, and the identifying restriction can be refuted by the reduced-form models. }

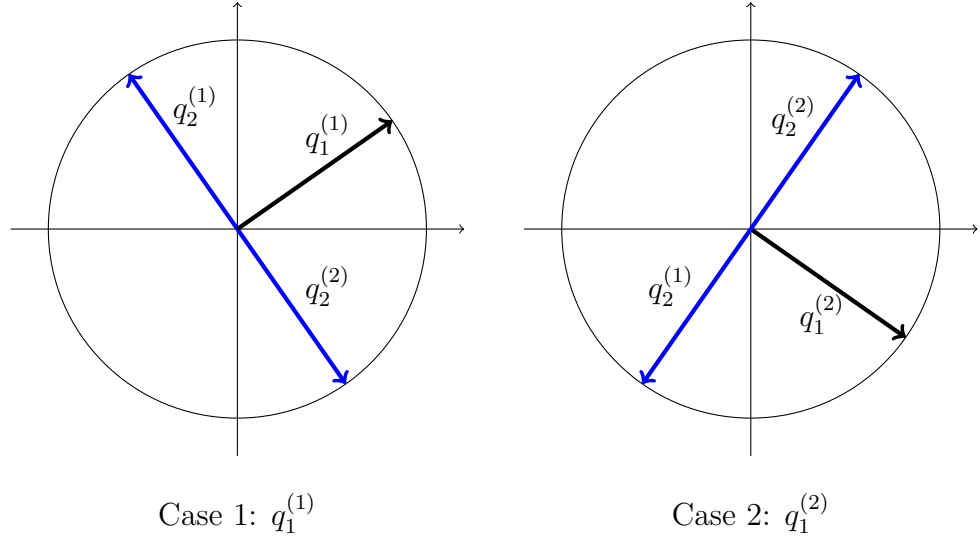
\begin{figure}
\caption{Identification of $q_2$ in the bivariate SVAR with non-zero restriction.}
\label{fig:Baby2}
\begin{center}
\begin{tikzpicture}[scale=2.5]
\draw [->] (0,0) -- (2.4,0);
\draw [->] (1.2,-1.2) -- (1.2,1.2);
\draw [blue, ultra thick][->] (1.2,0) -- ++(125:1);
\node [right] at (0.8,0.65) {$q_2^{(1)}$};
\draw [blue, ultra thick][->] (1.2,0) -- ++(305:1);
\node [right] at (1.5,-0.30) {$q_2^{(2)}$};
\draw [black, ultra thick][->] (1.2,0) -- ++(35:1);
\node [right] at (1.5,0.5) {$q_1^{(1)}$};
\draw (1.2,0) circle [radius=1];
\node at (1.2,-1.5) {Case 1: $q_1^{(1)}$};
\end{tikzpicture}
\hspace{0.5cm}
\begin{tikzpicture}[scale=2.5]
\draw [->] (0,0) -- (2.4,0);
\draw [->] (1.2,-1.2) -- (1.2,1.2);
\draw [black, ultra thick][->] (1.2,0) -- ++(325:1);
\node [right] at (1.4,-0.45) {$q_1^{(2)}$};
\draw [blue, ultra thick][->] (1.2,0) -- ++(235:1);
\node [right] at (0.6,-0.3) {$q_2^{(1)}$};
\draw [blue, ultra thick][->] (1.2,0) -- ++(55:1);
\node [right] at (1.25,0.6) {$q_2^{(2)}$};
\draw (1.2,0) circle [radius=1];
\node at (1.2,-1.5) {Case 2: $q_1^{(2)}$};
\end{tikzpicture}
\end{center}
{\scriptsize{\textit{Notes}: The left panel shows the identification of the $q_2^{(1)}$ and $q_2^{(2)}$ vectors (in blue), conditional on the identified $q_1^{(1)}$ (in black). Similarly, the right panel shows the identification of the $q_2^{(1)}$ and $q_2^{(2)}$ vectors (in blue), conditional on the identified $q_1^{(2)}$ (in black).}}
\end{figure}

The second column of $Q$, i.e. the unit-length vector $q_2$, can be pinned down through its orthogonality with $q_1$
\begin{equation}
\label{eq:biVARsysq2}
\left\{
\begin{array}{rcl}
q_2^\prime\,q_1 &=& 0\\
q_2^\prime\,q_2 &=& 1.
\end{array}
\right.
\end{equation}
If $q_1$ is only locally identified with two admissible vectors $q_1^{(1)}$ and $q_1^{(2)}$, Eq. (\ref{eq:biVARsysq2}) needs to be solved given both. Solving the system when $q_1 = q_1^{(1)}$ provides two solutions for $q_2$ that are depicted in the left panel of Figure \ref{fig:Baby2}. As the two solutions mirror each other across the origin, only one will satisfy the sign normalization restriction for the second shock. A similar picture is obtained when $q_1 = q_1^{(2)}$ (the right panel of Figure \ref{fig:Baby2}), and here too one of the solutions for $q_2$ can be ruled out by the sign normalization restriction. 

To summarize, an equality restriction with $c>0$ leads to local but non-global identification for $q_1$, and there are then two admissible $Q$ matrices, $Q_1=[q_1^{(1)},\,q_2^{(1)}]$ and $Q_2=[q_1^{(2)},\,q_2^{(2)}]$ given $\phi$. This implies that both $A_0=Q_1^\prime \Chol^{-1}$ and $A_0=Q_2^\prime \Chol^{-1}$ are admissible. In this example, we obtain two observationally equivalent $Q$ matrices, which is consistent with the upper bound on the number of observationally equivalent $Q$ matrices in Proposition \ref{prop:LocVsLack}.

\begin{figure}[ht]
        \caption{Impulse response functions related to the locally identified SVAR discussed in Section \ref{sec:geo}.}
  \label{fig:IRFbiVAR}
\begin{center}
\subfigure[{response of $y_{1t}$ to $\e_{1t}$}]{
                \includegraphics[scale=0.22]{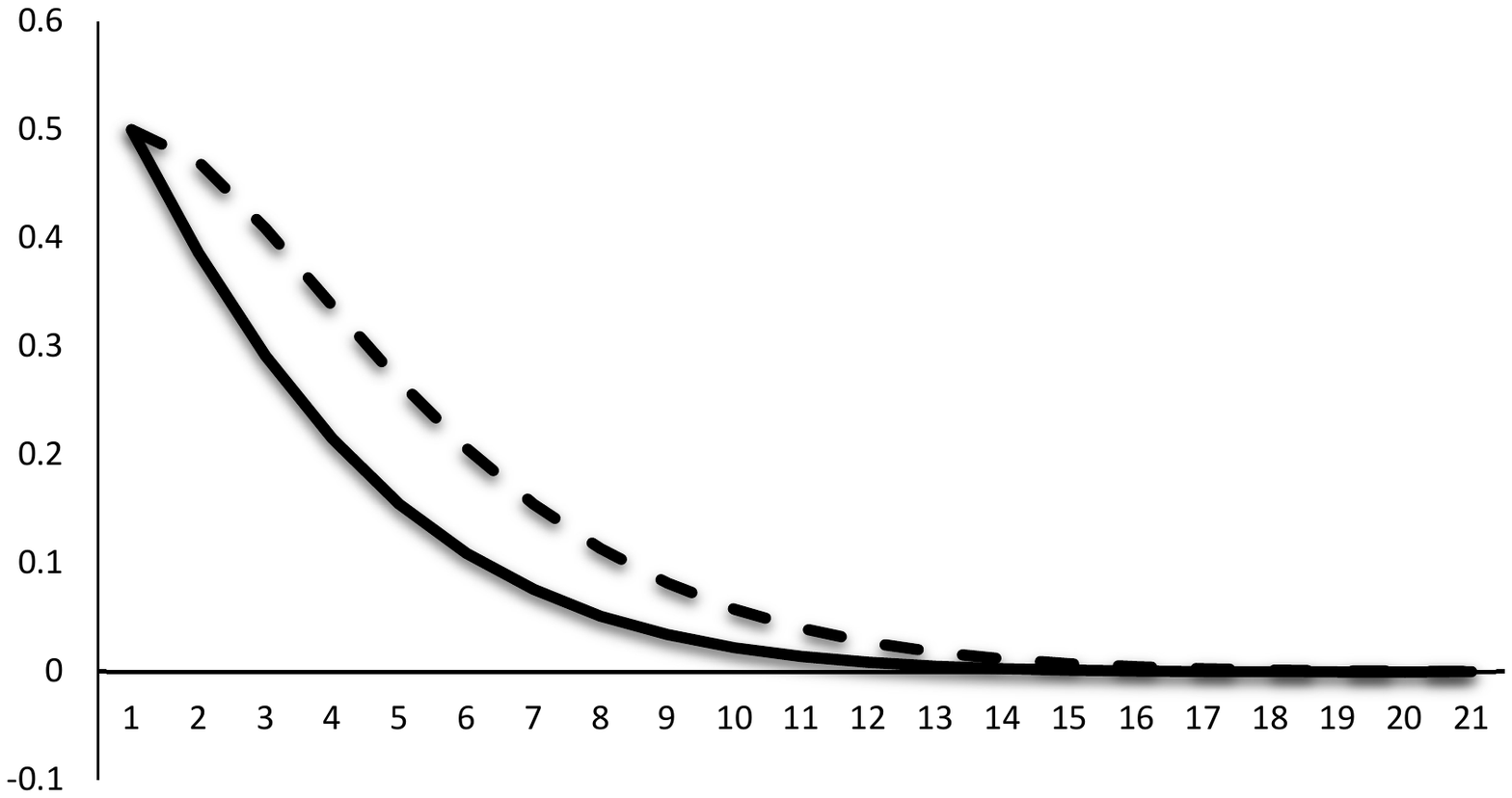}
                }               
\subfigure[response of $y_{1t}$ to $\e_{2t}$]{
                \includegraphics[scale=0.22]{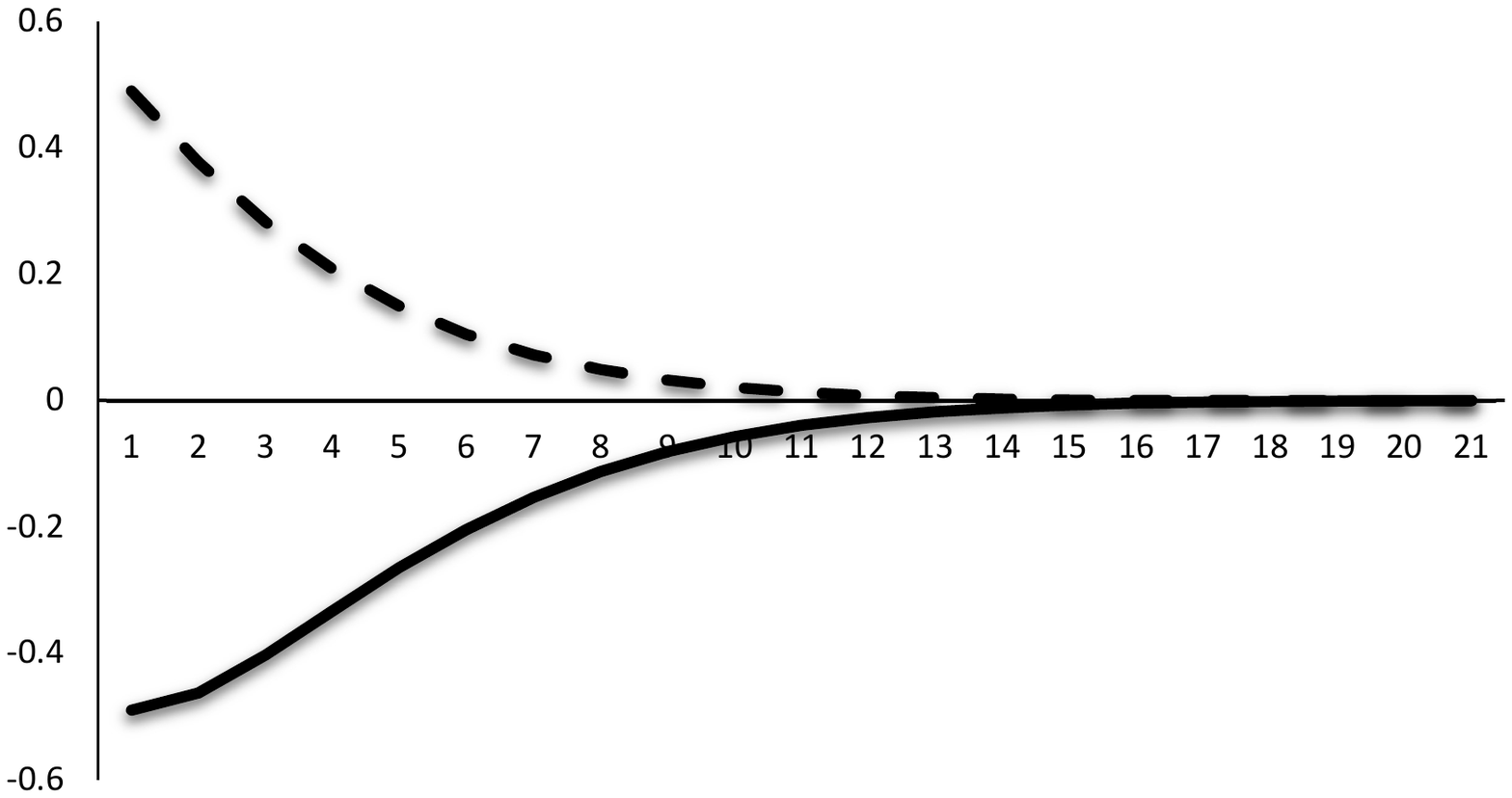}
                }               
\subfigure[response of $y_{2t}$ to $\e_{1t}$]{
                \includegraphics[scale=0.22]{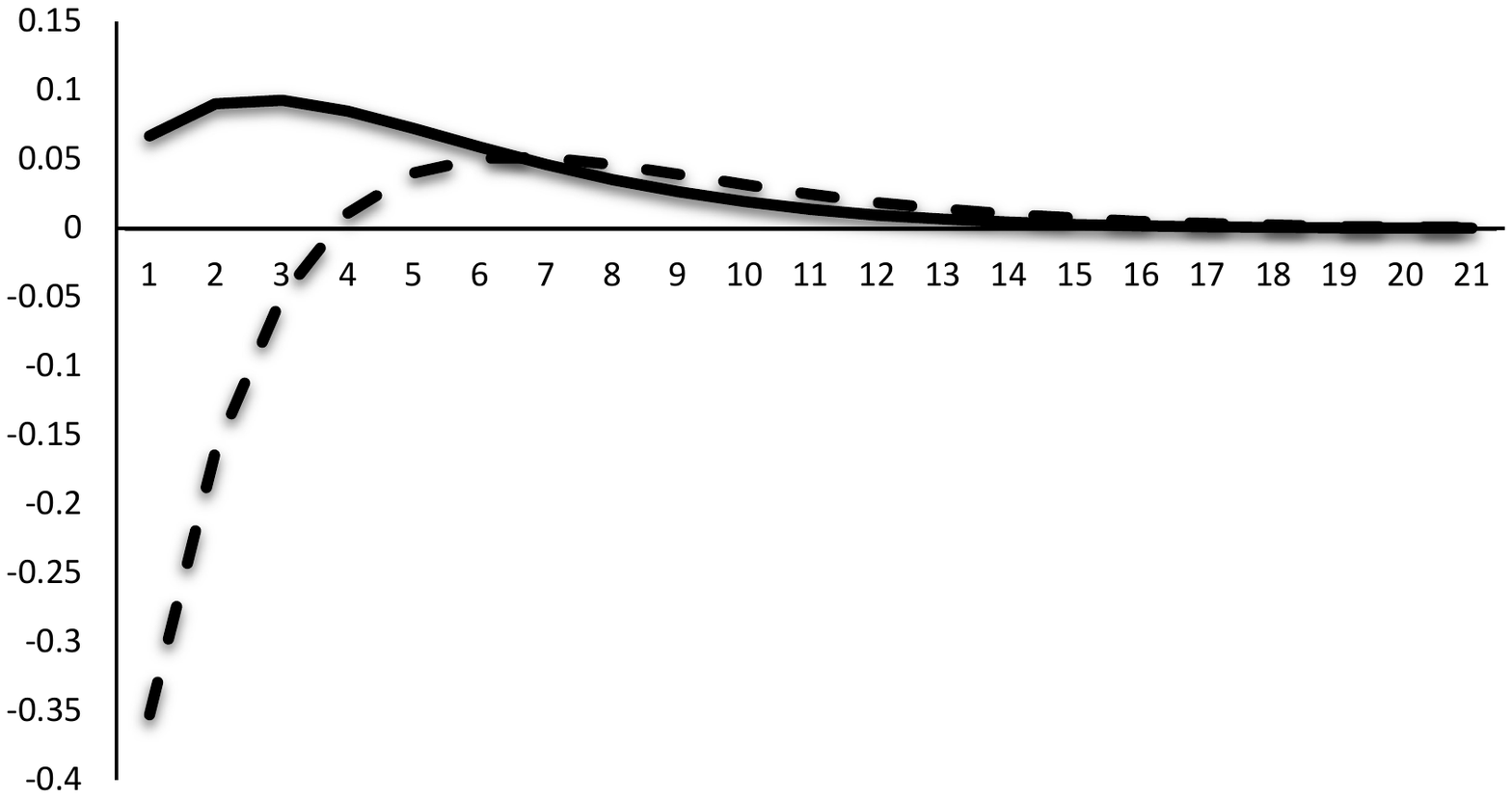}
                }
\subfigure[response of $y_{2t}$ to $\e_{2t}$]{
                \includegraphics[scale=0.22]{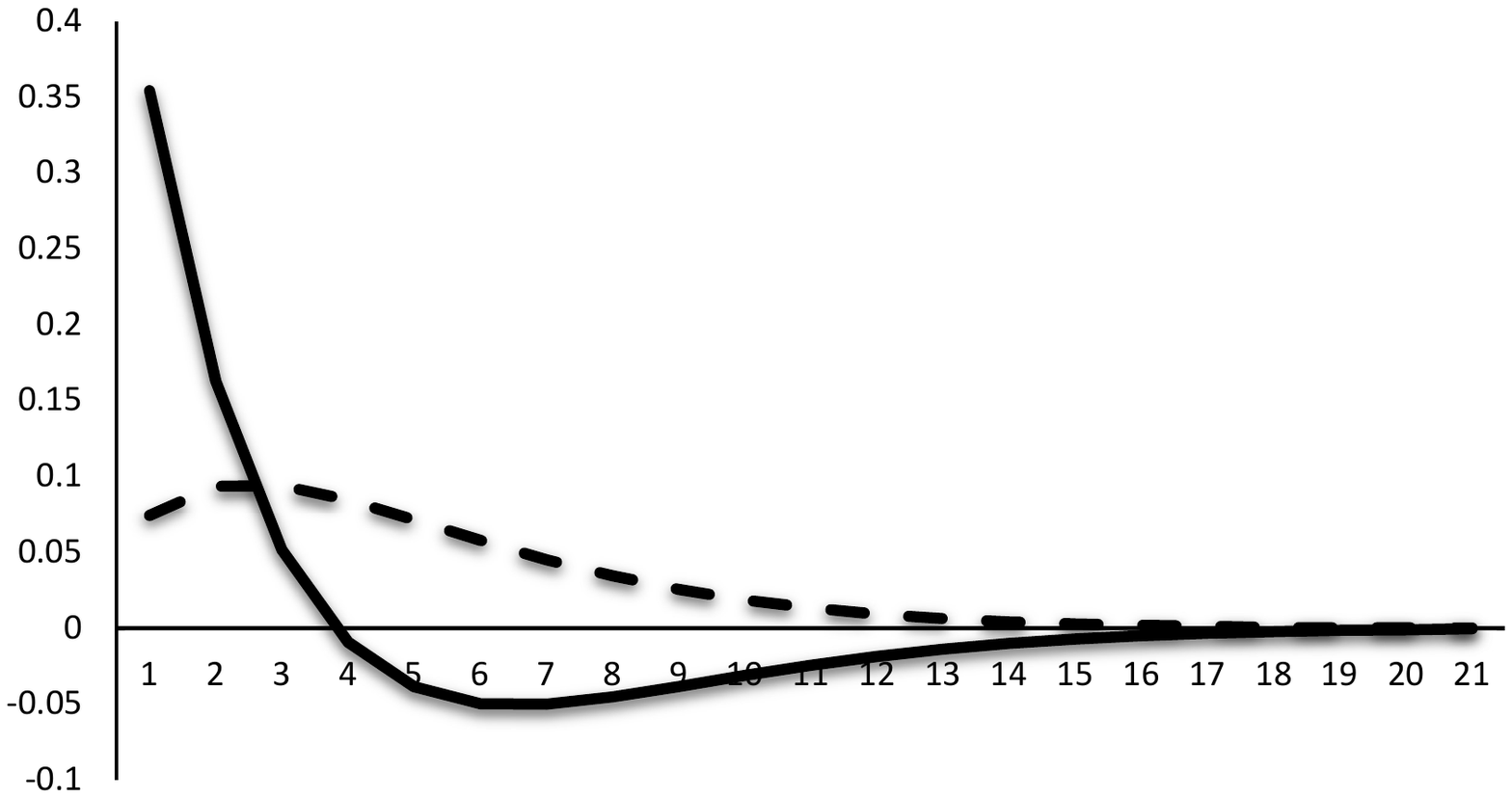}
                }               
\end{center}
\begin{minipage}{\textwidth}%
{\scriptsize{\textit{Notes}: The bivariate SVAR is characterized by a non-zero restriction. In solid lines we report the IRFs obtained through $Q_1=(q_1^{(1)},\,q_2^{(1)})$ while in dashed lines those obtained through $Q_2=(q_1^{(2)},\,q_2^{(2)})$.\par}}
\end{minipage}
\end{figure}

For a specific numerical illustration, let the bivariate VAR be characterized by constants that are zero and a single lag with reduced-form parameters
\begin{equation}
\label{eq:DGPbiVAR}
B_1=\left(
\begin{array}{cc}
0.8     &-0.2\\
0.1     & 0.6
\end{array}
\right)
\:,\hspace{1cm}
\Sigma=\left(
\begin{array}{cc}
0.49 & -0.14\\
-0.14 & 0.13
\end{array}
\right)
\:,\hspace{1cm}
\Chol=\left(
\begin{array}{cc}
0.7     & 0\\
-0.2    & 0.3
\end{array}
\right),\nonumber
\end{equation}
and consider imposing restriction $(A_0)_{[1,1]}^{-1}=0.5 \:\Longleftrightarrow\: (e_1^\prime \Chol)\,q_1 = 0.5$.
Following Eq. (\ref{eq:biVARq1}) and Eq. (\ref{eq:biVARq21}) - Eq. (\ref{eq:biVARq22}) in Appendix \ref{app:Geom}, we calculate the two admissible matrices $Q_1$ and $Q_2$
\begin{equation}
\label{eq:biVARQ}
Q_1=\left(
\begin{array}{cc}
0.714   & -0.700\\
0.700   & 0.714
\end{array}
\right)
\hspace{1cm}\text{and}\hspace{1cm}
Q_2=\left(
\begin{array}{cc}
0.714   & 0.700\\
-0.700 &        0.714
\end{array}
\right)\nonumber
\end{equation}
with associated admissible $A_{0}$ matrices
\begin{equation}
\label{eq:biVARA0}
A_0^{(1)}=\left(
\begin{array}{cc}
1.687   & 2.333\\
-0.320 & 2.381
\end{array}
\right)
\hspace{1cm}\text{and}\hspace{1cm}
A_0^{(2)}=\left(
\begin{array}{cc}
0.354   & -2.333\\
1.680 & 2.381
\end{array}
\right).\nonumber
\end{equation}
Based on these structural parameter values, Figure \ref{fig:IRFbiVAR} shows the impulse response of $y_t=(y_{1t},y_{2t})'$ to the structural shocks $\e_{1t}$ and $\e_{2t}$. Despite the simplicity of this example, it clearly illustrates the extent to which conclusions depend on the choice of observationally equivalent $Q$ matrices.
 
\subsection{Local identification in heteroskedastic SVARs}
\label{sec:IdHSVAR}

As discussed in Section \ref{sec:HSVAR}, the impulse responses in HSVAR are only locally identified in the absence of economic restrictions that can pin down a unique ordering of the eigenvalues in the decomposition (\ref{eq:CovarVARWB}). 
To refine the identified set of $A_0$ (i.e., the ordering of the entries in $\Lambda$) in this case, the literature has considered imposing the following types of identifying restrictions (in addition to the sign normalization restrictions on $Q$): 
\begin{enumerate}
\item Assume a certain ranking of the magnitudes of the volatility breaks among the structural shocks. 
\item Restrict the sign or shapes of the impulse response functions.
\item Determine the shock of interest by assessing the  correlations between statistically identified shocks with proxy variables of the structural shocks \citep{BPSS21}.
\end{enumerate}
If the eigenvalues in $\Lambda$ are all distinct and the first assumption can pin down a unique ordering of the eigenvalues in $\Lambda$, it yields global identification of $A_0$. 
On the other hand, it is rare in practice that one can credibly assume the complete ranking of the eigenvalues (i.e., magnitudes of the volatility ratios across the regimes). If a restriction on $\Lambda$ implies partial ranking of the eigenvalues, the multiplicity of $\Lambda$ remains and leads to local identification of $A_0$. 

The second type of restrictions is commonly imposed in practice, while as we show in our empirical application in Section \ref{sec:EmpApp}, the sign restrictions of the impulse responses are not guaranteed to yield global identification, and local identification of $A_0$ can persist. In such scenario, the methods for estimation and inference in locally-identified SVARs that we propose in this paper is useful for HSVARs as well.


\section{Computing identified sets of locally identified SVARs}
\label{sec:Estimation}

A common approach to estimating SVAR structural parameters is constrained maximum likelihood  (\citeauthor{AG}, \citeyear{AG}), with the maximization performed numerically given some initial values. The standard gradient-based algorithm stops once it reaches a local maximum, and does not check for the existence of other observationally equivalent parameter values. Hence, the conventional maximum likelihood procedure applied to an SVAR that is locally but not globally identified will select one of the observationally equivalent structural parameters in a nonsystematic way, limiting the credibility of the resulting estimates and inference. This section proposes computational methods that produce estimates of all the observationally equivalent admissible structural parameters, both for SVAR and for HSVAR models, respectively.


\subsection{A general computation procedure for locally identified SVARs}
\label{sec:Algorithm}

This section proposes computational methods that produce estimates of all the observationally equivalent $A$ matrices given the identifying restrictions. Our approach is first to obtain $\hat{\phi} = (\hat{B}, \hat{\Sigma})$, an estimate of the reduced-form parameters $\phi$, and then compute the identified set for $A_0$ given $\hat{\phi}$, $\mathcal{A}_0(\hat{\phi}| F,S)$, by solving a system of equations for the $Q$ matrix given $\hat{\phi}$.\footnote{\label{footnote:phis} For estimators of $\phi$, we consider (i) an unconstrained reduced-form VAR estimator for $\phi$ and (ii) an estimator for $\phi$ induced by a constrained maximum likelihood estimator for $A$ under the identifying restrictions (i.e., one of the locally identified structural parameter points maximizing the likelihood). In the Bayesian inference methods considered in Section \ref{sec:InfLocId}, we view $\hat{\phi}$ as a draw from the posterior of $\phi$.}   

In this section we propose a procedure to compute $\mathcal{A}_0(\hat{\phi}| F,S)$. The procedure is general and invokes a non-linear solver. An alternative procedure, that is more constructive and involves only elementary calculus, but the allowed type of identifying restrictions is more limited, is reported and described in Appendix \ref{app:AlgorithmNH}. Both algorithms deal with SVARs characterized by exactly $f=n(n-1)/2$ restrictions, and we presume the rank conditions in Proposition \ref{prop:LocIdent} or Proposition \ref{prop:SuffLocIdent} are ensured or have been checked empirically prior to implementation. This preliminary step is extremely important as, if the rank condition is not met, the model can be at most set identified. \cite{ARW18} and \cite{GK18}, among others, propose estimation and inference strategies when combining zero and sign restrictions in SVARs featuring departures from the rank condition we discuss in Proposition \ref{prop:GlobIdent}. In particular, in both papers, set identification arises when, at least for one $1\leq j \leq n$, the number of restrictions is less than what is required for global identification, i.e. $f_j<n-j$.\footnote{Interestingly, \cite{GK18} provide theoretical results for the identified sets to be convex, that simplifies the interpretation of the results.} However, this is not the only situation leading to set identification. More in general, whenever the rank condition for (local or global) identification is not met, either because $f<n(n-1)/2$ or because the restrictions are not appropriately chosen, set identification occurs. Although this goes beyond the aims of the present paper, it reinforces the message that our estimation and inference machinery work once a rank condition for local identification is met.    

Given $\hat{\phi}$, this method computes the orthogonal matrices subject to the identifying restrictions by solving a non-linear system of equations.\footnote{\cite{KK18QE} similarly check global identification of DSGE models by examining the solutions of a non-linear system of equations.} If the model is locally identified, then it yields at most $2^{n(n+1)/2}$ solutions for $Q$. Some of these will be discarded by normalization and sign restrictions. The remaining solutions for $Q$ are then used to span the identified set for $A_0$, $\mathcal{A}_0(\hat{\phi}| F,S)$, and its projection leads to the identified set of an impulse response $IS_{\eta}(\hat{\phi})$. All these steps are stated formally in the next algorithm. 

\begin{algo}
\label{algo:Estimation}
Consider an SVAR with normalization restrictions, equality restrictions as in Eq. (\ref{eq:GenFormRest}) and sign restrictions as in Eq. (\ref{eq:SignRest}), and assume $f=n(n-1)/2$ equality restrictions are imposed. Let $\hat{\phi}$ be a given estimator for $\phi$.
\begin{enumerate}
\item Solve the system of equations for $Q$:
        \begin{equation}
                \left\{\begin{array}{rcl}
                  \textbf{F}(\hat{\phi})\text{vec}\, Q-\bf{c} & = & \textbf{0}\\
                        Q'\,Q & = & I_n; 
                \end{array}\right.\label{eq.algorithm1}
        \end{equation}
\item If the set of real solutions for $Q$ is non-empty, then retain only those satisfying the normalization and sign restrictions to obtain $\mathcal{Q}_R(\hat{\phi})$. $\mathcal{A}_0(\hat{\phi}|F,S)$ is constructed accordingly by $ \big\{ A_0 = Q' \hat{\Sigma}_{tr}^{-1} : Q \in \mathcal{Q}_R(\hat{\phi})\big\}$.

\item It is possible that no real solution for $Q$ exists in Step 1. If so, we return $\mathcal{Q}_R(\hat{\phi})=\emptyset$, i.e., $\hat{\phi}$ is not compatible with the imposed identifying restrictions.
\end{enumerate}
\end{algo}

The crucial step in this algorithm is obtaining all the solutions to the equation system (\ref{eq.algorithm1}). This is a system of polynomial equations consisting of linear and quadratic equations.\footnote{\cite{Sturmfels02solvingsystems} provides a good overview of systems of polynomial equations with potential applications in statistics and economics. As we saw in Section \ref{sec:IdentCond}, this system can be also seen as a minimization problem of the quadratic objective function subject to the orthogonality constraints $Q'\,Q = I_n$. Noting that the orthogonality constraints generate the Stiefel manifold, we can consider applying algorithms for optimization on the Stiefel manifold. See  \cite{EAT98} and \cite{Manopt}.} Closed-form solutions do not seem available, but numerical algorithms to compute all the roots of the polynomial equations are. Matlab, for example, has the function \texttt{vpasolve}, an algorithm to find all the solutions of a system of non-linear equations.\footnote{An alternative approach could be to solve the system analytically through the Matlab function \texttt{solve}, and then approximate the roots numerically using the function \texttt{vpa}. For all cases investigated in our empirical analyses, the two strategies lead to the same set of results.} According to the Matlab documentation,\footnote{https://uk.mathworks.com/help/symbolic/vpasolve.html} \texttt{vpasolve} returns the complete set of solutions in the case of polynomial equations. The strength of this algorithm is its generality, but it is a black-box function.\footnote{Matlab solvers are not open source, and we fail to uncover the precise numerical algorithm \texttt{vpasolve} uses to find roots of nonlinear equation systems.}  

When non-homogeneous restrictions or restrictions across shocks are imposed, the model becomes observationally restrictive. Hence, when $\hat{\phi}$ is obtained from an unconstrained reduced-form VAR estimator, if $\hat{\phi}$ happens to be outside of $\Phi_R$, as highlighted in Proposition \ref{prop:LocVsLack}, then Step 3 of Algorithm \ref{algo:Estimation} becomes relevant. When the algorithm returns $\mathcal{Q}_R(\hat{\phi})=\emptyset$, the maximum likelihood reduced-form model suggests that some of the imposed identifying restrictions are misspecified. One can hence consider relaxing some of the imposed sign restrictions, or modify the value of $\textbf{c} \neq \textbf{0}$ if non-homogeneous restrictions are present. 


\subsection{A general computation procedure for locally identified heteroskedastic SVARs}
\label{sec:AlgorithmHSVAR}

In Section \ref{sec:HSVAR} and Section \ref{sec:IdHSVAR} we have seen that an HSVAR is globally identified once a specific ordering of the shifts in the variances, contained in the diagonal $\Lambda$ matrix, has been fixed. However, a problem compatible with the local identification issue might arise when labeling the structural shocks. Two or more shocks can be admissible as they produce economically reasonable effects on the variables of interests. Admissible $Q$ matrices can be obtained by flipping the ordering of their columns, as far as the corresponding impulse responses are in line with the reference economic theory.

The next algorithm extends Algorithm \ref{algo:Estimation} to HSVARs, and allows to compute $\mathcal{A}_0(\hat{\phi}| F,S)$. It is worth stressing that, as far as HSVARs are characterized by no equality restriction, then $\textbf{F}(\phi)=\bf{0}$, $\bf{c}=\bf{0}$, and, as a consequence, $f=0$. Sign restrictions, instead, are extremely useful for the identification of the admissible solutions. 

\begin{algo}
\label{algo:EstimationHSVAR}
    Consider an HSVAR characterized by two volatility regimes, with break date $T_B$ known. Let $\hat{\phi}=\big(\hat{B},\hat{\Sigma_1},\hat{\Sigma_2}\big)$ be a given estimator for $\phi=(B,\,\Sigma_1,\,\Sigma_2)$. Let $\hat{\Sigma}_{1,tr}$  be the lower triangular Cholesky decomposition of $\hat{\Sigma}_1$. 
    \begin{enumerate}
        \item Solve the eigen-decomposition problem
        \begin{equation}
        \label{eq:algorithmHSVAR}
            \hat{\Sigma}_{1,tr}^{-1} \hat{\Sigma}_2 \hat{\Sigma}_{1,tr}^{-1\prime} = Q\Lambda Q^\prime
        \end{equation}
        where $\Lambda$ is the diagonal matrix containing the eigenvalues and $Q$ the matrix whose columns are the related eigenvectors.
        \item Calculate $A_0^{-1}=\hat{\Sigma}_{1,tr}^{-1} Q_f$, where $Q_f$ is obtained from $Q$ by fixing a specific ordering of its columns, e.g. such that $\lambda_1>\lambda_2>\ldots>\lambda_n$. Obtain the structural parameters and the objects of interest, e.g. the impulse responses. Let $q_i$ be the \textit{i}-th column of $Q_f$.
        \item Collect all the indices $i$, $1\leq i \leq n$, such that $q_i$, or $-q_i$, generates objects of interest satisfying all the sign restrictions. Let $I(\hat{\phi})=\{i_1,\ldots,i_M\}$ be the set of all these indices, and $M(\hat{\phi})$ its cardinality. 
        \item Let, for simplicity, our shock of interest be positioned as first. The set of all admissible orthogonal matrices $\mathcal{Q}_R(\hat{\phi})$ can be obtained, starting from $Q_f$, by flipping $q_1$ and $q_k$, for any $k \in I(\hat{\phi})$. $\mathcal{A}_0(\hat{\phi}|F,S)$ is constructed accordingly by $\Big\{ A_0 = Q^\prime \hat{\Sigma}_{tr}^{-1} : Q \in \mathcal{Q}_R(\hat{\phi})\Big\}$. 
        \item If none of the columns of $Q$ are admissible, then $I(\hat{\phi})=\emptyset$, $M(\hat{\phi})=0$, and we return $\mathcal{Q}_R(\hat{\phi})=\emptyset$, i.e., $\hat{\phi}$ is not compatible with the imposed sign restrictions.
    \end{enumerate}
\end{algo}

When the algorithm returns $\mathcal{Q}_R(\hat{\phi})=\emptyset$, perhaps some of the imposed sign restrictions are too stringent. One can hence consider relaxing some of them or, in alternative, try a different VAR specification, maybe considering a different set of endogenous variables to better identify the shock of interest. On the contrary, if the algorithm returns with a non-empty $\mathcal{Q}_R(\hat{\phi})$, then the model is characterize by a set of observationally equivalent admissible solutions, like in locally-identified SVARs identified through equality and sign restrictions. As we will see in the next section, the approaches for doing inference on the objects of interest are exactly the same, for both SVAR and HSVAR. 


\section{Inference for locally identified SVARs}
\label{sec:InfLocId}

\subsection{Bayesian inference}
\label{sec:BI}

Standard Bayesian inference specifies a prior distribution for either the structural parameters $A$ (e.g., \citeauthor{ARW18}, \citeyear{ARW18}; \citeauthor{BH15}, \citeyear{BH15}; \citeauthor{LSZ96}, \citeyear{LSZ96}; \citeauthor{WZ03Gibbs}, \citeyear{WZ03Gibbs}), or the reduced-form parameters and rotation matrix $(\phi,Q)$ as a reparametrization of $A$ (e.g., \citeauthor{Uhlig2005}, \citeyear{Uhlig2005}). When identification is local, the likelihood for the joint parameter vector $A$ can have multiple modes, which means that the posterior for the structural parameters and impulse responses may also have multiple modes. This leads to computational challenges as commonly used Markov Chain Monte Carlo (MCMC) methods can fail to adequately explore the posterior when it is multi-modal. For instance, in the standard Metropolis-Hastings algorithm, the presence of multiple modes complicates the choice of proposal distribution. If the proposal distribution in the Metropolis-Hastings algorithm does not support some modes well, a lack of irreducibility of the Markov chain can lead it to fail to converge to the posterior. Similarly, in the standard Gibbs sampler, the presence of multiple modes in the posterior for $A$ leads its support to be almost disconnected, which can then lead a break down of irreducibility and the Gibbs sampler to fail to converge (see Example 10.7 in \citeauthor{RC04}, \citeyear{RC04}). 
By combining our constructive algorithms (either Algorithm \ref{algo:Estimation} or Algorithm \ref{algo:EstimationTrNonHomo}) for computing $IS_{\eta}(\phi)$ with the posterior sampling algorithm for $\phi$, we can overcome such 
computational challenges. 
  
We consider approximating the posterior for a scalar impulse response $\eta(\phi)$. Assume that the reduced-form parameters yield non-empty $\mathcal{Q}_R(\phi)$. Let $IS_{\eta}(\phi)$ consist of $M(\phi)\geq 1$ distinct points, 
\begin{equation}
IS_{\eta}(\phi) = \big\{\eta_{1}(\phi), \eta_2(\phi), \dots, \eta_{M(\phi)}(\phi) \big\}, \label{IS_Mphi}
\end{equation}
where we index the observationally equivalent impulse responses to satisfy $\eta_1(\phi)<\eta_2(\phi)< \cdots < \eta_{M(\phi)}(\phi)$. 

We follow the ``agnostic'' Bayesian approach of \citet{Uhlig2005}. The posterior for $\eta$ is induced by the posterior for $\phi$, $\pi_{\phi|Y}$, which is supported on $\Phi_{R} \equiv \{\phi: \mathcal{Q}_R(\phi) \neq \emptyset \}$, and $Q$ has a uniform prior supported only on the admissible set of rotation matrices $\mathcal{Q}_R(\phi)$ given $\phi \in \Phi_{R}$. Local identification with the $M(\phi)$-point identified set as in (\ref{IS_Mphi}) can be obtained by projecting the $M(\phi)$ admissible rotation matrices into the space of impulse responses 
if each of them leads to distinct values of impulse response. Hence, the uniform weights assigned over these rotation matrices imply that equal weights are assigned to the points in $IS_{\eta}(\phi)$. As a result, for $G \subset \mathbb{R}$, the posterior for $\eta$ can be expressed as
\begin{equation}
\pi_{\eta | Y}(\eta \in G) \propto E_{\phi|Y} \left[ \frac{1}{M(\phi)}\sum_{m=1}^{M(\phi)} 1\big\{ \eta_m(\phi) \in G \big\} \right]. \label{eta posterior}
\end{equation} 
Since the reduced-form VAR (or HVAR) likelihood is unimodal and concentrated around the maximum likelihood estimate, MCMC algorithms will perform well when sampling from $\pi_{\phi|Y}$. Hence, the posterior in Eq. (\ref{eta posterior}) can be approximated by combining a posterior sampler for $\phi$ with the algorithm for computing $\big\{ \eta_m(\phi): m= 1, \dots, M(\phi) \big\}$. 


\subsection{Frequentist-valid inference}
\label{sec:FVI}

Bayesian inference as considered above can be sensitive to the choice of prior even in large samples due to the lack of global identification. The standard Bayesian procedure (assuming a unique prior for the structural parameters) specifies an allocation of the prior belief over observationally equivalent impulse responses, $IS_{\eta}(\phi)$, conditional on $\phi$. This conditional belief given $\phi$ is not updated by the data and, as a result, the shape and heights of the posterior around the modes remain sensitive to its specification.\footnote{\label{footnote:sensitivity} If the impulse response functions are not globally identified, posterior sensitivity to a prior becomes a concern no matter whether the object of interest is marginal or joint posterior of the impulse responses.} In this section, we propose an asymptotically valid frequentist inference procedure for the impulse response identified set that can draw inferential statements which are robust to the choice of prior weights over the set of locally identified parameter values. 

Our approach is to project asymptotically valid frequentist confidence sets for the reduced-form parameters $\phi$ through the identified set mapping $IS_{\eta}(\phi)$. In standard set-identified models where the identified set is a connected interval with positive width, the projection approach to constructing the confidence set has appeared in the 2011 working paper version of \cite{MS12}, \cite{NT14}, \cite{KT13}, among others. This approach generally yields asymptotically valid (but conservative) confidence sets even when the identified set consists of discrete points. However, a challenge unique to the discrete identified set case is the computation of projection confidence sets for the impulse responses based on a finite number of grid points or draws of $\phi$ from their confidence set. In what follows, we propose methods to tackle this computational challenge.  

Let $CS_{\phi,\alpha}$ be an asymptotically valid confidence set for $\phi$ with coverage probability $\alpha \in (0,1)$. If the maximum likelihood estimator $\hat{\phi}$ is $\sqrt{T}$-asymptotically normal, the likelihood contour set $CS_{\phi, \alpha}$ is determined by the $\alpha$-th quantiles of the $\chi^2$ distribution with the degree of freedom $dim(B)+n(n-1)/2$. If the posterior for $\phi$ satisfies the Bernstein-von Mises property, that is the posterior for $\sqrt{T}(\phi - \hat{\phi})$ asymptotically coincides with the sampling distribution of the maximum likelihood estimator, the Bayesian highest density posterior region with credibility $\alpha$ can be used for $CS_{\phi,\alpha}$. The MCMC confidence set procedure developed by \cite{CCT18} can then be used to obtain draws of $\phi$ from the highest density posterior region with credibility $\alpha$. We follow this procedure in our empirical application below. The inference procedure below allows for any $CS_{\phi,\alpha}$ with asymptotically valid coverage, and takes draws or grids of $\phi$ from $CS_{\phi,\alpha}$ as given.

The projection confidence set is defined as
\begin{equation}
CS_{\eta, \alpha}^{p} = \bigcup_{\phi \in CS_{\phi, \alpha}} IS_{\eta}(\phi). \label{CSeta}
\end{equation}
We assume that $CS_{\phi,\alpha}$ is an asymptotically valid confidence set for $\phi$ in the sense that
\begin{equation}
\lim_{T \to \infty} p_{Y^T|\phi_0}(\phi_0 \in CS_{\phi,\alpha}) = \alpha, \notag 
\end{equation}
where $p_{Y^{T}|\phi_0}$ is the sampling distribution of the data with sample size $T$ and $\phi_0$ is the true value of $\phi$. Since $\{\phi_0 \in CS_{\phi,\alpha} \}$ implies $\{ IS_{\eta}(\phi_0) \subset CS^p_{\eta,\alpha} \}$, $CS^{p}_{\eta,\alpha}$ (and any set including $CS^p_{\eta,\alpha}$) is an asymptotically-valid but potentially conservative confidence set for $IS_{\eta}(\phi_0)$, 
\begin{equation}
\lim_{T \to \infty} p_{Y^T|\phi_0}(IS_{\eta}(\phi_0) \subset CS^p_{\eta,\alpha}) \geq \alpha. \notag
\end{equation}

Let $\{ \phi_{k} : k=1, \dots, K\}$ be a finite number of Monte Carlo draws  or grid points from  $CS_{\phi,\alpha}$. A sample analogue of the projection confidence set, $\bigcup_{k=1,\dots,K} IS_{\eta}(\phi_k)$, is less useful in approximating $CS^p_{\eta, \alpha}$, because each $IS_{\eta}(\phi_k)$ is a discrete set, whereas the underlying $CS^{p}_{\eta,\alpha}$ we want to approximate can be a union of disconnected intervals with positive widths. In addition, it is difficult to judge how many disconnected intervals $IS^p_{\eta,\alpha}$ has and where the possible gaps lie within $CS_{\eta, \alpha}^p$ from a finite number of  draws of $IS_{\eta}(\phi_k)$, $k=1, \dots, K$. Reporting the convex hull of $\bigcup_{k=1,\dots,K} IS_{\eta}(\phi_k)$ is simple, but it can lead to a connected confidence set that obscures the discrete feature of the identified set.

In what follows, we propose two different approaches for computing the projection confidence set for an impulse response given a set of Monte Carlo draws for $\phi$. We refer to the first as \textit{switching-label projection confidence sets}. It allows the labels indexing observationally equivalent impulse responses to vary across the horizons, and produces confidence sets that can capture multi-modality of the posterior distribution or the integrated  likelihood for each impulse response at each horizon. We refer to the second approach as \textit{fixed-label projection confidence sets}. It maintains unique labels for observationally equivalent structural parameters across the impulse responses and over horizons, i.e., the labels for observationally equivalent structural parameters are defined in terms of the modes of the posterior for $Q$. This approach may produce confidence sets that are wider than the switching-label projection confidence sets, but it can better capture and visualize dependence of the impulse responses over the horizons.

\subsubsection{Switching-label projection confidence sets}
\label{sec:SwitchingLabel}

The switching-label approach draws inference for each impulse response at each horizon one-by-one. We hence set $\eta(\phi)$ to a particular scalar impulse response. 

Maintaining the notation of the previous subsection, let $IS_{\eta}(\phi_k) = \big\{ \eta_1(\phi_k), \dots, \eta_{M(\phi_k)} (\phi_k) \big\}$, where $M(\phi_k)$ is the number of distinct points in the identified set at  $\phi=\phi_k$. We label these points in increasing order, $\eta_1(\phi_k) < \cdots < \eta_{M(\phi_k)}(\phi_k)$. Let $\bar{M}=\max_k M(\phi_k)$ be the largest cardinality of $IS_{\eta}(\phi_k)$ among the draws of $\phi_k$, $k=1, \dots, K$. $\bar{M}$ indicates the largest possible number of disconnected intervals of $CS^p_{\eta,\alpha}$. We view these intervals as clusters, each of which is indexed by $\tilde{m} \in \{1, \dots, \bar{M} \}$. Let $\tilde{K} = |\{\phi_k: M(\phi_k) = \bar{M} \}|$ be the number of $\phi$ draws that has the maximal number of observationally equivalent impulse responses and define estimates of the cluster-specific mean and variance by
\begin{align}
\mu_{\tilde{m}} &= \frac{1}{\tilde{K}} \sum_{\phi_k: M(\phi_k) = \bar{M}} \eta_{\tilde{m}}(\phi_k), \notag \\
\sigma^2_{\tilde{m}} & = \frac{1}{\tilde{K}-1} \sum_{\phi_k: M(\phi_k) = \bar{M}} \big(\eta_{\tilde{m}}(\phi_k) - \mu_{\tilde{m}} \big)^2,
\end{align}
for each $\tilde{m} = 1, \dots, \bar{M}$.  

For each $\phi_k$, $k=1,\dots, K$, we augment a binary vector of length $\bar{M}$, $D(\phi_k) = \big\{D_{\tilde{m}}(\phi_k) \in \{0 ,1 \}: \tilde{m}=1, \dots, \bar{M}\big\}$, which indicates whether or not any one point of $IS_{\eta}(\phi_k)$ can be associated with $\tilde{m}$-th cluster. The true $D(\phi_k)$ is not observed, so must be imputed by, for instance, maximizing the Gaussian log-likelihood criterion in the following manner. Let $\rho_{\phi_k}$ be an increasing injective map from $\big\{1, \dots, M(\phi_k) \}$ to $\{ 1, \dots, \bar{M} \big\}$, characterizing which cluster each $\eta_{m}(\phi_k)$, $m = 1, \dots, M(\phi_k)$, belongs to. Define
\begin{equation}
\label{eq:assign_label}
    \hat{\rho}_{\phi_k} \in \arg \min_{\rho_{\phi_k}} \sum_{m=1}^{M(\phi_k)} \frac{\left( \eta_{m}(\phi_k) - \mu_{\rho_{\phi_k}(m)} \right)^2}{\sigma^2_{\rho_{\phi_k}(m)}}, 
\end{equation}
which minimizes the sum of variance-weighted squared distances to the cluster-specific means. We then construct $D(\phi_k) = \big\{D_{\tilde{m}}(\phi_k): \tilde{m}=1, \dots, \bar{M}\big\} \in \{0, 1 \}^{\bar{M}}$ from the indicators for whether $\hat{\rho}_{\phi_k}$ maps any $m \in \big\{ 1, \dots, M(\phi_k) \big\}$ to $\tilde{m}$, i.e., $D_{\tilde{m}}(\phi_k) = 1\big\{ \exists\,\, m \mspace{10mu} \textrm{s.t.} \mspace{10mu} \rho_{\phi_k}(m) = \tilde{m} \big\}$. 
We then construct an interval for each cluster $\tilde{m} \in \{ 1, \dots, \bar{M}\}$ by
\begin{equation}
C_{\tilde{m}} = \left[ \min_{\phi_k: D_{\tilde{m}}(\phi_k)=1} \eta_{\hat{\rho}_{\phi_k}^{-1}(\tilde{m})}(\phi_k), \max_{\phi_k: D_{\tilde{m}}(\phi_k)=1} \eta_{\hat{\rho}_{\phi_k}^{-1}(\tilde{m})}(\phi_k) \right].
\end{equation}
An approximation of the projection confidence set is then formed by taking the union of $C_{\tilde{m}}$: 
\begin{equation}
\widehat{CS}^p_{\eta,\alpha} \equiv \bigcup_{\tilde{m}=1}^{\bar{M}} C_{\tilde{m}}. \label{eq:union_Cm} 
\end{equation}
Note $\widehat{CS}^p_{\eta,\alpha}$ obtained in this way includes all the $IS_{\eta}(\phi_k)$, $k=1, \dots, K$, and at the same time, can yield a collection of disconnected intervals. Moreover, if the maximum likelihood estimator for $\phi$ is consistent for $\phi_0$, $IS_{\eta}(\phi)$ is a continuous correspondence at $\phi_0$ and $M(\phi)$ is constant in an open neighborhood of $\phi_0$, it can be shown that $\widehat{CS}^p_{\eta,\alpha}$ converges to $IS_{\eta}(\phi_0)$ in the Hausdorff metric. Hence, $\widehat{CS}^p_{\eta,\alpha}$ can consistently uncover the true identified set consisting of potentially multiple points.

We construct $\widehat{CS}^p_{\eta,\alpha}$ separately for each impulse response at each horizon. Hence, the labeling of the clusters $\tilde{m} = 1,\dots, \bar{M}$ defined for one impulse response does not correspond to the labeling of the clusters defined for other impulse responses or horizons. For example, a particular impulse response function labeled as $\tilde{m}=1$ in one horizon can be labeled as $\tilde{m}=2$ in another horizon. We expect that switching-label projection confidence sets can visualize well the multi-modality of the marginal posterior for each impulse response.    

\subsubsection{Fixed-label projection confidence sets}
\label{sec:ConstantLabel}

In contrast to switching-label projection confidence sets, fixed-label projection confidence sets maintain fixed-labeling across impulse responses and over time horizons. For example, an impulse response function labeled as $\tilde{m}=1$ at one horizon is labeled as $\tilde{m}=1$ at other horizons.

To implement this procedure, we need to anchor the labels to a particular impulse response, say, the impulse response of $i^{\ast}$-th variable to $j^{\ast}$-th structural shock at a particular horizon $h=h^{\ast}$, denoted hereafter by $\eta^{\ast}(\phi,q_{j^{\ast}}) \equiv e_{i^{\ast}}' C_{h^{\ast}}(\phi) q_{j^{\ast}} $. Given a Monte Carlo draw of the reduced-form parameters, $\phi_{k}$, $k=1, \dots, K$, from $CS_{\phi, \alpha}$, let $q_{j^{\ast},m}(\phi_k)$, $m=1, \dots, M(\phi_k)$ be  observationally equivalent $q_{j^*}$ vectors  labeled according to the ordering of $\eta^*\big(\phi,q_{j^{\ast}}\big)$, i.e., $\eta^*\big(\phi_k, q_{j^{\ast},1}(\phi_k)\big) < \eta^*\big(\phi_k, q_{j^{\ast},2}(\phi_k)\big)<  \dots < \eta^*\big(\phi_k, q_{j^{\ast},M(\phi_k)}(\phi_k)\big)$. Similarly to the labeling procedure shown in Eq. (\ref{eq:assign_label}), we assign cluster identifier $\tilde{m} = 1, \dots, \bar{M}$ to $q_{j^{\ast},m}(\phi_k)$ by constructing $\hat{\rho}_{\phi_k}$ an increasing injective map from $\big\{1, \dots, M(\phi_k) \big\}$ to $\{ 1, \dots, \bar{M} \}$,
\begin{equation}
    \hat{\rho}_{\phi_k} \in \arg \min_{\rho_{\phi_k}} \sum_{m=1}^{M(\phi_k)} \frac{\left( \eta^*(\phi_k, q_{j^{\ast},m}) - \mu_{\rho_{\phi_k}(m)} \right)^2}{\sigma^2_{\rho_{\phi_k}(m)}}, \notag
\end{equation}
where $\mu_{\tilde{m}} = \frac{1}{\tilde{K}} \sum_{\phi_k: M(\phi_k) = \bar{M}} \eta^*(\phi_k,q_{j^{\ast},\tilde{m}})$ and $\sigma^2_{\tilde{m}} = \frac{1}{\tilde{K}-1} \sum_{\phi_k: M(\phi_k) = \bar{M}} \big(\eta^{\ast}(\phi_k,q_{j^{\ast},\tilde{m}}) - \mu_{\tilde{m}} \big)^2$. We then construct $D(\phi_k)$ in the same way as the switching-label projection confidence sets. 

For each impulse response $\eta(\phi,q_{j^{\ast}}) = e_{i}' C_{h}(\phi) q_{j^{\ast}}$, $i=1, \dots, n$, and $h=0, 1, \dots, $ we construct
\begin{equation}
C_{\tilde{m}} = \left[ \min_{\phi_k: D_{\tilde{m}}(\phi_k)=1} \eta(\phi_k,q_{j^{\ast},\hat{\rho}_{\phi_k}^{-1}(\tilde{m})}), \max_{\phi_k: D_{\tilde{m}}(\phi_k)=1} \eta(\phi_k,q_{j^{\ast},\hat{\rho}_{\phi_k}^{-1}(\tilde{m})}) \right] \notag
\end{equation} 
and form confidence sets by taking the union over $\tilde{m}$ as in Eq. (\ref{eq:union_Cm}). 

In contrast to the switching-label procedure, the fixed-label projection confidence sets keep the labeling of the observationally equivalent impulse responses $\hat{\rho}_{\phi_k}(m)$ fixed over variables $i=1, \dots, n$ and different horizons $h=0,1, \dots$. If the impulse response $\eta^{\ast}(\phi,q_{j^{\ast}})$ chosen to anchor the labels can tie the observationally equivalent impulse responses to different economic models or hypotheses, the labels can be interpreted as indexing the underlying economic model or hypothesis and kept invariant throughout the impulse response analysis. The fixed-label projection confidence sets approach is suitable in such a case, and allows us to track and compare the observationally equivalent impulse response functions across different models. 

\subsubsection{Robust Bayesian interpretation}
\label{sec:RobBayesInterpret}

If we obtain $\{ \phi_k : k=1,\dots, K \}$ as draws from the credible region of the posterior distribution for $\phi$, $\widehat{CS}^p_{\eta,\alpha}$ can be seen as an approximation of the set $C_{\eta,\alpha}$ satisfying
\begin{equation}
\pi_{\phi|Y}\big(IS_{\eta}(\phi) \subset C_{\eta,\alpha}\big) \geq \alpha. \notag
\end{equation}
In terms of the robust Bayesian procedure proposed in \citet{GK18}, $C_{\eta,\alpha}$ can be interpreted as a robust credible region with credibility $\alpha$; a set of $\eta$ on which a posterior distribution for $\eta$ assigns probability at least $\alpha$ irrespective of the choice of the unrevisable part of the prior $\pi_{Q|\phi}$. Our construction of the robust credible region can be conservative and is not guaranteed to provide the shortest one. We leave the construction of the shortest robust credible region for future research.

This link to robust Bayes inference also suggests that the \textit{range} of posterior probabilities (lower and upper probabilities) spanned by arbitrary conditional priors for $Q$ given $\phi$ can be computed straightforwardly if we can draw $\phi$ from the posterior. 
Let $\{ \phi_{\ell }: \ell=1,\dots,L \}$ be Monte Carlo draws from $\pi_{\phi|Y}$ and $H_0 \subset \mathbb{R}$ be a hypothesis of interest. 
By applying Corollary A.1 of \citet{GK18}, the range of posterior probabilities for $\{ \eta \in H_0 \}$ is given by the convex interval: 
\begin{equation}
        \label{posterior prob range}
        \pi_{\eta|Y}(H_0) \in \big[ \pi_{\eta|Y \ast} (H_0), \pi_{\eta|Y}^{\ast} (H_0) \big] \equiv 
        \big[ \pi_{\phi|Y} \big(IS_{\eta}(\phi) \subset H_0\big), \pi_{\phi|Y} \big(IS_{\eta}(\phi) \cap H_0 \neq \emptyset\big) \big]. 
\end{equation}
Since the algorithms given in Section \ref{sec:Estimation} exhaust all the locally identified parameter values in $IS_{\eta}(\phi)$, we can approximate the lower and upper bounds of the posterior probabilities in Eq. (\ref{posterior prob range}) for each hypothesis of interest by the Monte Carlo frequencies for $\big\{ IS_{\eta}(\phi) \subset H_0 \big\}$ and $\big\{ IS_{\eta}(\phi) \cap H_0 \neq \emptyset \big\}$, respectively,
\begin{equation}
        \big[ \hat{ \pi}_{\eta|Y \ast} (H_0), \hat{\pi}_{\eta|Y}^{\ast} (H_0) \big] \equiv 
        \left[ \frac{1}{L} \sum_{\ell=1}^L 1\big\{IS_{\eta}(\phi_{\ell}) \subset H_0 \big\}, \frac{1}{L} \sum_{\ell=1}^L 1 \big\{ IS_{\eta}(\phi_{\ell}) \cap H_0 \neq \emptyset\big\} \right]. \notag
\end{equation}

For a scalar impulse response, it is also straightforward to compute the range of posterior means. Let $\underline{\eta}(\phi) =\min\big\{ \eta \in IS_{\eta}(\phi) \big\}$ and $\bar{\eta}(\phi) = \max\big\{ \eta \in IS_{\eta}(\phi) \big\}$. Theorem 2 in \citet{GK18} shows that the range of posterior means is given by the connected interval $\big[E_{\phi|Y}\big(\underline{\eta}(\phi)\big), E_{\phi|Y}\big(\bar{\eta}(\phi)\big)\big]$, which can be approximated by Monte Carlo analogues based on draws $\{\phi_{\ell}: \ell=1, \dots, L \}$ from $\pi_{\phi|Y}$.  


\section{Monetary policy, real activity and credit spreads: evidence from a locally-identified HSVAR}
\label{sec:EmpApp}

Our empirical application builds on recent work of \cite{CH2019AEJ} that investigates the relationships between monetary policy, real activity and credit spreads, and emphasize the role of latters in identifying the monetary policy shocks. 
They construct a proxy for monetary policy shocks from the high frequency data by detecting the unexpected policy interventions announced in Federal Open Market Committee (FOMC) statements and perform proxy SVAR. Since the FOMC started issuing statements straight away after each meeting only in 1994, their sample of observations covers the period 1994-2007. 

We complement their investigation by considering a larger sample and a different identification approach. Our strategy is to exploit the change in volatility occurred between the great inflation and the great moderation periods and employ the identification approach of HSVAR. This strategy allows us to expand the sample to cover the observations before 1994. 

Following the specification in \cite{CH2019AEJ}, we consider a five-equation VAR for the federal funds rate ($i_t$), the log of manufacturing industrial production ($ip_t$), the unemployment rate ($u_t$), the log of the producer price index for finished goods ($p_t$) and the Moody's seasoned BAA corporate bond yield relative to the yield on ten-year treasury constant maturity ($baa_t$). The data cover the period 1954:07-2007:6, monthly frequency. For the period in common, they coincide with \citeauthor{CH2019AEJ}'s ones. Data source is FRED. 


Concerning the break in the variances, based on evidence presented in \cite{SW02}, we set our break date to the beginning of 1984, although all the results are robust to any alternative break dates within the 1983-1985 interval.\footnote{The iterative procedure of \cite{LMNS21} to detect the break date suggests the break date being November 1986. Our results are also robust to this choice of break date.} 

We specify the autoregressive part of the VAR to include a constant and 3 lags.\footnote{We obtain this specification according to the Hannan-Quinn information criterion. This choice is parsimonious from one side, and guarantees no significant autocorrelation on the residuals from the other side. These results are available from the authors upon request.} For simplicity, we first estimate the reduced form of the model through a feasible generalized least squares (FGLS) estimator.\footnote{See \cite{Lan_Lut2008} and \cite{BBKM} for all the details about the FGLS estimator of heteroskedastic VARs.} The time series included in the VAR and the reduced-form residuals are shown in Appendix \ref{sec:fig}.

\begin{table}[t]
\centering
\begin{threeparttable}
\caption{Estimated eigenvalues ($\lambda_j$) and eigenvectors ($Q$) for the heteroskedastic VAR estimated through FGLS}
\label{tab:TabLam}
\begin{tabular}{cc @{\hspace{2cm}} ccccccc}
\hline\hline
\multicolumn{1}{c}{\footnotesize{panel (a)}}& & & &\multicolumn{5}{c}{\footnotesize{panel (b)}}\\
\hline
 & & & &\multicolumn{5}{c}{\footnotesize{$\hat{Q}$}}\\
\multicolumn{1}{c}{\footnotesize{$\hat{\lambda}_j$}} & & & & $\hat{q}_1$ & $\hat{q}_2$ & $\hat{q}_3$ & $\hat{q}_4$ & $\hat{q}_5$\\
\hline
0.12	& & & &	-0.99	&	0.08	&	-0.01	&	-0.05	&	0.05	\\
0.30	& & & &	-0.08	&	-0.99	&	-0.03	&	-0.05	&	-0.04	\\
0.44	& & & &	-0.04	&	-0.01	&	-0.63	&	0.77	&	-0.14	\\
0.66	& & & &	0.03	&	-0.05	&	0.13	&	0.28	&	0.95	\\
1.16	& & & &	0.05	&	0.04	&	-0.77	&	-0.57	&	0.27	\\
\hline\hline
\end{tabular}
    \begin{tablenotes}
      \small
      \item {\scriptsize{\textit{Notes}: Panel (a) shows the estimated eigenvalues, $\hat{\lambda}_j$ for $j=1,\ldots,5$. Panel (b) shows the related estimated eigenvectors $\hat{q}_j$ for $j=1,\ldots,5$, forming the orthogonal rotation matrix $Q$.}}
    \end{tablenotes}

 \end{threeparttable}
\end{table}

Table \ref{tab:TabLam} presents estimates for the eigenvalues and eigenvectors in the $\Lambda$ and $Q$ matrices, respectively, where the eigenvalue estimates are sorted in the increasing order. 
The matrices of $\Lambda$ and $Q$ can be pinned down uniquely under the ordering and sign normalization restrictions, while in the absence of additional restrictions, the impulse response functions to a shock of interest is not globally identified because of indeterminacy of labeling among the structural shocks, i.e., any of the five column vectors of $Q$ generate observationally equivalent impulse response functions to the monetary policy shock.    

\begin{figure}[ht!]
        \caption{Impulse response of the federal funds rate, industrial production and unemployment to two potential monetary policy shocks.}
  \label{fig:IRF_GLS}
\begin{center}
    \subfigure{\includegraphics[angle=270,origin=c,scale=0.225]{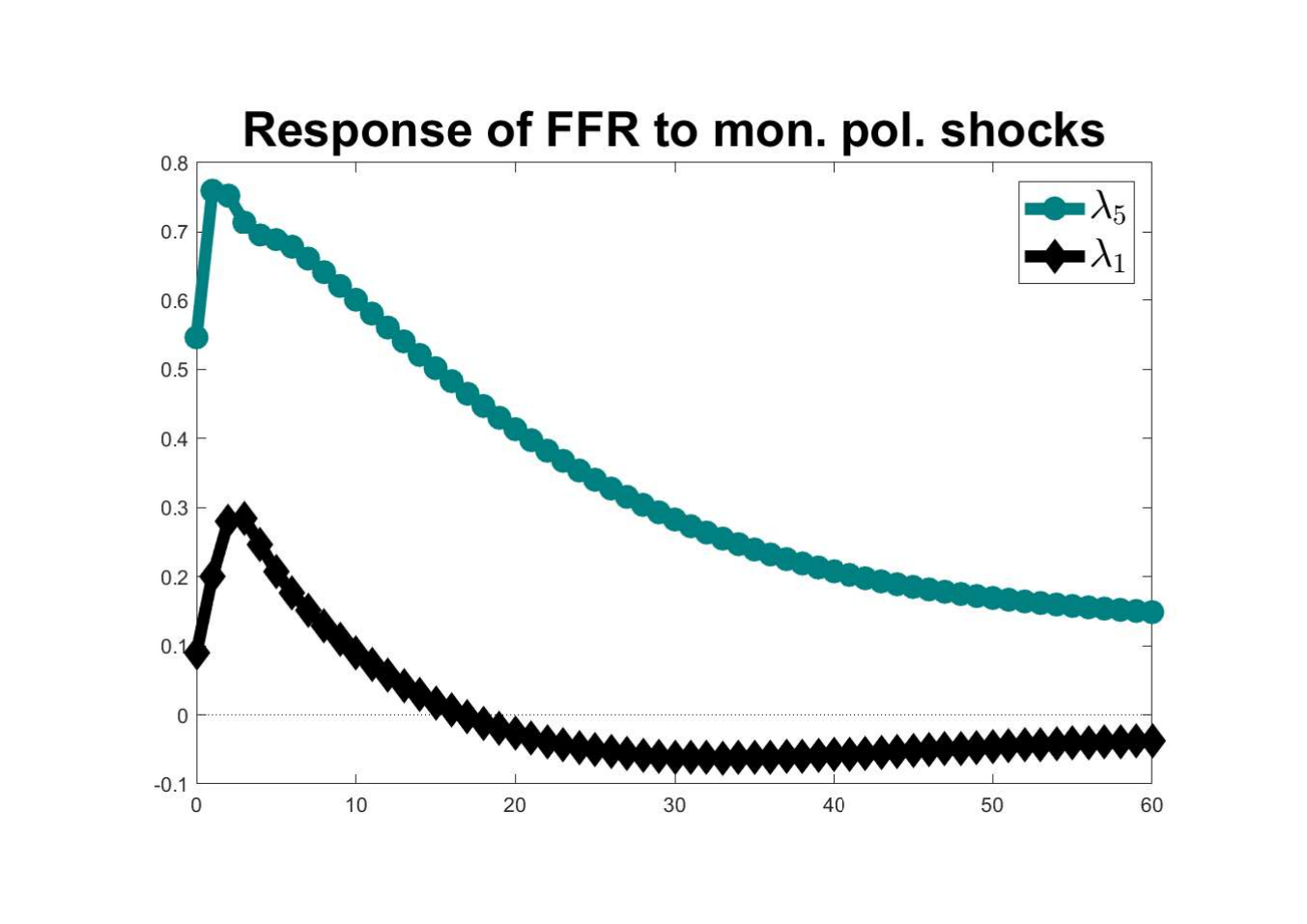}}
    \subfigure{\includegraphics[angle=270,origin=c,scale=0.225]{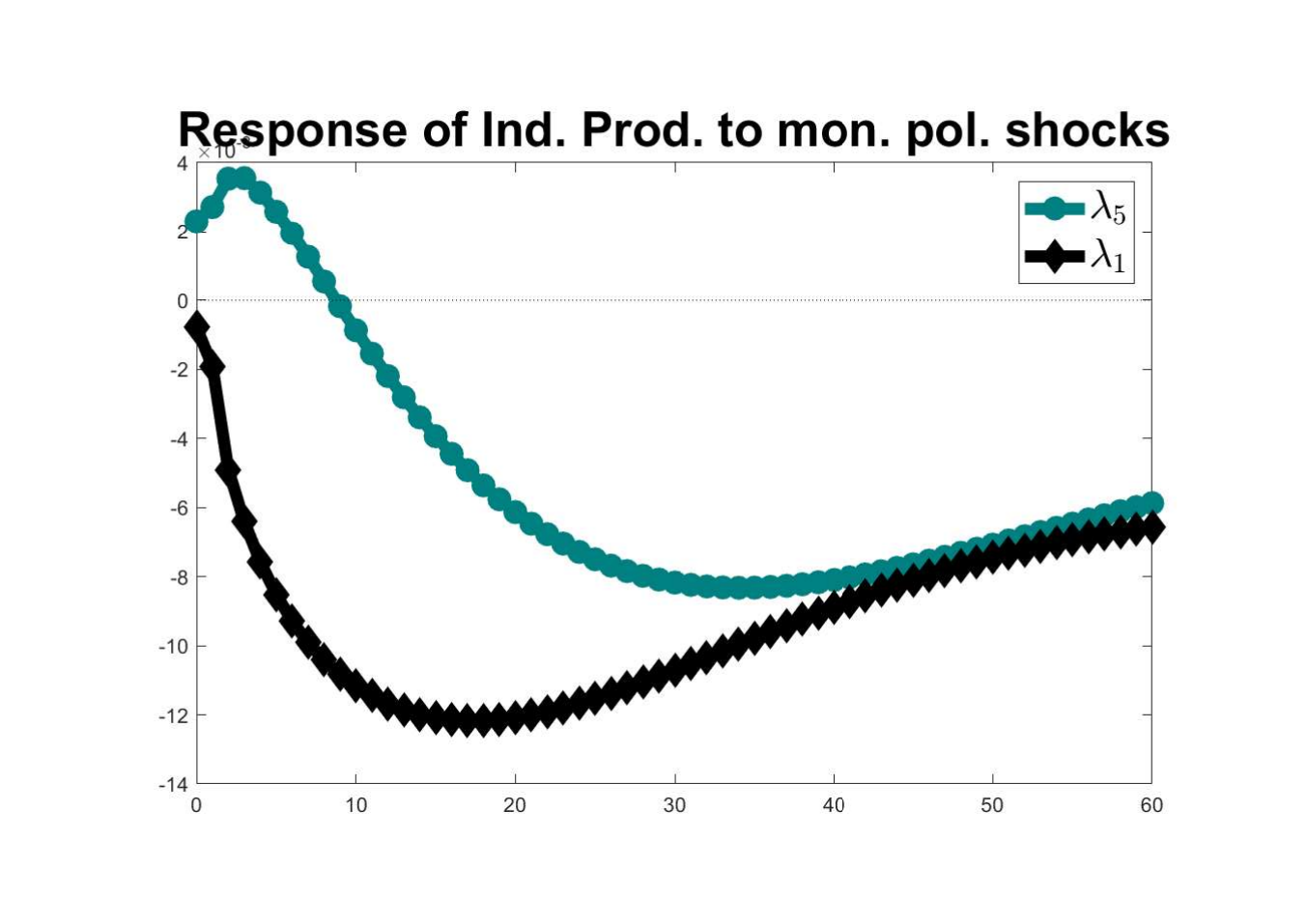}}
    \subfigure{\includegraphics[angle=270,origin=c,scale=0.225]{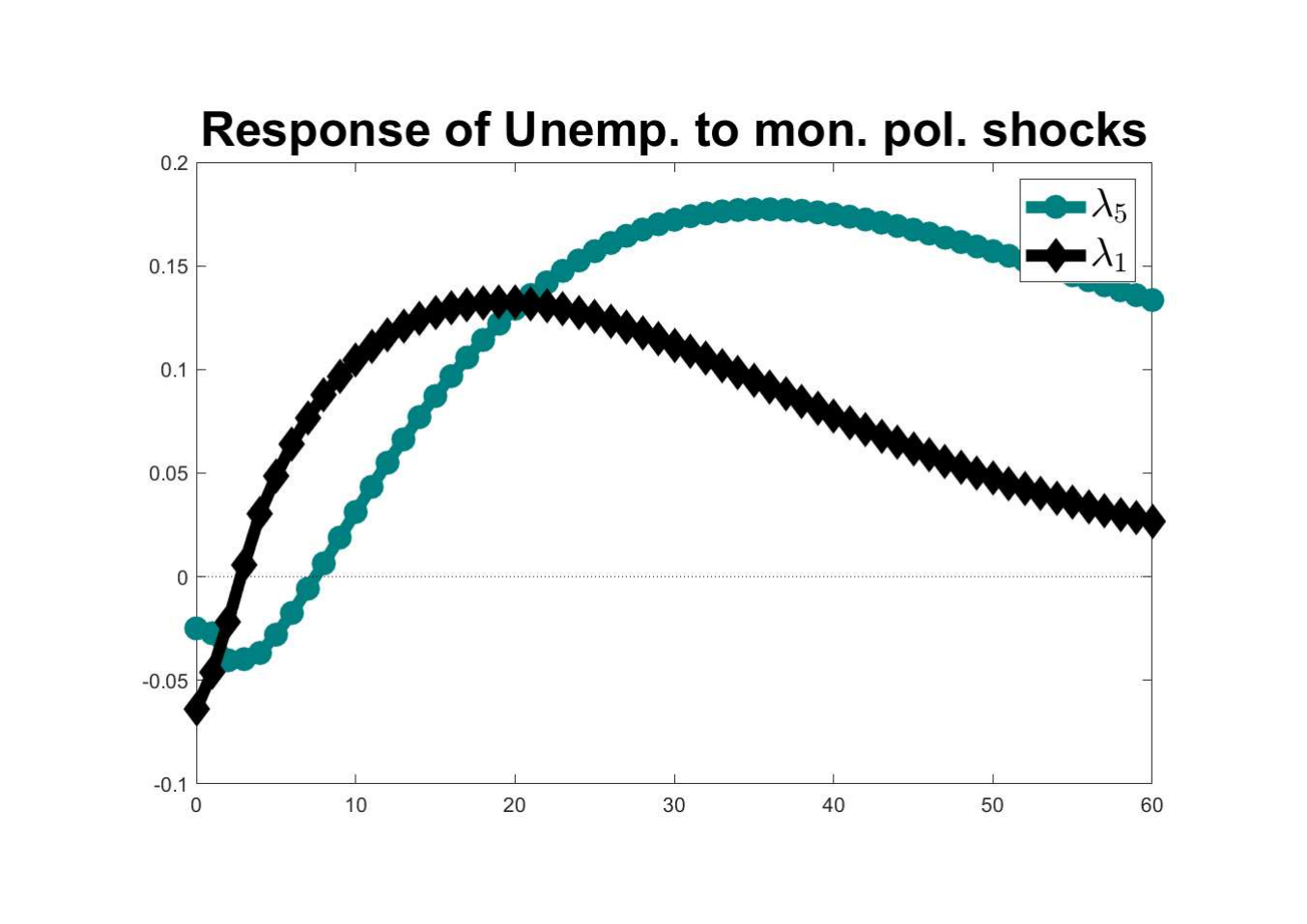}}
\end{center}
\end{figure}

To refine the identified set of the impulse responses, we impose sign restrictions such that the impulse response of the policy rate to monetary policy shock is persistently positive. In our specific case, there are two potential candidates that lead to impulse response functions on the federal funds rate $i_t$ being in line with those generally found in empirical contributions (\citeauthor{CH2019AEJ}, \citeyear{CH2019AEJ}, among many others). 
These impulse responses reported in the left panel of Figure \ref{fig:IRF_GLS} are corresponding to the two eigenvalues $\hat{\lambda}_1=0.12$ and $\hat{\lambda}_5=1.16$. If we had credible assumption about the direction of the change in volatility of monetary policy shocks across the regimes, it could be used to pin down the eigenvalue corresponding to the monetary policy shock, but to our knowledge, the literature does not provide a clear consensus on it.
As a consequence, if we rank first the monetary policy shock, the two rotation matrices $\hat{Q}_1=\big(\hat{q}_1,\:\hat{q}_2,\:\hat{q}_3,\:\hat{q}_4,\:\hat{q}_5\big)$ and $\hat{Q}_2=\big(\hat{q}_5,\:\hat{q}_2,\:\hat{q}_3,\:\hat{q}_4,\:\hat{q}_1\big)$ are observationally equivalent, and hence HSVAR is locally identified. Moreover, as reported in the middle and right panels of Figure \ref{fig:IRF_GLS}, if we look at the impact of these two potential monetary policy shocks on industrial production $ip_t$ and unemployment $u_t$, they produce similar recessionary responses of the two real business cycle indicators, suggesting that it is difficult to argue that one is more plausible than the other in terms of a restriction postulated by economic theory. 

Allowing for local identification of the impulse responses under these sign restrictions, we perform the inference methods proposed in Sections \ref{sec:Estimation} and \ref{sec:InfLocId} for the impulse responses to monetary policy shock. 
We first estimate the reduced-form parameters of the heteroskedastic VAR using a Bayesian approach as implemented in \cite{BBKM}, where we specify the reduced-form errors to follow Gaussian with a normal and inverse Wishart prior for $(B,\Sigma_1, \Sigma_2)$ conditional on the known break $T_B$. 
We obtain posterior draws of $(B,\Sigma_1, \Sigma_2)$ by Gibbs sampling.  See \cite{BBKM} for the details of implementation. 
Given each draw of reduced-form parameters, we compute observationally equivalent impulse responses by running Algorithm \ref{algo:EstimationHSVAR} in Section \ref{sec:AlgorithmHSVAR}. After discarding 500 draws for burn-in, we obtain 2000 posterior draws of the reduced-form parameters and, at every draw, the identified set $\mathcal{Q}_R(\hat{\phi})$ is non-empty and consists of two orthogonal matrices that satisfy the eigen-decomposition (\ref{eigen decomposition}) and they correspond to different permutations of the eigenvalues and eigenvectors. 

\begin{figure}[ht!]
        \caption{Impulse response of the federal funds rate, industrial production, unemployment and credit spread to a contractionary monetary policy shock in the locally-identified HSVAR: No sign restrictions.}
  \label{fig:IRF_HSVAR_NoSign}
\begin{center}
                \subfigure{
                \includegraphics[angle=0,origin=c,scale=0.3]{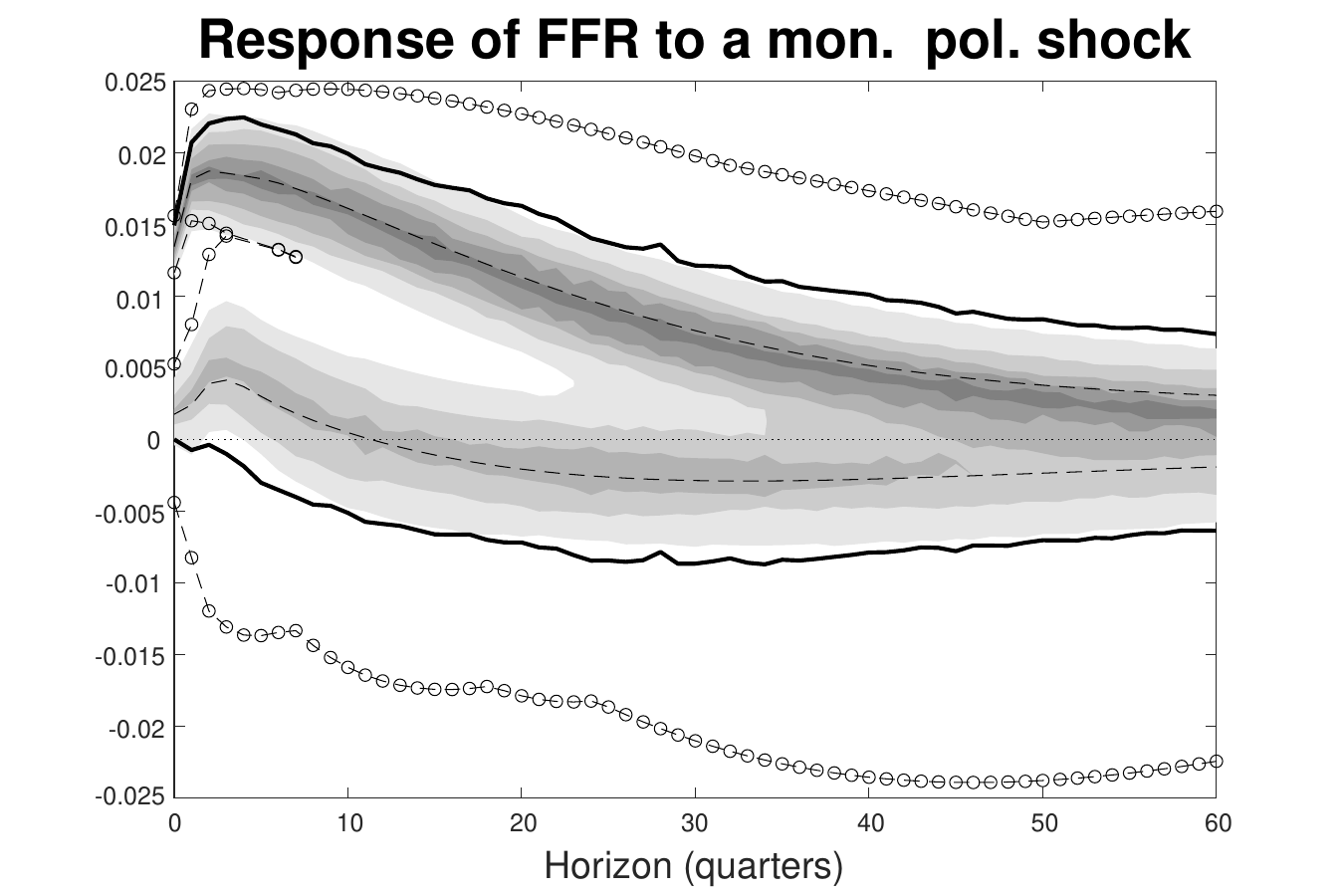}}
                \hspace{1.5cm}
                \subfigure{
                \includegraphics[angle=0,origin=c,scale=0.3]{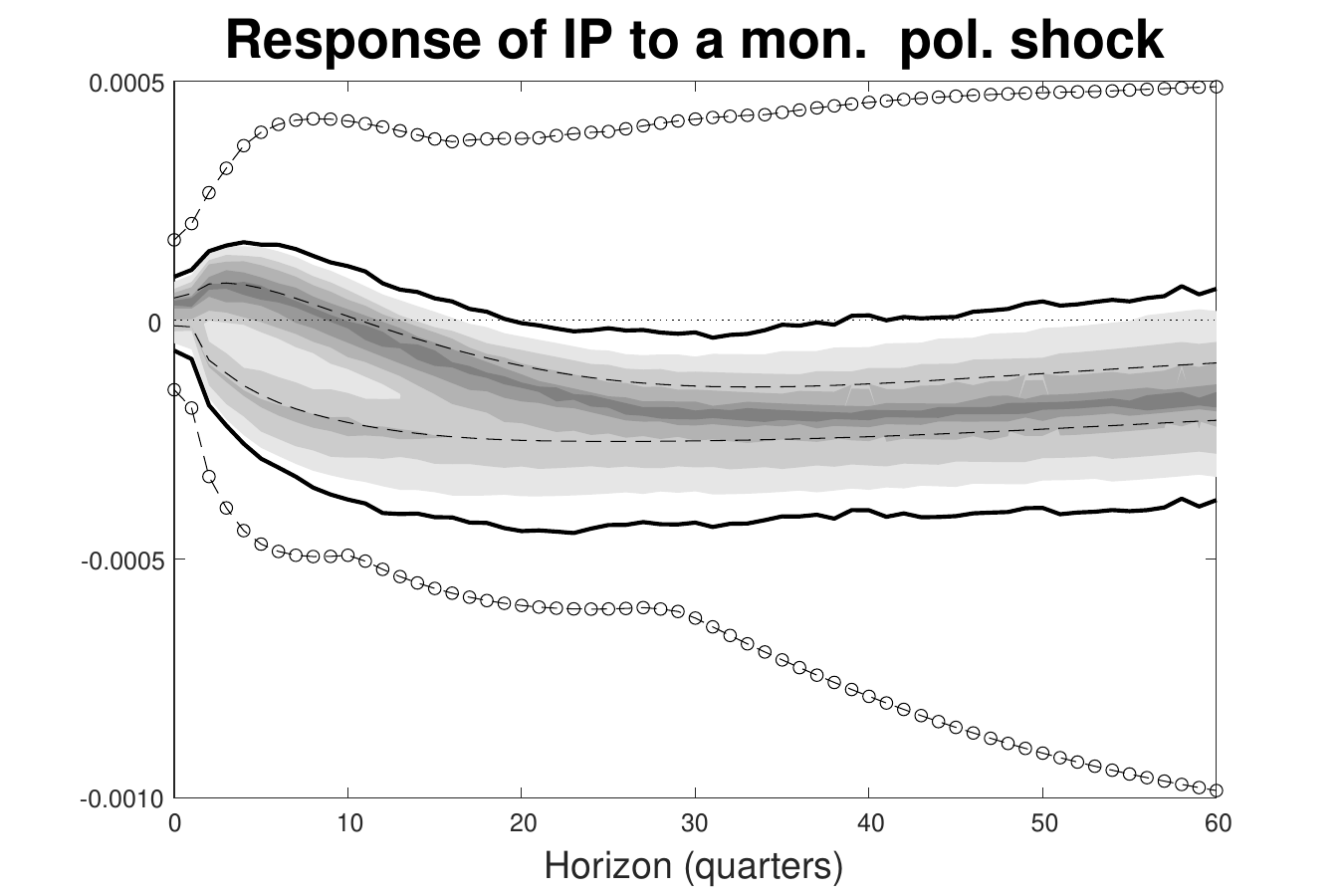}}
                \subfigure{
                \includegraphics[angle=0,origin=c,scale=0.3]{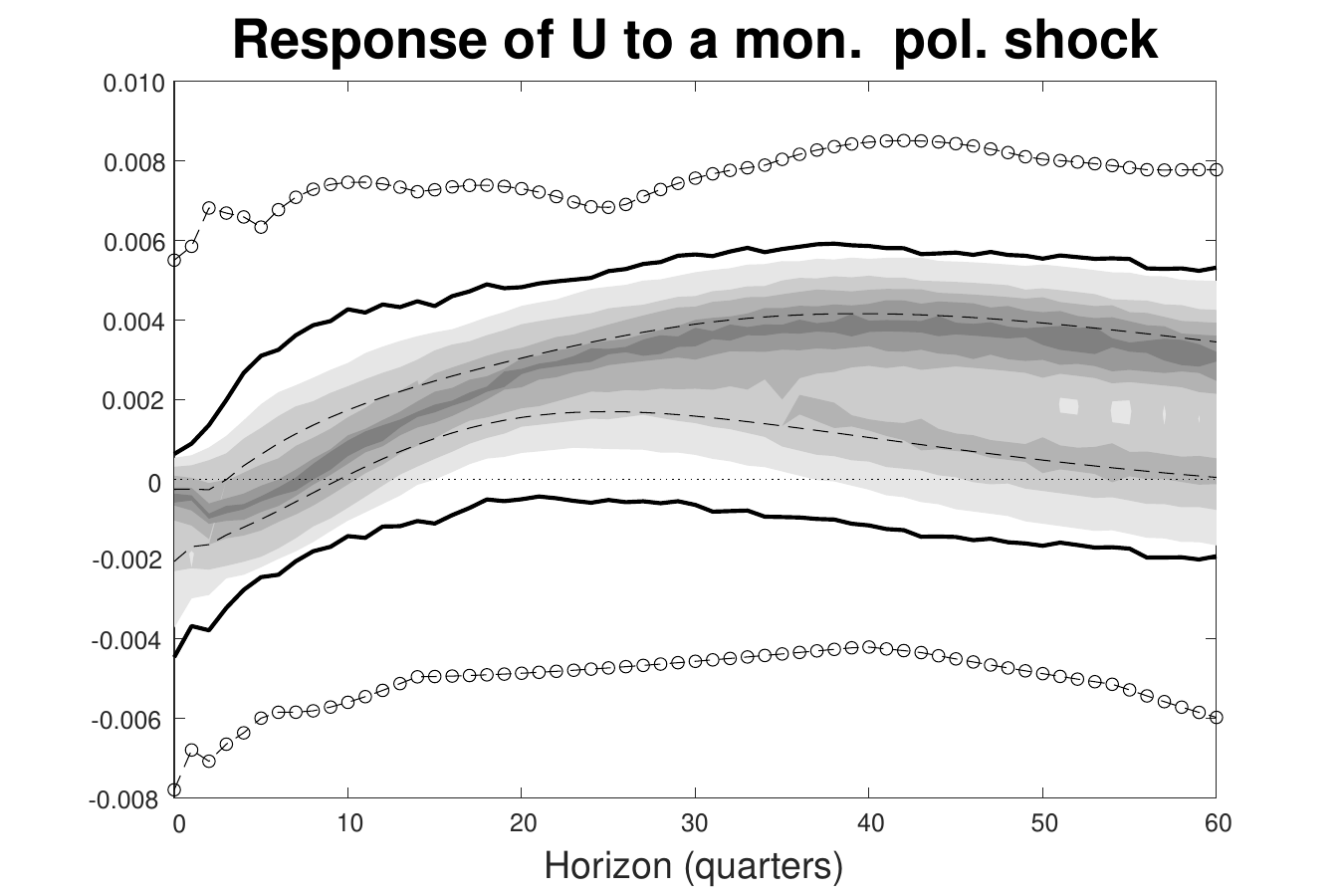}}
                \hspace{1.5cm}
                \subfigure{
                \includegraphics[angle=0,origin=c,scale=0.3]{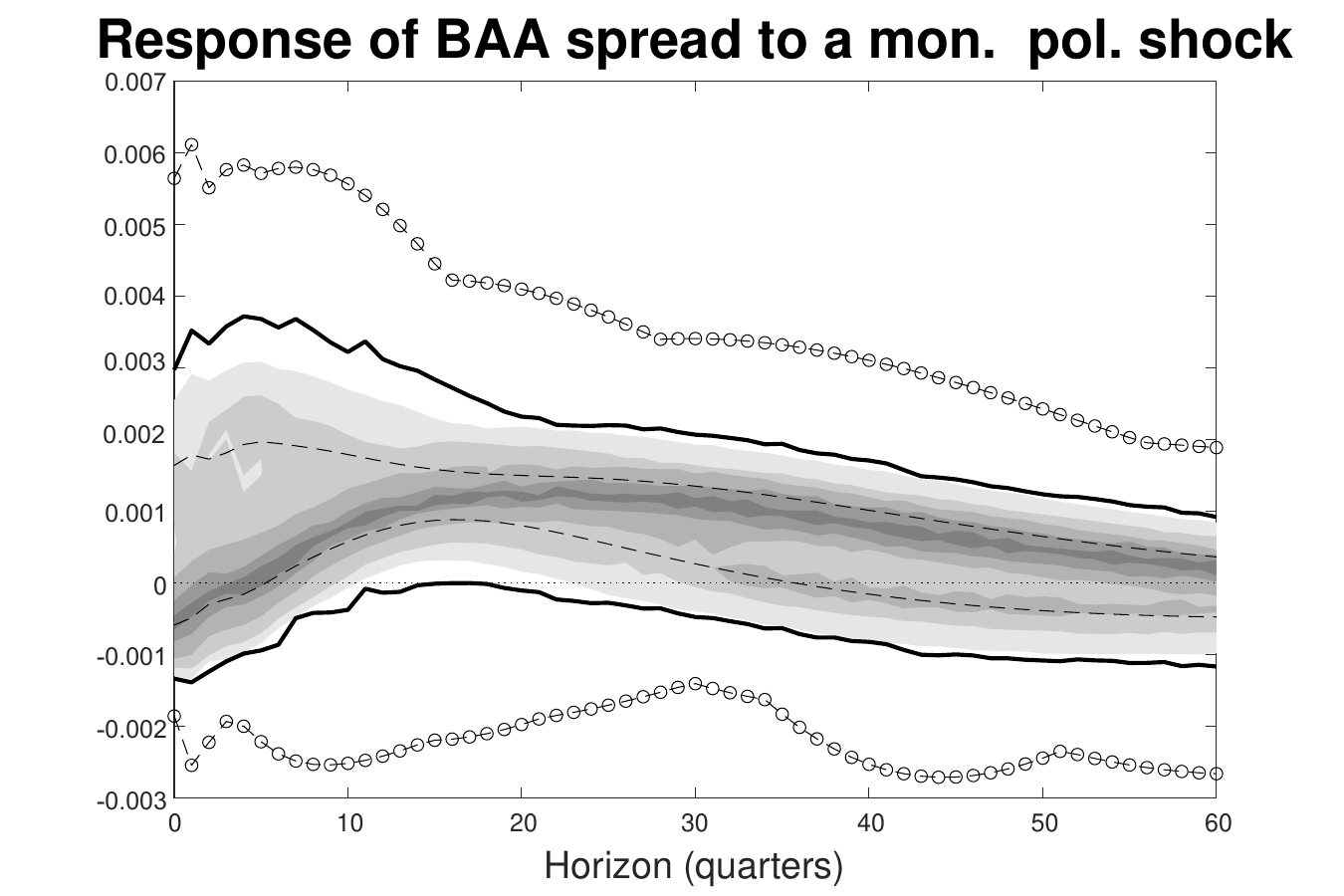}}
\end{center}
\begin{minipage}{\textwidth}%
{\scriptsize{\textit{Notes}: Each graph reports the posterior highest density regions at $90\%$, $75\%$, $50\%$, $25\%$ and $10\%$ in gray scale. The upper and lower bounds of the frequentist confidence sets with coverage $90\%$ obtained through the fixed-label projection approach are plotted by the dotted-circle lines. The dotted lines plot the set of posterior means, while the solid curves are the upper and lower bounds of the robust credible regions with credibility $90\%$.\par}}
\end{minipage}
\end{figure}

Figure \ref{fig:IRF_HSVAR_NoSign} shows the inference results for the impulse responses to monetary policy shock obtained by the Bayesian, robust Bayesian, and frequentist approaches. The Bayesian posteriors marginalized at each horizon are shown as heat plots with gray scales. From lightest to darkest, we report the highest density regions with credibility $90\%$, $75\%$, $50\%$, $25\%$ and $10\%$.\footnote{The highest posterior density regions are computed by slicing the marginal posterior density approximated with kernel smoothing of the posterior draws of the impulse responses.} For the Bayesian approach, we have used a uniform conditional prior (equal weights) over each admissible $Q$ given the reduced-form parameters $\phi$. Robust Bayesian inference that views the set of admissible $Q$ matrices given $\phi$ as the identified set reports the set of posterior means (dotted lines) and the bounds of the robust credible regions (solid lines) with credibility 90\%. Frequentist-valid confidence intervals with coverage 90\% are plotted by the dotted-circle lines. They are obtained by retaining the $90\%$ of the draws of $\phi$ with the highest likelihood value and then calculating the fixed-label projection confidence sets presented in Section \ref{sec:ConstantLabel}. The related switching-label projection confidence sets are practically indistinguishable. 

As the posterior distributions show, local identification with two admissible solutions generates bimodal posterior distributions for the impulse responses, which are particularly visible in the response of the federal funds rate and industrial production for the first 10 - 20 quarters. 
In line with Figure \ref{fig:IRF_GLS}, the modes share positive responses for the federal funds rate but with different magnitudes and persistence. In contrast, bimodality of the posteriors of the impulse responses of industrial production and unemployment is less evident. The posteriors indicate recessionary responses to the monetary policy shock in the short-run, though the robust-Bayesian and frequentist confidence intervals are wide. The credit spread responds positively and significantly to the monetary policy shock at least in the medium run. These empirical results are robust and in line with the economic theory despite the lack of global identification. 

In summary, with our identification strategy and the extended sample, we have local identification with two admissible impulse responses but our empirical findings incorporating local identification support the main empirical results obtained by \cite{CH2019AEJ}. 


\section{Conclusion}
\label{sec:conclusion}

This paper analyzes SVARs that attain local identification but may fail to attain global identification. We identify the class of identifying restrictions that delivers local but non-global identification. This is characterized by non-recursive and/or across-shock equality restrictions. Similar situations might appear also in SVARs identified through heteroskedasticity, non-normality or through external instruments. Exploiting the geometric structure of the identification problem, we propose a novel way to analyze and exhaustively compute the observationally equivalent impulse responses. The novel analytical and computational insights also contribute to the development of a posterior sampling algorithm for Bayesian inference and projection-based frequentist-valid inference in the presence of locally identified parameters.  

Our approach for estimation and inference can become a reference in all cases in which, mainly for the complexity of the problem, identification can be addressed only locally, like in many DSGE models, or in generalized method of moments (GMM) or minimum distance (MD) estimation strategies. To the best of our knowledge, \cite{Sentana23} is the only work addressing the issue of local identification, that he calls \textit{finite underidentification}, in a GMM framework. In his contribution, he provided interesting empirically relevant cases where local identification might happen, and, in line with our standpoint, stresses the importance of considering all the admissible solutions, both in terms of efficiency gains and as a strategy for yielding underidentification tests. Extending our computational and inference approaches to these estimation strategies, as well as to locally identified DSGE models, is a promising avenue for future research.  

\newpage


\renewcommand{\thesection}{\Alph{section}}\setcounter{section}{0}

\setlength{\baselineskip}{12pt} 
\bibliographystyle{ecta}
\bibliography{SVAR}

@Article{AH95,
  author      = {K. G. Abraham and J. C. Haltiwanger},
  title       = {Real Wages and the Business Cycle},
  journal     = {Journal of Economic Literature},
  year        = {1995},
  volume      = {33},
  number      = {3},
  pages       = {1215--1264},
  month       = {September},
}

@Book{Giannini92,
  title     = {Topics in structural {VAR} econometrics},
  publisher = {Springer-Verlag},
  year      = {1992},
  author    = {Carlo Giannini},
}

@Book{AG,
  author =       {Gianni Amisano and Carlo Giannini},
  title =        {Topics in structural {VAR} econometrics},
  publisher =    {Springer-Verlag},
  year =         {1997},
  edition =      {2nd},
}

@TechReport{BK_SVARWB,
  author  = { (Emanuele Bacchiocchi and Toru Kitagawa},
  title = {SVARs with breaks: Identification and inference},
  year= {2022},
  type    = {Working Paper},
  institution = {mimeo},
}

@TechReport{BBKM,
  author  = {Emanuele Bacchiocchi and Andrea Bastianin and Toru Kitagawa and Elisabetta Mirto},
  title = {Partially identified heteroskedastic SVARs: Identification and inference},
  year= {2022},
  type    = {Working Paper},
  institution = {mimeo},
}

@Article{Bacchiocchi17,
  author      = {Emanuele Bacchiocchi},
  title       = {On the identification of interdependence and contagion of financial crises},
  journal     = {Oxford Bulletin of Economics and Statistics},
  year        = {2017},
  volume      = {79},
  number      = {6},
  pages       = {1148-1175},
  month       = {July},
  institution = {Department of Economics, Management and Quantitative Methods at Universit{\`a} degli Studi di Milano},
  type        = {mimeo},
}

@ARTICLE{BF15,
  author = {Emanuele Bacchiocchi and Luca Fanelli},
  title = {Identification in Structural Vector Autoregressive Models with Structural
	Changes with an Application to U.S. Monetary Policy},
  journal = {Oxford Bulletin of Economics and Statistics},
  year = {2015},
  volume = {77},
  number = {6},
	pages = {761--779}
}

@article{Bernanke86,
	author = {Bernanke, Ben},
	title = {Alternative explanations of the money-income correlation},
	journal = {Carnegie-Rochester Conference Series on Public Policy},
	year = 1986,
	volume = 25,
  number = 4,
	pages = {49--99}
}

@article{BlanQuah1,
	author = {Blanchard, Olivier and Quah, Danny},
	title = {The Dynamic Effects of Aggregate Demand and Aggregate Supply Shocks},
	journal = {American Economic Review},
	year = 1989,
	volume = 79,
           number = 4,
	pages = {655--73}
}

@article{BlanPerotti02,
	author = {Blanchard, Olivier and Perotti, Roberto},
	title = {An empirical characterization of the dynamic effects of changes in government spending and taxes on output},
	journal = {The Quarterly Journal of Economics},
	year = 2002,
	volume = 117,
           number = 4,
	pages = {1329--1368}
}

@incollection{BW86,
  author      = "Blanchard, O and Watson, M.",
  title       = "Are Business Cycles All Alike?",
  editor      = "R J Gordon",
  booktitle   = "The American Business Cycle",
  publisher   = "NBER and Chicago Press",
  year        = "1986",
  pages       = "123--179",
}

@ARTICLE{BPSS21,
AUTHOR={Markus Brunnermeier and Darius Palia and Karthik A. Sastry and Christopher A. Sims},
TITLE={Feedbacks: Financial Markets and Economic Activity},
JOURNAL={American Economic Review},
YEAR=2021,
VOLUME={111},
NUMBER={6},
PAGES={1845--1879},
MONTH={June},
}

@TechReport{TraynorCaron2005,
  author = 	 {Tim Traynor and Richard J. Caron},
  title = 	 {The Zero Set of a Polynomial},
  institution =  {Department of Mathematics and Statistics, University of Windsor, Windsor, ON Canada},
  year = 	 {2005},
  key = 	 {WSMR Report 05-02},
  OPTtype = 	 {},
  OPTnumber = 	 {},
  OPTaddress = 	 {},
  month = 	 {May},
  OPTnote = 	 {},
  OPTannote = 	 {}
}

@Article{DK11,
  author  = {Davis, L. W. and Kilian, L.},
  title   = {Estimating the Effect of a Gasoline Tax on Carbon Emissions},
  journal = {Journal of Applied Econometrics},
  year    = {2011},
  volume  = {26},
  number  = {7},
  pages   = {1187--1214},
  month   = {November},
}

@Article{DW21,
  author  = { (Thorsten Drautzburg and Jonathan Wright},
  title = {Refining Set-Identification in VARs through Independence},
  year= {2023},
  journal = {Journal of Econometrics},
  volume  = {forthcoming},
}

@ARTICLE{DM80,
AUTHOR={Angus Deaton and John Muellbauer},
TITLE={An Almost Ideal Demand System},
JOURNAL={American Economic Review},
YEAR=1980,
VOLUME={70},
NUMBER={3},
PAGES={312--326},
MONTH={June},
optNOTE={available at http://ideas.repec.org/a/aea/aecrev/v81y1991i4p819-40.html}
}

@Article{Lucchetti06,
  author  = {Lucchetti, Riccardo},
  title   = {Identification of Covariance Structures},
  journal = {Econometric Theory},
  year    = {2006},
  volume  = {22},
  number  = {02},
  pages   = {235--257},
  month   = {April},
}

@Article{GMR2017JoE,
  author={Gourieroux, C. and Monfort, A. and Renne, J.P.},
  title={Statistical inference for independent component analysis: Application to structural VAR models},
  journal={Journal of Econometrics},
  year=2017,
  volume={196},
  number={1},
  pages={111--126},
  month={January}
}

@Article{Johansen95,
  author = 	 {S{\o}ren Johansen},
  title = {Identifying restrictions of linear equations with
  applications to simultaneous equations and cointegration},
  journal = 	 {Journal of Econometrics},
  year = 	 {1995},
  volume = 	 {69},
  number = 	 {1},
  pages = 	 {112--132},
  month={September},
}

@Article{Kilian10,
  author = 	{Lutz Kilian},
  title = {Explaining Fluctuations in Gasoline Prices: A Joint Model of the Global
		Crude Oil Market and the U.S. Retail Gasoline Market},
  journal = 	 {The Energy Journal},
  year = 	 {2010},
  volume = 	 {31},
  number = 	 {2},
  pages = 	 {87--112},
  month={September},
}

@Article{Kilian98,
  author={Lutz Kilian},
  title={Small-Sample Confidence Intervals For Impulse Response Functions},
  journal={The Review of Economics and Statistics},
  year={1998},
  volume={80},
  number={2},
  pages={218--230},
  month={May}
 }

@Article{LL2021JBES,
  author={Markku Lanne and J Luoto},
  title={GMM Estimation of Non-Gaussian Structural Vector Autoregression},
  journal={Journal of Business \& Economic Statistics},
  year=2019,
  volume={39},
  number={1},
  pages={69--81},
  month={September}
}

@Article{Lan_Lut2008,
  author={Lanne, Markku and L\"utkepohl, Helmut},
  title={Identifying monetary policy shocks via changes in volatility},
  journal={Journal of Money, Credit and Banking},
  year=2008,
  volume={40},
  number={6},
  pages={1131--1149},
  month={September}
}

@Article{Lewis21Restud,
  author  = {D. J. Lewis},
  title   = {Identifying Shocks via Time-Varying Volatility},
  journal = {Review of Economic Studies},
  year    = {2021},
  volume={88},
  number={6},
  pages={3086--3124},
  month={November}
}

@Article{Lewis22Restat,
  author  = {D. J. Lewis},
  title   = {Robust Inference in Models Identified via Heteroskedasticity},
  journal = {Review of Economics and Statistics},
  year    = {2022},
  volume  = {104},
  number={3},
  pages={510--524},
}

@Book{LutBook06,
  title     = {New Introduction to Multiple Time Series},
  publisher = {Springer},
  year      = {2006},
  editor    = {L\"utkepohl, Helmut},
}

@BOOK{magnus88,
	author={Magnus, Jan},
	title={Linear Structures},
	publisher={Charles Griffin \& Co.},
	year= 1988
}

@Book{mn07,
  title     = {Matrix differential calculus with applications in statistics and econometrics},
  publisher = {John Wiley \& Sons},
  year      = {2007},
  author    = {Magnus, Jan and Neudecker, Heinz},
  edition   = {Third},
}

@Article{Rigobon03,
  author={Roberto Rigobon},
  title={{Identification Through Heteroskedasticity}},
  journal={The Review of Economics and Statistics},
  year=2003,
  volume={85},
  number={4},
  pages={777--792},
  month={November}
}

@Article{RWZ10,
  author  = {Rubio-Ramirez, Juan and Waggoner, Dan and Zha, Tao},
  title   = {Structural Vector Autoregressions: Theory of Identification and Algorithms for Inference},
  journal = {Review of Economic Studies},
  year    = {2010},
  volume  = {77},
  number  = {2},
  pages   = {665--696},
  month   = {04},
}

@Article{Sen_Fio01,
  author={Sentana, Enrique and Fiorentini, Gabriele},
  title={Identification, estimation and testing of conditionally heteroskedastic factor models},
  journal={Journal of Econometrics},
  year=2001,
  volume={102},
  number={2},
  pages={143--164},
  month={June}
}

@article{sims80,
	author = {Sims, Christopher A.},
	title = {Macroeconomics and Reality},
	journal = {Econometrica},
	year = 1980,
	volume = 48,
	pages = {1-48}
}

@article{sims86,
	author = {Sims, Christopher A.},
	title = {Are Forecasting Models Usable for Policy Analysis?},
	journal = {Minneapolis Federal Reserve Bank Quarterly Review},
	year = 1986,
	volume = 10,
	pages = {2-16}
}

@TechReport{Sims22,
  author  = {C Sims},
  title   = {SVAR Identification through Heteroskedasticity with Misspecified Regimes},
  year    = {2020},
  type    = {Unpublished},
	institution = {Princeton University},
}

@Article{SimsZha06AER,
  author  = {C.A. Sims and T. Zha},
  title   = {Were There Regime Switches in U.S. Monetary Policy?},
  journal = {American Economic Review},
  year    = {2006},
  volume  = {96},
	number  = {1},
  pages   = {54--81},
}

@Article{Uhlig2005,
  author  = {Harald Uhlig},
  journal = {Journal of Monetary Economics},
  title   = {What are the effects of monetary policy on output? Results from an agnostic identification procedure},
  year    = {2005},
  issn    = {0304-3932},
  number  = {2},
  pages   = {381 -- 419},
  volume  = {52},
  doi     = {http://dx.doi.org/10.1016/j.jmoneco.2004.05.007},
}

@TechReport{BL18,
  author      = {E. Bacchiocchi and R. Lucchetti},
  title       = {Structure-based SVAR identification},
  institution = {University of Milan},
  year        = {2018},
  type        = {mimeo},
}

@Article{HWZ07,
  author  = {J. Hamilton and D. Waggoner and T. Zha},
  title   = {Normalization in econometrics},
  journal = {Econometric Reviews},
  year    = {2007},
  volume  = {26},
  number  = {2-4},
  pages   = {221-252},
}

@TechReport{RWZ08,
  author      = {Rubio-Ramirez, Juan and Waggoner, Dan and Zha, Tao},
  title       = {Structural Vector Autoregressions: Theory of Identificatifion and Algorithms for Inference},
  institution = {Federal Reserve Bank of Atlanta},
  year        = {2008},
}

@Article{GK18,
  author  = {Raffaella Giacomini and Toru Kitagawa},
  journal = {Econometrica},
  title   = {Robust Bayesian Inference for Set-identified Models},
  year    = {2021},
  volume  = {89},
  number  = {4},
  pages   = {1519--1556},
}

@Book{Hamilton94,
  title     = {Time series analysis},
  publisher = {Princeton University Press, Princeton},
  year      = {1994},
  author    = {J. D. Hamilton},
}

@Book{Fisher66,
  title     = {The identification problem in econometrics},
  publisher = {McGraw-Hill, Inc.},
  year      = {1966},
  author    = {F M Fisher},
}

@Article{Kelly75,
  author  = {J S Kelly},
  title   = {Linear cross-equation constraints and the identification problem},
  journal = {Econometrica},
  year    = {1975},
  volume  = {43},
  pages   = {125-140},
}

@Article{WZ03,
  author  = {D. Waggoner and T. Zha},
  title   = {Likelihood preserving normalization in multiple equation models},
  journal = {Journal of Econometrics},
  year    = {2003},
  volume  = {114},
  number  = {2},
  pages   = {329-347},
}

@Article{WZ03Gibbs,
  author  = {D. Waggoner and T. Zha},
  title   = {A Gibbs Sampler for Structural Vector Autoregressions},
  journal = {Journal of Economic Dynamics and Control},
  year    = {2003},
  volume  = {28},
  number  = {2},
  pages   = {349-366},
}

@Article{CEE99,
  author  = {L.J. Christiano and M. Eichenbaum and C.L. Evans},
  title   = {Monetary policy shocks: What have we learned and to what end?},
  journal = {Handbook of Macroeconomics},
  year    = {1999},
  volume  = {1},
  number  = {Part A},
  pages   = {65-148},
}

@Article{MR2013,
  author  = {K. Mertens and M. Ravn},
  title   = {The dynamic effects of personal and corporate income tax changes in the United States},
  journal = {American Economic Review},
  year    = {2013},
  volume  = {103},
  number  = {4},
  pages   = {1212--1247},
}

@Article{LL10JBES,
  author  = {M. Lanne and H. L\"{u}tkepohl},
  title   = {Structural vector autoregressions with nonnormal residuals},
  journal = {Journal of Business \& Economic Statistics},
  year    = {2010},
  volume  = {28},
  number  = {1},
  pages   = {159--168},
}

@Article{LMS17JoE,
  author  = {M. Lanne and M. Meitz and P. Saikkonen},
  title   = {Identification and estimation of non-gaussian structural vector identification},
  journal = {Journal of Econometrics},
  year    = {2017},
  volume  = {196},
  number  = {2},
  pages   = {288--304},
}

@Article{ARW21,
  author  = {Jonas E. Arias and Juan F. Rubio-Ram\'{i}rez and Daniel F. Waggoner},
  title   = {Inference in Bayesian Proxy-SVARs},
  journal = {Journal of Econometrics},
  year    = {2021},
  volume  = {225},
  number  = {1},
  pages   = {88--106},
}

@Article{SW18,
  author  = {James H. Stock and Mark W. Watson},
  title   = {Identification and Estimation of Dynamic Causal Effects in Macroeconomics},
  journal = {Economic Journal},
  year    = {2018},
  volume  = {28},
  number  = {610},
  pages   = {917-948},
}

@Article{SW12,
  author  = {James H. Stock and Mark W. Watson},
  title   = {Disentangling the Channels of the 2007-09 Recession},
  journal = {Brookings Papers on Economic Activity},
  year    = {2012},
  number  = {1},
  pages   = {81-135},
}

@Article{AF19,
  author  = {Giovanni Angelini and Luca Fanelli},
  title   = {Exogenous uncertainty and the identification of Structural Vector Autoregressions with external instruments},
  journal = {Journal of Applied Econometrics},
  year    = {2019},
  volume  = {34},
  number  = {6},
  pages   = {951-971},

}

@Article{Rothenberg71ECTA,
  author    = {T.J. Rothenberg},
  title     = {Identification in parametric models},
  journal   = {Econometrica},
  year      = {1971},
  volume    = {39},
  pages     = {577-591},
}

@Book{Dhrymes78,
  title     = {Introductory Econometrics},
  publisher = {Springer-Verlag},
  year      = {1978},
  author    = {P.J. Dhrymes},
}

@Article{LSZ96,
  author    = {Leeper, E.M. and Sims, C. and Zha, T.},
  title     = {What Does Monetary Policy Do?},
  journal   = {Carnegie Rochester Conference Series on Public Policy},
  year      = {1996},
  volume    = {2},
  pages     = {1-78},
}

@Article{SZ99,
  author    = {Sims, C. and Zha, T.},
  title     = {Error Bands for Impulse Responses},
  journal   = {Econometrica},
  year      = {1999},
  volume    = {67},
  pages     = {1113-1155},
}

@Article{Faust98,
  author    = {J. Faust},
  title     = {The Robustness of identified VAR Conclusions about Money},
  journal   = {Carnegie Rochester Conference Series on Public Policy},
  year      = {1998},
  volume    = {49},
  pages     = {207-244},
}

@Article{CanovadeNicolo02JME,
  author  = {F. Canova and G. {de Nicol{\'o}}},
  title   = {Monetary Disturbances Matter for Business Fluctuations in the G-7},
  journal = {Journal of Monetary Economics},
  year    = {2002},
  volume  = {49},
  pages   = {1131-1159},
}

@Article{MU09,
  author    = {Mountford, A. and Uhlig, H.},
  title     = {What Are the Effects of Fiscal Policy Shocks?},
  journal   = {Journal of Applied Econometrics},
  year      = {2009},
  volume    = {24},
  number    = {6},
  pages     = {960-992},
}

@Article{MSG13,
  author    = {E. Granziera and H.R. Moon and F. Schorfheide},
  title     = {Inference for VARs Identified with Sign Restrictions},
  journal   = {Quantitative Economics},
  year      = {2018},
  volume    = {9},
  number    = {3},
  pages     = {1087-1121},
}

@Article{KT13,
  author    = {B. Kline and E. Tamer},
  title     = {Bayesian Inference in a Class of Partially Identified Models},
  journal   = {Quantitative Economics},
  year      = {2016},
  volume    = {7},
  number    = {2},
  pages     = {329-366},
}

@Article{CCT18,
  author  = {X. Chen and T. Christensen and E. Tamer},
  title   = {Monte Carlo Confidence Sets for Identified Sets},
  journal = {Econometrica},
  year    = {2018},
  volume  = {86},
  number  = {6},
  pages   = {1965--2018},
}

@Article{MS12,
  author    = {{H.R.} Moon and F. Schorfheide},
  title     = {Bayesian and Frequentist Inference in Partially Identified Models},
  journal   = {Econometrica},
  year      = {2012},
  volume    = {80},
  number    = {2},
  pages     = {755-782},
}

@Article{LS13,
  author    = {Y. Liao and A. Simoni},
  title     = {Bayesian Inference for Partially Identified Smooth Convex Models},
  journal   = {Journal of Econometrics},
  year      = {2019},
  volume    = {211},
  number  	= {2},
  pages   	= {338--360},
}

@Article{GKV18,
  author  = {R. Giacomini and T. Kitagawa and A. Volpicella},
  title   = {Uncertain Identification},
  journal = {Quantitative Economics},
  year    = {2020},
  volume  = {13},
  number  = {1},
  pages   = {95-123},
}

@Article{GKU19,
  author  = {R. Giacomini and T. Kitagawa and H. Uhlig},
  title   = {Estimation under Ambiguity},
  journal = {Cemmap working paper},
  year    = {2019},
}

@Article{NT14,
  author  = {Andriy Norets and Xun Tang},
  title   = {Semiparametric Inference in Dynamic Binary Choice Models},
  journal = {Review of Economic Studies},
  year    = {2014},
  volume  = {81},
  number  = {3},
  pages   = {1229-1262},
}

@Article{BH15,
  author    = {C. Baumeister and J.D. Hamilton},
  title     = {Sign Restrictions, Structural Vector Autoregressions, and Useful Prior Information},
  journal   = {Econometrica},
  year      = {2015},
  volume    = {83},
  number    = {5},
  pages     = {1963-1999},
}

@InProceedings{Hausman83,
  author    = {Jerry A. Hausman},
  title     = {Specification and Estimation of Simultaneous Equation Models},
  booktitle = {Handbook of Econometrics, Vol. 1},
  year      = {1983},
  editor    = {Zvi Griliches and Michael D. Intriligator},
  pages     = {391-448},
}

@Article{AS11,
  author    = {S.B. Aruoba and F. Schorfheide},
  title     = {Sticky Prices versus Monetary Frictions: An Estimation of Policy Trade-offs},
  journal   = {American Economic Journal: Macroeconomics},
  year      = {2011},
  volume    = {3},
  number    = {1},
  pages     = {60-90},
}

@incollection{Kilian13,
  author      = "Lutz Kilian",
  title       = "Structural vector autoregressions",
  editor      = "Nigar Hashimzade and Michael A. Thornton",
  booktitle   = "Handbook of Research Methods and Applications in Empirical Macroeconomics",
  publisher   = "Edward Elgar Publishing",
  address     = "Northampton",
  year        = 2013,
  pages       = "515--554",
  chapter     = 22,
}

@Book{KLbook,
  title     = {Structural Vector Autoregressive Analysis},
  publisher = {Cambridge University Press},
  year      = {2017},
  author    = {Lutz Kilian and Helmut L\"{u}tkepohl},
}

@Article{Cochrane06,
  author  = {J.H. Cochrane},
  title   = {Identication and Price Determination with Taylor Rules: A Critical Review},
  journal = {unpublished manuscript},
  year    = {2006},
}

@Article{FWZ07DSGE,
  author  = {M. Fukac and D. Waggoner and T. Zha},
  title   = {Local and Global Identication of DSGE Models: A Simultaneous-equation Approach},
  journal = {unpublished manuscript},
  year    = {2007},
}

@Book{Canova05Book,
  title         = {Methods for Applied Macroeconomic Research},
  publisher     = {Princeton, NJ: Princeton University Press},
  year          = {2005},
  author        = {Fabio Canova},
}

@Article{CEV06NBER,
  author  = {L. J. Christiano and M. Eichenbaum and R. Vigfusson},
  title   = {Assessing Structural VARs},
  journal = {NBER Working Paper No. 12353},
  year    = {2006},
}

@Article{FRSW07AER,
  author  = {J. Fernandez-Villaverde and J. Rubio-Ramirez and T. Sargent and M. Watson},
  title   = {ABCs (and Ds) of Understanding VARs},
  journal = {American Economic Review},
  year    = {2007},
  volume  = {97},
  number  = {3},
  pages   = {1021--1026},
}

@Article{Ravenna07JME,
  author  = {F. Ravenna},
  title   = {Vector Autoregressions and Reduced Form Representations of DSGE Models},
  journal = {Journal of Monetary Economics},
  year    = {2007},
  volume  = {54},
  number  = {7},
  pages   = {2048--2064},
}

@Article{Iskrev10JME,
  author  = {Nikolay Iskrev},
  title   = {Local identification in DSGE models},
  journal = {Journal of Monetary Economics},
  year    = {2010},
  volume  = {57},
  pages   = {189--202},
}

@TechReport{ASZ19,
  author      = {Majid M. Al-Sadoon and Piotr Zwiernik},
  title       = {The Identification Problem for Linear Rational Expectations Models},
  institution = {mimeo},
  year        = {2019},
}

@Article{KK18QE,
  author  = {A. Kociecki and M. Kolasa},
  title   = {Global Identification of Linearized DSGE Models},
  journal = {Quantitative Economics},
  year    = {2018},
  volume  = {9},
  number  = {3},
  pages   = {1243--1263},
}

@Article{KomNg11ECTA,
  author  = {I. Komunjer and S. Ng},
  title   = {Dynamic Identification of Dynamic Stochastic General Equilibrium Models},
  journal = {Econometrica},
  year    = {2011},
  volume  = {79},
  number  = {6},
  pages   = {1995--2032},
}

@Article{QuTka12QE,
  author  = {Z. Qu and D. Tkachenko},
  title   = {Identification and frequency domain quasimaximum likelihood estimation of linearized dynamic stochastic general equilibrium models},
  journal = {Quantitative Economics},
  year    = {2012},
  volume  = {3},
  number  = {1},
  pages   = {95--132},
}

@Article{EAT98,
  author  = {A. Edelman and T.A. Arias and S.T. Smith},
  title   = {The geometry of algorithms with orthogonality constraints},
  journal = {SIAM Journal on Matrix Analysis and Applications},
  year    = {1998},
  volume  = {20},
  number  = {2},
  pages   = {303--353},
}

@InProceedings{Sturmfels02solvingsystems,
  author    = {Bernd Sturmfels},
  title     = {Solving Systems of Polynomial Equations},
  booktitle = {American Mathematical Society, CBMS Regional Conference Series, n. 97},
  year      = {2002},
}

@Article{Stiefel,
  author  = {E. Stiefel},
  title   = {Richtungsfelder und fernparallelismus in n-dimensionalem mannig faltigkeiten},
  journal = {Commentarii Math. Helvetici},
  year    = {1935-1936},
  volume  = {8},
  pages   = {305--353},
}

@Article{Manopt,
  author  = {Nicolas Boumal and Bamdev Mishra and P. A. Absil and Rodolphe Sepulchre},
  title   = {Manopt, a Matlab Toolbox for Optimization on Manifolds},
  journal = {Journal of Machine Learning Research},
  year    = {2014},
  volume  = {15},
  pages   = {1455--1459},
}

@Book{RC04,
  title     = {Monte Carlo Statistical Methods},
  publisher = {Springer},
  year      = {2004},
  author    = {Christian P. Robert and George Casella},
  edition   = {2nd},
}

@Article{ARW18,
  author    = {J.E. Arias and J.F. Rubio-Ram\'{i}rez and D.F Waggoner},
  title     = {Inference Based on SVARs Identified with Sign and Zero Restrictions: Theory and Applications},
  journal   = {Econometrica},
  year      = {2018},
  volume    = {86},
  number    = {2},
  pages     = {685-720},
}

@Article{GKR19,
  author      = {Raffaella Giacomini and Toru Kitagawa and Matthew Read},
  title       = {Robust Bayesian Inference in Proxy SVARs},
  journal     = {Journal of Econometrics},
  year        = {2022},
  volume      = {228},
  number    	= {1},
  pages     	= {107--126},
}

@Article{GafarovOlea14,
  author    = {B. Gafarov and M. Meier and J.L. Montiel-Olea},
  title     = {Delta-method Inference for a Class of Set-identified SVARs},
  journal   = {Journal of Econometrics},
  year      = {2018},
  volume    = {203},
  number    = {2},
  pages     = {316-327},
}

@Book{Spivak65,
  title     = {Calculus on Manifolds: A Modern Approach to Classical Theorems of Advanced Calculus},
  publisher = {New York, New York: The Benjamin/Cummings Publishing Company},
  year      = {1965},
  author    = {M Spivak},
}

@TechReport{BKglob20,
  author      = {Emanuele Bacchiocchi and Toru Kitagawa},
  institution = {mimeo},
  title       = {On global identication in Structural Vector Autoregressions},
  year        = {2021},
}

@Article{Sentana23,
  author      = {Enrique Sentana},
  title       = {Finite underidentification},
  journal 		= {Journal of Econometrics},
  year    		= {2025},
  pages   = {forthcoming},
}

@Article{PMW19,
  author      = {M Plagborg-M{\o}ller and C K Wolf},
  title       = {Instrumental Variable Identification of Dynamic Variance Decompositions},
  journal 		= {Journal of Political Economy},
  year    		= {2022},
  volume  		= {130},
  number  = {8},
  pages   = {2164--2202},
}

@Article{CH2019AEJ,
  author  = {D. Caldara and E. Herbst},
  title   = {Monetary Policy, Real Activity, and Credit Spreads: Evidence from Bayesian Proxy SVARs},
  journal = {American Economic Journal: Macroeconomics},
  year    = {2019},
  volume  = {11},
  number  = {1},
  pages   = {157--192},
}

@Article{SW02, 
  author  = {J. H. Stock and M. W. Watson},
  title   = {Has the Business Cycle Changed and Why?},
  journal = {NBER Macroeconomics Annual},
  year    = {2002},
  volume  = {17},
  number  = {1},
  pages   = {159--218},
}

@Article{LMNS21,
  author  = {H L\"{u}tkepohl and M Meitz and A Net\v{s}unajev and P Saikkonen},
  title   = {Testing identification via heteroskedasticity in structural vector autoregressive models},
  journal = {The Econometrics Journal},
  year    = {2021},
  volume  = {24},
  number  = {1},
  pages   = {1-22},
}

@Article{SRP23,
  author  = {T Schlaak and M Rieth and M Podstawski},
  title   = {Monetary policy, external instruments, and heteroskedasticity},
  journal = {Quantitative Economics},
  year    = {2023},
  volume  = {14},
  number  = {1},
  pages   = {161-200},
}
\addcontentsline{toc}{section}{Bibliography}


\appendix


\newpage
\section{Appendix: Some analytical results on the geometry of identification}
\label{app:Geom}

Let the data generating process be the bivariate VAR defined in Section \ref{sec:geo} with the identifying restriction 
\begin{equation}
\label{eq:biVARconstrApp}
(A_0)_{[1,1]}^{-1}=c \:\Longleftrightarrow\: (e_1^\prime \Chol)\,q_1 = c
\end{equation}
where $c>0$ is a known (positive) scalar and $e_1$ is the first column of the $(2\times 2)$ identity matrix. The non-homogeneous restriction in Eq. (\ref{eq:biVARconstrApp}) affects the orthogonal matrix $Q$ as $\sigma_1^\prime\,q_1=c$, with $\sigma_1$ denoting the first column of $\Chol^\prime=\left(\begin{array}{cc} \sigma_{1,1}&\sigma_{2,1}\\0&\sigma_{2,2}\end{array}\right)$. 

The vector $q_1$ must satisfy the two equations
\begin{equation}
\label{eq:biVARsysq1App}
\left\{
\begin{array}{rcl}
\sigma_1^\prime\,q_1 &=& c\\
q_1^\prime\,q_1 &=& 1
\end{array}
\right.\nonumber
\end{equation}
By simple algebra, the two solutions are
\begin{equation}
\label{eq:biVARq1}
q_1^{(1)}=\left(
\begin{array}{c}
c/\sigma_{1,1}\\
\\
+\sqrt{\frac{\sigma_{1,1}^2-c^2}{\sigma_{1,1}^2}}
\end{array}
\right)\hspace{0.8cm}\text{and}\hspace{0.8cm}
q_1^{(2)}=\left(
\begin{array}{c}
c/\sigma_{1,1}\\
\\
-\sqrt{\frac{\sigma_{1,1}^2-c^2}{\sigma_{1,1}^2}}
\end{array}
\right).
\end{equation}
These two possible solutions are represented in Figure \ref{fig:Baby1}. If $(\sigma_{1,1}^2<c^2)$, the straight (vertical) line does not intersect the unit circle, and no real solution is admissible. The SVAR, although identified, does not admit any real solution given the reduced-form parameters $\phi$. If, instead, $(\sigma_{1,1}^2=c^2)$, i.e. the vertical red line is tangent to the unit circle, we continue to have global identification, although the imposed restriction is not coherent with those derived by RWZ. In all other situations, there will be two solutions that, \textit{a priori}, can be admissible despite the sign normalization restriction. This is the case depicted in Figure \ref{fig:Baby1}.

Concerning the sign normalization restriction, if we choose the standard practice of the diagonal elements in $A_0$ to be positive, for the first equation, it consists in $q_1^\prime\,\tilde{\sigma}_1\,\geq\,0$, where $\tilde{\sigma}_1$ is the first column of $\Choli=1/(\sigma_{1,1}\sigma_{2,2})\left(\begin{array}{cc} \sigma_{2,2}&0\\-\sigma_{2,1}&\sigma_{1,1}\end{array}\right)$. Through elementary algebra, we obtain that
\begin{equation}
\label{eq:SignBaby}
q_1^\prime\,\tilde{\sigma}_1\,\geq\,0 \hspace{1cm} \Longleftrightarrow \hspace{1cm} \frac{q_{1,1}}{\sigma_{1,1}}\geq\frac{q_{1,2}\sigma_{2,1}}{\sigma_{1,1}\sigma_{2,2}}
\end{equation}
where $q_{1,1}$ and $q_{1,2}$ are the two generic elements of $q_1$, i.e. $q_{1}=(q_{1,1}\,,\,q_{1,2})^\prime$. Suppose, first, that from the data we have $\sigma_{2,1}<0$. In this case, if we substitute in the values of $q_{1,1}$ and $q_{1,2}$ obtained for $q_1^{(1)}$ in the left-hand side of Eq. (\ref{eq:biVARsysq1App}), the sign normalization condition for the first equation becomes
\begin{equation}
\label{eq:SignBabyq1}
\frac{c}{\sigma_{1,1}^2}\geq\frac{\sigma_{2,1}}{\sigma_{1,1}\sigma_{2,2}}\sqrt{\frac{\sigma_{1,1}^2-c^2}{\sigma_{1,1}^2}}
\end{equation}
As the left-hand side is always positive and the right-hand side always negative, this is always satisfied. If, instead, we substitute the values $q_{1,1}$ and $q_{1,2}$ obtained for $q_1^{(2)}$ in the right-hand side of Eq. (\ref{eq:biVARsysq1App}), the sign normalization condition for the first equation becomes
\begin{equation}
\label{eq:SignBabyq2}
\frac{c}{\sigma_{1,1}^2}\geq-\frac{\sigma_{2,1}}{\sigma_{1,1}\sigma_{2,2}}\sqrt{\frac{\sigma_{1,1}^2-c^2}{\sigma_{1,1}^2}}
\end{equation}
that is also satisfied when $c^2 \geq \frac{1}{2}\frac{\sigma_{1,1}^2\sigma_{2,1}^2}{\sigma_{2,2}^2}$. If this is the case, both solutions $q_1^{(1)}$ and $q_1^{(2)}$ are admissible, leading to local identification. The situation is very similar when $\sigma_{2,1}>0$.

If, instead, $c=0$ as in the standard RWZ setup, the two $q_1$ vectors in Eq. (\ref{eq:biVARsysq1App}) become $q_1^{(1)}=(0\,,\,1)^\prime$ and $q_1^{(2)}=(0\,,\,-1)^\prime$. If, as before, we suppose $\sigma_{2,1}<0$, the sign normalization for $q_1^{(1)}$ in Eq. (\ref{eq:SignBabyq1}) reduces to $0\geq \sigma_{2,1}/(\sigma_{1,1}\sigma_{2,2})$, which is always true. The sign normalization for $q_1^{(2)}$ in Eq. (\ref{eq:SignBabyq2}) is $0\geq -\sigma_{2,1}/(\sigma_{1,1}\sigma_{2,2})$ which, in contrast, is never true. The case where $\sigma_{2,1}>0$, is exactly the same, but with inverted results. One of the two solutions, thus, will be always ruled out by the sign normalization, and global identification is guaranteed.  

The second column of $Q$, the unit-length vector $q_2$, although not restricted, can be pinned down through the its orthogonality to $q_1$
\begin{equation}
\label{eq:biVARsysq2App}
\left\{
\begin{array}{rcl}
q_2^\prime\,q_1 &=& 0\\
q_2^\prime\,q_2 &=& 1.
\end{array}
\right.
\end{equation}
However, given that there are two admissible vectors $q_1^{(1)}$ and $q_1^{(2)}$, the system Eq. (\ref{eq:biVARsysq2App}) must be solved for both. This can be done with simple algebra, yielding the two solutions
\begin{equation}
\label{eq:biVARq21}
q_2^{(1)}=\left(
\begin{array}{c}
+\sqrt{\frac{(\sigma_{1,1}^2-c^2)(2c^2-\sigma_{1,1}^2)}{c^4}}\\
\\
-\sqrt{\frac{2c^2-\sigma_{1,1}^2}{c^2}}
\end{array}
\right)\hspace{0.8cm}\text{and}\hspace{0.8cm}
q_2^{(1)}=\left(
\begin{array}{c}
-\sqrt{\frac{(\sigma_{1,1}^2-c^2)(2c^2-\sigma_{1,1}^2)}{c^4}}\\
\\
+\sqrt{\frac{2c^2-\sigma_{1,1}^2}{c^2}}
\end{array}
\right)
\end{equation}
One of the two, precisely which depends on the reduced-form parameters, will be eliminated by the sign normalization  restriction. This case is represented in the left panel of Figure \ref{fig:Baby2}, together with $q_1^{(1)}$.

The other possibility, represented in the right panel of Figure \ref{fig:Baby2}, is when we solve the system conditional on $q_1^{(2)}$, obtaining
\begin{equation}
\label{eq:biVARq22}
q_2^{(2)}=\left(
\begin{array}{c}
+\sqrt{\frac{(\sigma_{1,1}^2-c^2)(2c^2-\sigma_{1,1}^2)}{c^4}}\\
\\
+\sqrt{\frac{2c^2-\sigma_{1,1}^2}{c^2}}
\end{array}
\right)\hspace{0.8cm}\text{and}\hspace{0.8cm}
q_2^{(2)}=\left(
\begin{array}{c}
-\sqrt{\frac{(\sigma_{1,1}^2-c^2)(2c^2-\sigma_{1,1}^2)}{c^4}}\\
\\
-\sqrt{\frac{2c^2-\sigma_{1,1}^2}{c^2}}
\end{array}
\right)
\end{equation}
where, as before, one of the two solutions is ruled out by the sign normalization restriction.


\newpage
\section{Appendix: Further results on local identification}
\label{app:FurResId}

In this appendix we provide a new result on local identification for SVAR models. We consider a set of equality restrictions $\textbf{F}(\phi,Q)$
satisfying the triangular identification scheme in Definition \ref{def:tri}.

\begin{prop}[RWZ sufficient condition for checking local identification]
\label{prop:SuffRWZ}
        Consider an SVAR with triangular identifying restrictions of the form Eq. (\ref{eq:GenFormRest}). 
        The SVAR is locally identified at $A=\left(A_0,A_+\right) \in \AR$ if, for $i=1,\ldots,n$, 
        \begin{equation}
        \label{eq:SuffRWZ}
                M_i(Q)\equiv
                \left(\begin{array}{c}
                \underset{(n-i)\times n}{F_{ii}(\phi)}\cdot\underset{n\times n}{Q}\\
                \left(\begin{array}{ccc}
                \underset{i\times i}{I_i}&&\underset{i\times (n-i)}{0}
                \end{array}\right)
                \end{array}\right)
        \end{equation}
        is of rank $n$.
\end{prop}

\begin{proof} See Appendix \ref{app:Proofs}. \end{proof}

Proposition \ref{prop:SuffRWZ} reconciles our condition for local identification of triangular SVARs with the general rank condition for global identification provided by RWZ (their Theorem 1). In particular, under a triangular identification scheme, the RWZ condition for global identification developed for the case of homogeneous restrictions implies local identification, even though we allow non-homogeneous and across shock restrictions.


\newpage
\section{Appendix: Proofs} 
\label{app:Proofs}

This appendix collects proofs for all propositions reported in this paper. We make use of the following matrices. $K_n$ is the $n^2\times n^2$ commutation matrix as defined in \cite{mn07} and $N_n = 1/2 (I_{n^2} + K_n)$. Let $\tilde{D}_n$ be the $n^2 \times n(n-1)/2$ full-column rank matrix $\tilde{D}_n$ defined in \cite{magnus88} such that for any $n(n-1)/2$-dimensional vector $v$,$\tilde{D}_n\,v \equiv \text{vec }(H)$ holds, where $H$ is an $n\times n$ skew-symmetric matrix ($H = -H^\prime$). See Appendix \ref{app:Dnt} for explicit constructions of $\tilde{D}_n$ for $n=2,3,4$. 

\subsection*{Proof of Proposition \ref{prop:LocIdent}: necessary and sufficient condition for local identification} 

Fixing $\phi$, a matrix $Q$ satisfies the identifying restrictions if:
\begin{eqnarray}
    \textbf{F}(\phi)\:\ve \, Q &=& \textbf{c}\label{eq:Sys1}\\
    Q^\prime Q &=& I_n\label{eq:Sys2}
\end{eqnarray}
which is a system of quadratic equations. Eq. (\ref{eq:Sys1}) consists of  $f=f_1+\cdots+f_n$ linear and non-homogeneous equations. Eq. (\ref{eq:Sys2}) is a set of quadratic equations stating that the columns of $Q$, the vectors $(q_1,\,\ldots\,,q_n)$, must be orthogonal and of unit length.

The system can be solved locally as:
\begin{eqnarray}
    \textbf{F}(\phi)\:\ve \, d Q &=& 0\nonumber\\
    dQ^\prime \,Q + Q^\prime d Q &=& 0,\nonumber
\end{eqnarray}
which, using the Kronecker product and its properties, becomes
\begin{eqnarray}
    \textbf{F}(\phi)\:\ve \, d Q &=& 0\nonumber\\
    \bigg[(Q^\prime \otimes I_n) + (I_n\otimes Q^\prime)\bigg] \ve d Q &=& 0.\nonumber
\end{eqnarray}
Moreover, using the commutation matrix $K_n$, we have
\begin{eqnarray}
    \textbf{F}(\phi)\:\ve \, d Q &=& 0\nonumber\\
    \bigg[K_n(I_n\otimes Q^\prime) + (I_n\otimes Q^\prime)\bigg] \ve d Q &=& 0, \nonumber
\end{eqnarray}
and recalling $N_n = 1/2 (I_{n^2} + K_n)$, we obtain
\begin{eqnarray}
    \textbf{F}(\phi)\:\ve \, d Q &=& 0\nonumber\\
    2N_n(I_n\otimes Q^\prime) \ve d Q &=& 0.\nonumber
\end{eqnarray}
The Jacobian matrix can, thus, be defined as
\begin{equation}
    J(Q)=
    \left(\begin{array}{c}
    \textbf{F}(\phi)\\
    2N_n(I_n\otimes Q^\prime)
    \end{array}\right)\label{eq:J}
\end{equation}
Following \cite{mn07}, a sufficient condition for local identification of $Q$ at the point $Q=Q_0$ is that $J(Q_0)$ has full column rank. If there exists a neighborhood of $Q_0$ such that $J(Q_0)$ is of full column rank, this condition becomes necessary too.

The condition regarding the rank of Eq. (\ref{eq:J}) can be further simplified. Given that $Q$ is invertible (it is orthogonal), the rank of $J(Q)$ is unchanged if we post-multiply Eq. (\ref{eq:J}) by $(I_n\otimes Q^{-1\prime})=I_n\otimes Q$. Checking whether $J(Q)$ is of full column rank, thus, corresponds to checking whether the system of equations 
\begin{eqnarray}
    \textbf{F}(\phi)(I_n\otimes Q)\,x&=&0\nonumber\\
    2N_n\,x&=&0\nonumber
\end{eqnarray}
admits the null vector $x$ as the unique solution. However, as in \cite{magnus88}, the second equation can be solved as $x=\tilde{D}_n z$, with $z$ a $n(n-1)/2\times 1$ vector. Substituting this solution into the first equation leads to the rank condition in Eq. (\ref{eq:RankCond}) of Proposition \ref{prop:LocIdent}. Since $\tilde{D}_n$ is a matrix of full column rank $n(n-1)/2$, a necessary condition for the rank condition Eq. (\ref{eq:RankCond}) is that the number of rows of $F(\phi)$, $f$, is greater than or equal to $n(n-1)/2$. This completes the proof of (i). 

\vspace{0.3cm}
Before moving to claim (ii), we derive the rank condition proposed in Corollary \ref{corol:RankCond}. Starting from the rank condition in Eq. (\ref{eq:RankCond}), we have that:
\begin{eqnarray}
    \rk \:\bigg(\textbf{F}(\phi)\big(I_n\otimes Q\big)\tilde{D}_n\bigg)
    & = &
    \rk \:\bigg(Z\Big(I_n \otimes z(\phi)^\prime\Big)\Big(I_n\otimes Q\Big)\tilde{D}_n\bigg)\nonumber\\
    & = &
    \rk \:\bigg(Z\Big(I_n \otimes z(\phi)^\prime Q\Big)\tilde{D}_n\bigg)\nonumber\\
    & = &
    \rk \:\bigg(Z\Big(I_n \otimes h(A_0,A_+)^\prime\Big)\tilde{D}_n\bigg)\nonumber
\end{eqnarray}
that proves the result.

\vspace{0.3cm}
To show claim (ii), we first recall that, from Eq. (\ref{eq:RestrFunction}), the restrictions can be written as
\begin{equation*}
    Z_j\:h(A)^\prime \:e_j - c_j = 0
\end{equation*}
where $A\equiv(A_0,A_+)$. Collecting all the restrictions for $j\in\{1,\ldots,n\}$, we can write
\begin{equation*}
    Z\:\ve\big[h(A)^\prime\big] - c = 0.
\end{equation*}
where the selection matrix $Z$, of dimension $f\times nk$, collects all the $Z_j$ matrices, as well as potential further restrictions across shocks.  
An explicit form for the restrictions is
\begin{equation}
    \label{eq:gamma}
    \ve\big[h(A)^\prime\big] = Z_{\perp}\,\gamma + \tilde{c}
\end{equation}
with $Z_\perp$, of dimension $nk\times (nk-f)$, being the orthogonal complement of $Z$, and $\tilde{c}$ is a constant vector such that $Z\,\tilde{c}=c$ holds. The $(nk-f)\times 1$ vector $\gamma$ one-to-one corresponds to $h(A)$ subject to the restrictions. By Condition \ref{cond:reg}, the range of $h(A)$ spanned by $A \in \mathcal{A}$ is a set of positive measure in $\mathbb{R}^{nk}$. Since $\rk(Z_\perp)=nk-f$, the space of $\gamma$ parameters is a set in $\mathbb{R}^{nk-f}$ with a positive measure in $\mathbb{R}^{nk-f}$. Here, $\gamma$ is introduced to index $h(A)$ in its restricted parameter space. 

Now, let
\begin{eqnarray}
    l_r\big(h(A)\big) & \equiv & Z\Big(I_n \otimes h(A)\Big)^\prime\tilde{D}_n\nonumber\\
                      & = & Z\Big(I_n \otimes \mathrm{vec}_{n\times k}^{-1} (Z_\perp \gamma + \tilde{c})\Big)^\prime\tilde{D}_n\nonumber\\
                      & \equiv & l_r(\gamma),\label{eq:lr}\nonumber
\end{eqnarray}
where the last equality holds being $h(A)$ expressed as a function of the free parameters $\gamma$ as for Eq. (\ref{eq:gamma}), and where $\mathrm{vec}_{n\times k}^{-1}(\cdot)$ is the inverse function of the $\ve$ operator, that specifically can be defined as
\begin{equation*}
    \mathrm{vec}_{n\times k}^{-1}(x)=\big((\ve I_k)^\prime \otimes I_n\big)\big(I_k\otimes x\big)\,\in \Re^{n\times k},   
\end{equation*}
for any vectors $x\in \Re^{nk}$. Following the same steps of algebra, we also have
\begin{eqnarray}
    l_d\big(h(A)\big) & \equiv & \tilde{D}_n^{\prime}\Big(I_n \otimes h(A)\Big)Z^\prime Z\Big(I_n \otimes h(A)\Big)^\prime\tilde{D}_n\nonumber\\
                      & = & \tilde{D}_n^{\prime}\Big(I_n \otimes \mathrm{vec}_{n\times k}^{-1} (Z_\perp \gamma + \tilde{c})\Big)Z^\prime Z\Big(I_n \otimes \mathrm{vec}_{n\times k}^{-1} (Z_\perp \gamma + \tilde{c})\Big)^\prime\tilde{D}_n\nonumber\\
                      & = & l_d(\gamma).\label{eq:ld}\nonumber
\end{eqnarray}
Then, $\rk \Big(l_r\big(\gamma\big)\Big) = \rk \Big( l_d\big(\gamma\big)\Big)$, such that the rank condition in Eq. (\ref{eq:RankCond}) is equivalent to
\begin{equation}
    \label{eq:RankCondDet}
    \mathrm{det}\Big( l_d\big(\gamma\big)\Big) \neq 0.
\end{equation}
The function $\mathrm{det}(l_d(\gamma))$ is a polynomial of $\gamma$, and it is well known that a polynomial is either identically zero or non-zero almost everywhere in the space of $\gamma$, i.e. $\Re^{nk-f}$ (see \citeauthor{Johansen95}, \citeyear{Johansen95}, or \citeauthor{TraynorCaron2005}, \citeyear{TraynorCaron2005}).
Now, given the definition of the restrictions
\begin{eqnarray}
    \ve\big[h(A)^\prime\big]=Z_\perp \gamma + \tilde{c} \hspace{0.2cm} \Longrightarrow \hspace{0.2cm} \gamma & = & \big(Z_\perp^\prime Z_\perp\big)^{-1}Z_\perp^\prime \Big[\ve\big[h(A)^\prime - \tilde{c}\Big]\nonumber\\
    & = & v\big(h(A)\big)\nonumber
\end{eqnarray}
where $v\big(h(A)\big)$ is a function from the space of $h(A)$ to the space of $\gamma$, whose first derivative with respect to $h(A)$ has full rank equal to $nk-f$, i.e. the dimension of the space of $\gamma$. Now, combining the previous result on the zero set of a polynomial with Lemma 2 in RWZ (proved in \citeauthor{Spivak65}, \citeyear{Spivak65}), we can say that the statement ``either there are no $\gamma$ in $\Re^{nk-f}$ for which the rank condition holds or it is satisfied for almost every $\gamma$ in $\Re^{nk-f}$'' does correspond to an equivalent condition of $h(A_0,A_+)$ on the space of all $n\times k$ matrices satisfying the restrictions.

Finally, let 
\begin{equation}
    \label{eq:W}
    \mathcal{W} = \Big\{(A_0,A_+)\in \mathcal{A}_R \hspace{0.2cm} : \hspace{0.2cm}
        \mathrm{det} \left( l_d\big(h(A)\big) \right) = 0 \Big\}
\end{equation}
and let $\tilde{\mathcal{W}} = h(\mathcal{W})$. The result of the proposition can be restated in terms of $\mathcal{W}$, rather than in terms of $\mathcal{K}$. In this respect, either $\mathcal{W}=\mathcal{A}_R$, or $\mathcal{W}$ is of measure zero in $\mathcal{A}_R$.
From the definition of $\mathcal{W}$ and $\mathcal{K}$, and using Condition \ref{cond:reg}, we can see that $\mathcal{W} = h^{-1}\big(\tilde{\mathcal{W}}\big)$. The result is finally obtained by using Lemma 2 in RWZ. $\square$

\subsection*{Proof of Corollary \ref{prop:LocIdentSubset}: necessary and sufficient condition for local identification of a subset of shocks} 

Fixing $\phi$, a matrix $Q = \big[\:Q_1\:|\:Q_2\:\big]$ satisfies the identifying restrictions if:
\begin{eqnarray}
        \textbf{F}_{11}(\phi)\:\textbf{q}_1 &=& \textbf{c}_1\nonumber\\
        \textbf{F}_{22}(\phi)\:\textbf{q}_2 &=& \textbf{c}_2\nonumber\\
        Q_1^\prime Q_1 &=& I_s\nonumber\\
        Q_2^\prime Q_2 &=& I_{n-s}\nonumber\\
        Q_1^\prime Q_2 &=& 0_{s\times(n-s)}\nonumber
\end{eqnarray}
which is a system of quadratic equations, with $\textbf{q}_1 = \ve\:Q_1$ and $\textbf{q}_2 = \ve\:Q_2$. 
Similarly to the steps followed in the proof of Proposition \ref{prop:LocIdent}, differentiating and using a bit of algebra, we obtain the system of equations
\begin{equation}
\label{eq:SysSubset}
        \left(\begin{array}{cc} 
                \textbf{F}_{11}(\phi) & 0\\
                2N_{ns}\,(I_n\otimes Q_1^\prime)\\
                K_{s(n-s)}\,(I_s\otimes Q_2^\prime) & (I_{(n-s)}\otimes Q_1^\prime)\\
                0 & \textbf{F}_{22}(\phi)\\
                0 & 2N_{n(n-s)}\,(I_n\otimes Q_2^\prime)
        \end{array}\right)
        \left(\begin{array}{c} d \textbf{q}_1 \\ d \textbf{q}_2 \end{array}\right) = 0.
\end{equation}
However, as the interest is only on $\textbf{q}_1$, the part of the Jacobian that matters is
\begin{equation}
\label{eq:JacobSubset}  
        J_1(Q_1) = \left(\begin{array}{c} \textbf{F}_{11}(\phi)\\ 2N_{ns}\,(I_n\otimes Q_1^\prime) \end{array}\right)
        \nonumber
\end{equation}
that, in order for $\textbf{q}_1$ to be locally identified, must have full column rank equal to $ns$. $\square$

\subsection*{Proof of Proposition \ref{prop:SuffLocIdent}: local identification in triangular SVARs} 

Assume that the rank condition of Proposition \ref{prop:LocIdent} holds at parameter point $A=\left(A_0,A_+\right) \in \AR$, and 
let $\phi$ be the corresponding reduced-form parameter. Since local identification holds at $A$, there is no observationally equivalent 
parameter point in a neighborhood of $A$. In other words, no infinitesimal rotation of the orthogonal matrix $Q$ generates observationally 
equivalent and admissible structural parameters in the neighborhood of $A$. Any infinitesimal rotation can be represented by $(I_n+H)$, 
where $H$ is an $n\times n$ skew-symmetric matrix (see \citeauthor{Lucchetti06} \citeyear{Lucchetti06}) whose \textit{i}-th column we 
denote by $h_i$. 

Projecting on $q_1$, an admissible structural parameter lying in a local neighborhood of $A$ has to satisfy
\begin{equation}
\nonumber
        F_{11}(\phi)\bigg[Q\left(I_n+H\right)\bigg]e_1=c_1 \hspace{0.5cm} \Longrightarrow \hspace{0.5cm} F_{11}(\phi)q_1+F_{11}QHe_1=c_1 
        \hspace{0.5cm}\Longrightarrow\hspace{0.5cm} F_{11}(\phi)Qh_1=0,
\end{equation}
where $e_i$ is the \textit{i}-th column of the identity matrix $I_n$, and the last equation follows from the fact that $F_{11}(\phi)q_1=c_1$.
The system $F_{11}(\phi)Qh_1=0$ is characterized by $n-1$ equations and an $n$-dimensional vector of unknowns $h_1$. The first element of $h_1$ is zero by definition (the elements on the main diagonal of a skew-symmetric matrix are equal to zero). Hence, we have 
\begin{equation}
\label{eq:sysh1}
        \left(\begin{array}{c}F_{11}(\phi)Q\\e_1^\prime \end{array}\right)h_1=0.
\end{equation}
This linear equation system has  $h_1=0$ as its unique solution if and only if $\left(\begin{array}{c}F_{11}(\phi)Q\\e_1^\prime \end{array}\right)$ is of rank $n$, or, equivalently,  $F_{11}(\phi)(q_2\,\ldots\,q_n)$ is of full rank (equal to $n-1$). Since the model is locally identified by assumption, $h_1=0$ has to be the only solution of Eq. (\ref{eq:sysh1}). Hence, $\rk\bigg(F_{11}(\phi)(q_2\,\ldots\,q_n)\bigg)=n-1$ must hold, implying that $\rk(F_{11}(\phi))=n-1$.

For $q_2$, given a $q_1$ vector solving $F_{11}(\phi)q_1 = c_1$, we have the following system:
\begin{equation}
\nonumber
        \left\{\begin{array}{rcl}F_{21}(\phi)q_1+F_{22}(\phi)q_2&=&c_2\\q_1^\prime q_2 &=& 0, \end{array}\right.
\end{equation}
Considering again an infinitesimal rotation
\begin{equation}
\nonumber
        \left\{\begin{array}{rcl}F_{21}(\phi)q_1+F_{22}(\phi)Q(I_n+H)e_2&=&c_2\\q_1^\prime Q(I_n+H)e_2 &=& 0 \end{array}\right.
        \hspace{0.3cm}\Longrightarrow\hspace{0.3cm}
        \left\{\begin{array}{rcl}F_{21}(\phi)q_1+F_{22}(\phi)Qe_2+F_{22}(\phi)Qh_2&=&c_2\\q_1^\prime Qh_2 &=& 0 \end{array}\right.,
\end{equation}
but, given the restrictions, $F_{21}(\phi)q_1+F_{22}(\phi)q_2=c_2 \hspace{0.3cm}\Longrightarrow\hspace{0.3cm} F_{22}(\phi)Qh_2=0$,
which allows the system to be written as
\begin{equation}
\label{eq:sysh2}
        \left\{\begin{array}{rcl}F_{22}(\phi)Qh_2&=&0\\q_1^\prime Qh_2 &=& 0 \end{array}\right.
        \hspace{0.5cm}\Longrightarrow\hspace{0.5cm}
        \left(\begin{array}{c}F_{22}(\phi)Q\\q_1^\prime Q\end{array}\right)h_2=0.
\end{equation}
Similarly to the argument for $h_1$ above, and noting that the first two entries of  $h_2$ are zero, we can represent the linear equations as
\begin{equation}
\nonumber
        \left(\begin{array}{c}F_{22}(\phi)Q\\q_1^\prime Q \\e_2^\prime\\e_1^\prime\end{array}\right)h_2=0.
\end{equation}
Since $q_1^\prime Q = e_1^\prime$, the last equation in this system is redundant. 
Thus, in order for $h_2=0$ to be the unique solution,  $\left(\begin{array}{c}F_{22}(\phi)\\q_1^\prime \end{array}\right)$ must 
have full raw rank (equal to $n-1$).

To obtain the sequential rank conditions of Proposition \ref{prop:SuffLocIdent}, we repeat this argument further for $i=3,4,\ldots,n$.  

Next, we show the reverse implication. For each column of $Q$, we consider a system of equations of the form,
\begin{equation}
\left\{\begin{array}{rcl}
\tilde{F}_{ii}(\phi)\,q_i & = & (c_i', 0, \dots, 0)' \\
q_i^\prime\,q_i & = & 1,
\end{array}\right. \notag
\end{equation}
sequentially for $i=1, \dots, n$, where $\tilde{F}_{ii}(\phi)$ is as defined in the statement of Proposition \ref{prop:SuffLocIdent}. If $\rk \bigg(\tilde{F}_{ii}(\phi)\bigg)=n-1$, the system of equations represents the intersection between a straight line and the unit circle  in $\Re^n$, 
which has at most two distinct solutions. Hence, any admissible $Q$ matrices are isolated points, so the SVAR is locally identified. The rank condition of Eq. (\ref{eq:RankCond}) follows by Proposition \ref{prop:LocIdent}. $\square$

\subsection*{Proof of Proposition \ref{prop:LocVsLack}: number of admissible $Q$'s}
We split the proof into five cases based on the type of equality restrictions. 
The first three cases are triangular identification schemes. The remaining two are non-triangular.

We first consider cases with triangular restrictions. That is, the variables are ordered to satisfy
\begin{equation}
f_1\geq f_2\geq \ldots \geq f_n.\label{eq:ordering}
\end{equation}

\vspace{0.4cm}
\noindent
\textit{Case 1: Recursive restrictions but no restrictions across shocks}
\vspace{0.3cm}

Under recursive restrictions and no restrictions across-shocks we fall into the global identification case and the sequential determination procedure of RWZ pins down a unique admissible $Q$ matrix. 

\vspace{0.4cm}
\noindent
\textit{Case 2: triangular restrictions but no restrictions across shocks}
\vspace{0.3cm}

Under triangular restrictions, consider solving for the admissible $Q$ matrices column by column by exploiting the sequential rank conditions Eq. (\ref{eq:Fiit}). For the first column $q_1$, we have
\begin{equation}
\label{eq:SysCase2q1}
    \left\{\begin{array}{rcl}
    F_{11}(\phi)\,q_1 & = & c_1\\
    q_1^\prime\,q_1 & = & 1
    \end{array}\right.
\end{equation}
Given that $F_{11}(\phi)$ has full row rank, the set of solutions of $q_1$ for the first equations can be spanned by any $n\times 1$ vector $t_1 \in \mathbb{R}$,  
\begin{eqnarray}
\label{eq:qABtransf}
    q_1 & = & F_{11}(\phi)^\prime\bigg(F_{11}(\phi)F_{11}(\phi)^\prime\bigg)^{-1}c_1 + \bigg(I_n-F_{11}(\phi)^\prime\bigg(F_{11}(\phi)F_{11}(\phi)^\prime\bigg)^{-1}F_{11}(\phi)\bigg)t_1\nonumber\\
                & \equiv & d_1 + B_1\,t_1
\end{eqnarray}
Since the $(n\times n)$ matrix $B_1$ has rank $n-f_1 = 1$, it can be decomposed as $B_1 = \alpha_1\beta_1^\prime$, where $\alpha_1$ is a basis for $\text{span}\,(B_1)$, i.e. the column space of $B_1$, and both $\alpha_1$ and $\beta_1$ are non-zero $ n\times 1$ vectors. We can hence write 
\begin{equation}
\label{eq:qtransf}
    q_1=d_1+\alpha_1\,z_1
\end{equation}
with $z_1=\beta_1^\prime\,t_1$, being any scalar. The second (quadratic) equation in system  (\ref{eq:SysCase2q1}) becomes
\begin{eqnarray}
q_1^\prime q_1 & = & (d_1+\alpha_1\,z_1)^\prime\,(d_1+\alpha_1\,z_1)\nonumber\\
& = & d_1^\prime d_1+2d_1^\prime\alpha_1z_1+ \alpha_1^\prime\alpha_1z_1^{2}=1\nonumber\\
& \Rightarrow & \lambda_1+2 \xi_1z_1+\omega_1z_1^2=0\nonumber
\end{eqnarray}
where $\lambda_1=d_1'd_1 - 1$, $\xi_1 = d_1'\alpha_1$ and $\omega_1 = \alpha_1' \alpha_1$ are all functions of the reduced form parameters. There are hence three possibilities: 
\begin{enumerate}
\item If $\xi_1^2-\lambda_1\omega_1>0, $ we have two real solutions. It may be that none, one, or both satisfy the sign normalization restriction for $q_1$.
\item If $\xi_1^2-\lambda_1\omega_1=0$, we have a single real solution. It may or may not satisfy the sign normalization restriciton.
\item If $\xi_1^2-\lambda_1\omega_1<0$, we have no real solution, implying that $\phi$ is not compatible with the imposed restrictions.
\end{enumerate}
In summary, at most there are two admissible $q_1$'s. Denote them by $q_1^{(1)}$ and $q_1^{(2)}$ (allowing $q_1^{(1)} = q_1^{(2)}$). 

Given an admissible $q_1 \in \{q_1^{(1)}, q_1^{(2)} \}$, consider obtaining an admissible second column vector $q_2$ by solving
\begin{equation}
\label{eq:SysCase2q2}
    \left\{\begin{array}{rcl}
    F_{22}(\phi)\,q_2 & = & c_2\\
    q_1^\prime\,q_2 & = & 0\\
    q_2^\prime\,q_2 & = & 1
    \end{array}\right.
\end{equation}
with $\rk\,((F_{22}(\phi)',q_{1}))=n-1$. This system can be transformed as
\begin{equation}
\label{eq:SysCase2q2t}
    \left\{\begin{array}{rcl}
    F_{22}(\phi)\,q_2 & = & c_2\\
    q_1^\prime\,q_2 & = & 0\\
    q_2^\prime\,q_2 & = & 1
    \end{array}\right.\Longrightarrow
    \left\{\begin{array}{rcl}
    \left(\begin{array}{c}F_{22}(\phi)\\q_1^{\prime}\end{array}\right)\,q_2 & = & \left(\begin{array}{c}c_2\\0\end{array}\right)\\
    q_2^\prime\,q_2 & = & 1
    \end{array}\right.\Longrightarrow
    \left\{\begin{array}{rcl}
    \tilde{F}_{22}(\phi)\,q_2 & = & \tilde{c}_2\\
    q_2^\prime\,q_2 & = & 1
    \end{array}\right.
\end{equation}
where $\tilde{F}_{22}(\phi)= (F_{22}'(\phi),q_1)' $ and 
$\tilde{c}_2=(c_2',0)'$. Given the assumption $\rk\,(\tilde{F}_{22}(\phi))=n-1$, Eq. (\ref{eq:SysCase2q2t}) can be solved in the same way as the system for $q_1$. We can hence obtain at most two admissible $q_2$ vectors for each of $q_1=q_1^{(1)}$ and $q_1 =q_1^{(2)}$. So far there are at most four admissible vectors for the first two columns of $Q$.

We repeat this argument for $i=3, \dots, n$. Given that there are at most $2^{i-1}$ admissible constructions of $(q_1, \dots, q_{i-1})$, and at each admissible $(q_1, \dots, q_{i-1})$, we solve for $q_i$ 
\begin{equation}
\label{eq:SysCase2qi}
\left\{\begin{array}{rcl}
\tilde{F}_{ii}(\phi)\,q_i & = & \tilde{c}_i\\
q_i^\prime\,q_i & = & 1,
\end{array}\right.
\end{equation}
where
\begin{equation}
\nonumber
\tilde{F}_{ii}(\phi)= (F_{ii}(\phi)', q_1, \dots, q_{i-1})' \hspace{1cm}
\text{and} \hspace{1cm} \tilde{c}_i=(c_i', 0, \dots,0) .
\end{equation}
Again, finding an admissible $q_i$ given $(q_1, \dots, q_{i-1})$ boils down to solving a quadratic equation, so there are at most two solutions for $q_i$, implying that there are at most $2^i$ admissible constructions of $(q_1, \dots, q_{i-1}, q_{i})$. At $i=n$, 
we obtain at most $2^n$ admissible $Q$ matrices.

\vspace{0.4cm}
\noindent
\textit{Case 3: triangular restrictions and restrictions across shocks}
\vspace{0.3cm}

The triangular restrictions imply that $\textbf{F}(\phi)$ is lower block-triangular, i.e. $F_{ij} = 0$ for $j>i$, and $f_i = n-i$ for all $i = 1, \dots,n$.
The case where $i=1$ is identical to the initial step in \textit{Case 2} above, so we have at most two admissible $q_1$ vectors. For $i>1$ we exploit the sequential structure of the restrictions and obtain each admissible $q_i$ sequentially given $(q_1,\ldots,q_{i-1})$ obtained in the preceding steps. The only difference with respect to
\textit{case 2} is that, once $(q_1,\ldots,q_{i-1})$ is given, the system of equations in 
Eq. (\ref{eq:SysCase2qi}), will be characterized by
\begin{equation}
\label{eq:case3}
    \tilde{F}_{ii}(\phi)=(F_{ii}'(\phi), q_1, \dots, q_{i-1})',  \hspace{0.5cm}
    \text{and} \hspace{0.5cm} \tilde{c}_i=\Big(\big(c_i-F_{i1}(\phi)q_1-\cdots-F_{i,i-1}(\phi)q_{i-1}\big)',0, \dots,0\Big)'. 
\end{equation}
Repeating the argument of \textit{Case 2}, we conclude there are at most $2^n$ admissible $Q\in \O$.

\vspace{0.4cm}

We now move to the cases with non-triangular identifying restrictions.

\vspace{0.4cm}

\noindent
\textit{Case 4: Non-triangular restrictions and no restrictions across shocks}
\vspace{0.3cm}

If $f_1=n-1$, we can proceed as in \textit{Case 2} and globally or locally identify $q_1$, depending on the restrictions at hand. If, instead, $f_1<n-1$, we can only identify the basis spanning a subspace in $\Re^n$ of dimension $n-f_1$ containing $q_1$. The system of equations characterizing $q_1$ is given by
\begin{equation}
\label{eq:SysCase3}
\left\{\begin{array}{rcl}
F_{11}(\phi)\,q_1 & = & c_1\\
q_1^\prime\,q_1 & = & 1.
\end{array}\right.
\end{equation}
Following the analysis of \textit{Case 2}, we can represent an admissible $q_1$ by $q_1=d_1+\alpha_1 z_1$, where $z_1=\beta_1^\prime\,t_1 \in \mathbb{R}^{n - f_1}$, $\alpha_1$ is a non-zero $n \times (n-f_1)$ matrix, $\beta_1$ is a non-zero $(n-f_1)\times n$ matrix, and $t_1 \in \mathbb{R}^n$. 
Given this representation of $q_1$, the second (quadratic) equation in system (\ref{eq:SysCase3}) becomes
\begin{eqnarray}
q_1^\prime q_1 & = & d_1^\prime d_1+2d_1^\prime\alpha_1z_1+z_1^\prime\alpha_1^\prime\alpha_1z_1=1\nonumber\\
& \Rightarrow &  \lambda_1+2\xi_1'z_1+z_1^\prime \omega_1z_1=0,\nonumber
\end{eqnarray}
where $\lambda_1 = d_1'd_1 - 1$, $\xi_1 = \alpha_1' d_1$, and $\omega_1 = \alpha_1' \alpha_1$. The set of real roots of this quadratic equation in $z_1$, if non-empty, is a singleton or a hyper-ellipsoid in $\mathbb{R}^{n-f_1}$ with its radius given by the constant term in the completion of squares (if nonnegative).

Assuming an admissible $q_1$ exists, consider the equation system for $q_2$,
\begin{equation}
\label{eq:SysCase3_2}
\left\{\begin{array}{rcl}
q_2 & = & d_2+\alpha_2\,z_2\\
q_2^\prime\,q_2 & = & 1,
\end{array}\right.
\end{equation}
whose set of roots, if non-empty, is again a singleton or a $n-f_2$-dimensional hyper-ellipsoid. In addition, we have the following orthogonality restriction between $q_1$ and $q_2$,
\begin{eqnarray}
q_1^\prime q_2 & = & d_1^\prime d_2+d_1^\prime\alpha_2z_2+z_1^\prime\alpha_1^\prime d_2+z_1^\prime\alpha_1^\prime\alpha_2z_2\nonumber\\
& \equiv &  \lambda_{1,2}+\xi_{1,2}'z_2+z_1^\prime \xi_{2,1}+z_1^\prime \omega_{1,2}z_2=0.\nonumber
\end{eqnarray}
Enumerating these equations for all $i = 1, \dots, n$, we obtain the following system of equations:
\begin{equation}
\label{eq:SysCase3_3}
\left\{\begin{array}{rcl}
z_1^\prime \omega_1 z_1 + 2 \xi_1'z_1 + \lambda_1 & = & 0\\
z_2^\prime \omega_2 z_2 + 2 \xi_2'z_2 + \lambda_2 & = & 0\\
\vdots &&\\
z_n^\prime \omega_n z_n + 2 \xi_n'z_n + \lambda_n & = & 0\\
z_1^\prime \omega_{1,2} z_2 + \xi_{1,2}'z_2 + z_1^\prime \xi_{2,1} + \lambda_{1,2} & = & 0\\
z_1^\prime \omega_{1,3} z_3 + \xi_{1,3}'z_3 + z_1^\prime \xi_{3,1} + \lambda_{1,3} & = & 0\\
\vdots &&\\
z_{n-1}^\prime \omega_{n-1,n} z_n + \xi_{n-1,n}'z_n + z_{n-1}^\prime \xi_{n,n-1} + \lambda_{n-1,n} & = & 0.
\end{array}\right.
\end{equation}
The number of equations is $n+n(n-1)/2=n(n+1)/2$. The number of unknowns, contained in $z_1,z_2,\ldots,z_n$, is 
\begin{equation}
(n-f_1)+(n-f_2)+\ldots+(n-f_n)  \leq n^2-n(n-1)/2=n(n+1)/2,
 \end{equation}
where the inequality follows by the order condition stated in Proposition \ref{prop:LocIdent}, $\sum_{i=1}^n f_i \geq n(n-1)/2$. Hence, we have a system of $n(n+1)/2$ equations with at most $n(n+1)/2$ unknowns. Moreover, each one is a quadratic equation and, importantly, given the rank condition for local identification is satisfied, each of the solutions has to be an isolated point. B\'ezout's theorem gives that the maximum number of solutions is the product of the polynomial degree of all the equations, so the number of solutions is at most $2^{n(n+1)/2}$.

\vspace{0.4cm}
\noindent
\textit{Case 5: Non-triangular and across-shocks restrictions}
\vspace{0.3cm}

In this case analysis of identification requires considering all equations jointly. We will have a system of equations of the form
\begin{equation}
\label{eq:SysCase4}
\left\{\begin{array}{rcl}
\textbf{F}(\phi)\text{vec}\, Q & = & \bf{c}\\
q_1^\prime\,q_1 & = & 1\\
q_2^\prime\,q_2 & = & 1\\
\vdots&&\\
q_n^\prime\,q_n & = & 1\\
q_1^\prime\,q_2 & = & 0\\
\vdots&&\\
q_{n-1}^\prime\,q_n & = & 0.
\end{array}\right.
\end{equation}
This system consists of  $n^2$ equations with $n^2$ unknowns (the elements in $Q$). The first $n(n-1)/2$ equations are linear and the latter $n(n+1)/2$ equations are all quadratic. By B\'ezout's theorem, the maximum number of solutions is $2^{n(n+1)/2}$.$\square$

\subsection*{Proof of Proposition \ref{prop:SuffRWZ}: RWZ sufficient condition for checking local identification} 

The result is a by-product of Proposition \ref{prop:SuffLocIdent}. As observed in Eq. (\ref{eq:sysh1}), the first column $q_1$ is locally identified if and only if $\left(\begin{array}{c}F_{11}(\phi)Q\\e_1^\prime\end{array}\right)$ has full column rank equal to $n$. When moving to the identification of $q_2$, from the system (\ref{eq:sysh2}), and recalling that the first two elements of $h_2$ are zero, we have no admissible infinitesimal rotation (i.e. $h_2=0$) if 
\begin{equation}
\nonumber
        \rk\left(\begin{array}{c}
        \underset{(n-2)\times n}{F_{22}(\phi)}\cdot\underset{n\times n}{Q}\\
        \left(\begin{array}{ccc}
        \underset{2\times 2}{I_2}&&\underset{2\times (n-2)}{0}
        \end{array}\right)
        \end{array}\right)=n.
\end{equation}
Repeating this argument for the remaining columns of $Q$, we obtain the result in the proposition. $\square$


\newpage
\section{Appendix: The $\Dtn$ matrix} 
\label{app:Dnt}

A skew-symmetric (square) matrix $A$ satisfies $A^\prime=-A$. Let $\tilde{v}(A)$ be a 
vector containing the $n(n-1)/2$ \textit{essential} elements of $A$. When $A$ is skew-symmetric, it is possible to expand the elements of $\tilde{v}(A)$ to obtain $\ve A$.
$\tilde{D}_n$, thus, can be defined to be the $n^2\times n(n-1/2)$ matrix with the property that
\begin{equation}
\label{eq:DtnDef}
\nonumber
\Dtn \, \tilde{v}(A) = \ve A
\end{equation} 
for any skew symmetric $n\times n$ matrix $A$. For a formal definition and  properties of $\Dtn$, see \cite{magnus88}. 
Here, we present $\Dtn$ for $n=2$, $n=3$ and $n=4$:

\begin{equation}
\label{eq:Dtn}
\nonumber
\begin{array}{c}
\tilde{D}_2=\left(\begin{array}{r}0\\1\\-1\\0\end{array}\right),\hspace{1cm}
\tilde{D}_3=\left(\begin{array}{rrr}
0&0&0\\
\circled{1}&0&0\\
0&\circled{1}&0\\
\hdashline
-1&0&0\\
0&0&0\\
0&0&\circled{1}\\
\hdashline
0&-1&0\\
0&0&-1\\
0&0&0\\
\end{array}\right), \hspace{1cm}
\tilde{D}_4=\left(\begin{array}{rrrrrr}
0&0&0&0&0&0\\
\circled{1}&0&0&0&0&0\\
0&\circled{1}&0&0&0&0\\
0&0&\circled{1}&0&0&0\\
\hdashline
-1&0&0&0&0&0\\
0&0&0&0&0&0\\
0&0&0&\circled{1}&0&0\\
0&0&0&0&\circled{1}&0\\
\hdashline
0&-1&0&0&0&0\\
0&0&0&-1&0&0\\
0&0&0&0&0&0\\
0&0&0&0&0&\circled{1}\\
\hdashline
0&0&-1&0&0&0\\
0&0&0&0&-1&0\\
0&0&0&0&0&-1\\
0&0&0&0&0&0\\
\end{array}\right)
\end{array},
\end{equation}
where we have circled the elements selecting the last $n-i$ columns, $i=1,\ldots,n$, of the $F_{ii}(\phi)Q$ matrix in the proof of Proposition \ref{prop:SuffRWZ}. 

Finally, as can be seen from $\tilde{D}_2$, $\tilde{D}_3$ and $\tilde{D}_4$, the matrix $\Dtn$ is always of full column rank $n(n-1)/2$.


\newpage
\section{Computational procedure for locally identified SVARs with triangular restrictions}
\label{app:AlgorithmNH}

If the identifying restrictions imposed allow the sequential determination of the column vectors of $Q$ as exploited in the identification arguments in the previous sections, we can modify Algorithm \ref{algo:Estimation}. In this section, we consider triangular SVARs, as covered in Proposition \ref{prop:SuffLocIdent}.

Let $Q_{1:i}$, $1 \leq i \leq n$, be a $n \times i$ matrix whose column vectors are orthonormal (i.e. it consists of the first $i$ column vectors of $Q$). Given $\phi$, define $\tilde{F}_{11}(\phi)=F_{11}(\phi)$ and the following matrices sequentially for $i=2, \dots, n$,
\begin{align}
\tilde{F}_{ii}(\phi)=\left(\begin{array}{c}F_{ii}(\phi)\\ Q_{1:(i-1)}(\phi)' \end{array}\right), \label{eq:Fii}
\end{align}
where $Q_{1:(i-1)}(\phi)$ satisfies the identifying restrictions for the first $(i-1)$ orthogonal vectors, i.e., $(F_{j1}(\phi), \dots F_{jj}(\phi))  \text{vec} Q_{1:(i-1)}(\phi)  = c_j$ holds for $j=1,\dots,(i-1)$. For $i=1,\dots,n$, we define a $(n-1)\times 1$ vector, 
\begin{equation}
\tilde{c}_i(\phi)=\left( \begin{array}{c}c_{i}-\left(F_{i1}(\phi), \dots, F_{i(i-1)}(\phi) \right)  \text{vec}  Q_{1:(i-1)}(\phi)\\ 0 \\\vdots\\0\end{array}\right)
\end{equation}
Then, for $i=1,\ldots,n$, define
\begin{eqnarray}
d_i (\phi) & = & \tilde{F}_{ii}(\phi)^\prime\bigg(\tilde{F}_{ii}(\phi)\tilde{F}_{ii}(\phi)^\prime\bigg)^{-1}\tilde{c}_i(\phi) \label{eq:Ai}, \\
B_i (\phi) & = & \bigg(I_n-\tilde{F}_{ii}(\phi)^\prime\bigg(\tilde{F}_{ii}(\phi)\tilde{F}_{ii}(\phi)^\prime\bigg)^{-1}\tilde{F}_{ii}(\phi)\bigg),\label{eq:Bi}
\end{eqnarray}
and let $\alpha_i(\phi)$ be a $n \times 1$ basis vector of the linear space spanned by the vectors in $B_i(\phi)$. Note that $B_i(\phi)$ is the $n \times n$ matrix projecting onto the linear space orthogonal to the row vectors of $\tilde{F}_{ii}(\phi)$. Hence, given the rank of $\tilde{F}_{ii}(\phi)$ is $n-1$, $B_i (\phi)$ has a rank of 1, so $\alpha_i(\phi)$ is unique up to sign, and $\tilde{F}_{ii}(\phi) \alpha_i(\phi)=0$ holds. 

Consider the $n \times 1$ vector, $x= d_i(\phi) + z \alpha_{i}(\phi)$, $z \in \mathbb{R}$. Due to the way $d_i(\phi)$ and $\alpha_i(\phi)$ are constructed, $\tilde{F}_{ii}(\phi)x = \tilde{c}_i$ holds. That is, by choosing $z$ so that $x$ is a unit-length vector, we can obtain $q_i$ vectors satisfying $\tilde{F}_{ii}(\phi)q_i = \tilde{c}_i$. Solving for $x$ is simple as it requires only finding the roots of a quadratic equation (see Eq. (\ref{eq:QuadEqNHAlgo}) and Eq. (\ref{eq:QuadEqNHAlgo2}) in Algorithm \ref{algo:EstimationTrNonHomo} below).  Given $\phi$, we repeat this process for every $i=1, \dots, n$ to determine the $q_i$ vectors sequentially, and compute all the $Q$ matrices satisfying the equality restrictions $\textbf{F}(\phi,Q)=\textbf{0}$. $\mathcal{A}_0(\phi| F,S)$ and $IS_{\eta}(\phi)$ can then be obtained by retaining the $Q$ that satisfy the normalization and sign restrictions. We summarize this computational procedure in the next algorithm.    

\begin{algo}
\label{algo:EstimationTrNonHomo}
Consider an SVAR satisfying the normalization restrictions, the equality restrictions Eq. (\ref{eq:GenFormRest}), and the sign restrictions Eq. (\ref{eq:SignRest}), where the imposed equality restrictions satisfy the sufficient condition for local identification given in Proposition \ref{prop:SuffLocIdent}. Let $\hat{\phi}$ be a given estimator for $\phi$. In the description of the algorithm below, we omit the argument $\hat{\phi}$ as far as it does not give rise confusion. 

Let $\textbf{b}=(b_1,\dots,b_n) \in \{0,1 \}^n$ be a bit vector which will be used to index each of the at most $2^n$ possible solutions for the $Q$ matrices. Beginning with $\textbf{B} = \{0,1 \}^n$, we will map each $\textbf{b} \in \textbf{B}$ to a possible solution of $Q$, check if it is feasible or not, and refine $\textbf{B}$ accordingly. The following algorithm describes this process in detail:    
\begin{enumerate}
\item Solve for $z \in \mathbb{R}$ in
\begin{equation}
        \label{eq:QuadEqNHAlgo}
                d_1' d_1+2d_1' \alpha_1z+\alpha_1' \alpha_1 z^2 =1,
        \end{equation}
and denote the two solutions by $z_{1}^{b_1}$, $b_1 \in \{0, 1 \}$. 
\begin{enumerate}
\item If they are real, then define $q_{1}^{b_1}=d_1+\alpha_1 z_1^{b_1}$, $b_1 \in \{0,1 \}$. Let $\textbf{B}_1 \subset \{0,1 \}$ be the set of $b_1$ such that $q_1^{b_1}$ satisfies the sign normalization and sign restrictions for $q_1$. If $\textbf{B}_1$ is empty (i.e., no $q_{1}^{b_1}$ satisfies the sign normalization and sign restrictions for $q_1$), then stop and conclude $\mathcal{Q}_R(\hat{\phi}) = \emptyset$
 

\item If the roots of Eq. (\ref{eq:QuadEqNHAlgo}) are not real, then stop and return $\mathcal{Q}_R(\hat{\phi}) = \emptyset$. 
\end{enumerate}

\item This step iterates sequentially for $i=2, \dots, n$, given $\textbf{B}_{i-1} \subset \{ 0,1 \}^{i-1}$.  
\begin{enumerate}
\item  For each $(b_1, \dots, b_{i-1}) \in \textbf{B}_{i-1}$, construct $\textbf{B}_i(b_1b_2 \cdots b_{i-1}) \subset \{0,1 \}^i$ by performing the following subroutines:
\begin{enumerate}
\item Construct $\tilde{F}_{ii}$ from Eq. (\ref{eq:Fii}) by setting $Q_{1:{i-1}} = [q_{1}^{b_1},q_{2}^{b_1b_2}, \dots, q_{i-1}^{b_1 \cdots b_{i-1}}]$, and obtain $d_{i}$ and $\alpha_i$ accordingly. Then,solve for $z \in \mathbb{R}$ in
\begin{equation}
d_i' d_i+2d_i' \alpha_iz+\alpha_i' \alpha_i z^2 =1, \label{eq:QuadEqNHAlgo2}
\end{equation}
 and denote the two solutions by $z_{i}^{b_1b_2\cdots b_{i}}$, $b_i \in \{0, 1 \}$. 
\item If they are real, define $q_{1}^{b_1b_2 \cdots b_i}=d_i+\alpha_i z_1^{b_1 b_2 \cdots b_i}$, $b_i \in \{0,1 \}$. Let $\textbf{B}_i(b_1b_2 \cdots b_i)$ be the set of $(b_1, b_2, \dots, b_i)\in \{0,1 \}^i$ such that $q_{i}^{b_1b_2, \cdots b_{i}}$ satisfies the sign normalization and sign restrictions for the $i$-th column vector of $Q$. This can be empty if no $q_{i}^{b_1b_2, \cdots b_{i}}$ satisfies them. 

\item If the roots of Eq. (\ref{eq:QuadEqNHAlgo2}) are not real, return $\textbf{B}_i(b_1b_2 \cdots b_{i-1}) = \emptyset$. 
\end{enumerate}
\item Construct $\textbf{B}_i = \bigcup_{(b_{1}, \dots, b_{i-1}) \in \textbf{B}_{i-1}} \textbf{B}_{i}(b_1 \cdots b_{i-1})$. If $\textbf{B}_i \neq \emptyset$, go back to the beginning of Step 2. 

\item If $\textbf{B}_i = \emptyset$, then stop and return $\mathcal{Q}_R(\hat{\phi}) = \emptyset$.
\end{enumerate}
\item We obtain
\begin{equation}
\mathcal{Q}_R(\hat{\phi}) = \left\{ \left(q_{1}^{b_1}, q_2^{b_1b_2}, \dots, q_n^{b_1 b_2 \cdots b_n} \right) : \textbf{b} \in \textbf{B}_n \right\}. \notag
\end{equation}
\end{enumerate}
\end{algo}

Algorithm \ref{algo:EstimationTrNonHomo} computes the set of all admissible $Q\in \mathcal{Q}_R(\hat{\phi})$. In the description of the algorithm, they are indexed by the bit vectors $\textbf{b} \in \textbf{B}_n$. The algorithm is constructive and guaranteed to compute all the admissible $Q$ matrices. Projecting this set of admissible matrices onto the impulse response of interest, we obtain a plug-in estimate of the identified set  $IS_{\eta}(\hat{\phi})$.

Algorithm \ref{algo:EstimationTrNonHomo} is more constructive than Algorithm \ref{algo:Estimation}, but it restricts the set of equality restrictions to be triangular. Algorithm \ref{algo:EstimationTrNonHomo} can be extended to a class of models involving non-triangular identifying restrictions (e.g., examples in Sections \ref{sec:NonRec} and \ref{sec:proxySVAR}) by incorporating steps that solve a certain system of quadratic equations. Such an algorithm is rather involved to present, so we do not include it in this paper. Algorithm \ref{algo:Estimation} can be certainly applied to a general class of models with non-triangular identifying restrictions.


\newpage
\section{Figures of the HSVAR application}
\label{sec:fig}

\begin{figure}[ht!]
 \caption{Data used in the HSVAR model for the impact of the monetary policy shocks (July 1955-June 2007)}
\begin{center}
  \includegraphics[origin=c,scale=0.6]{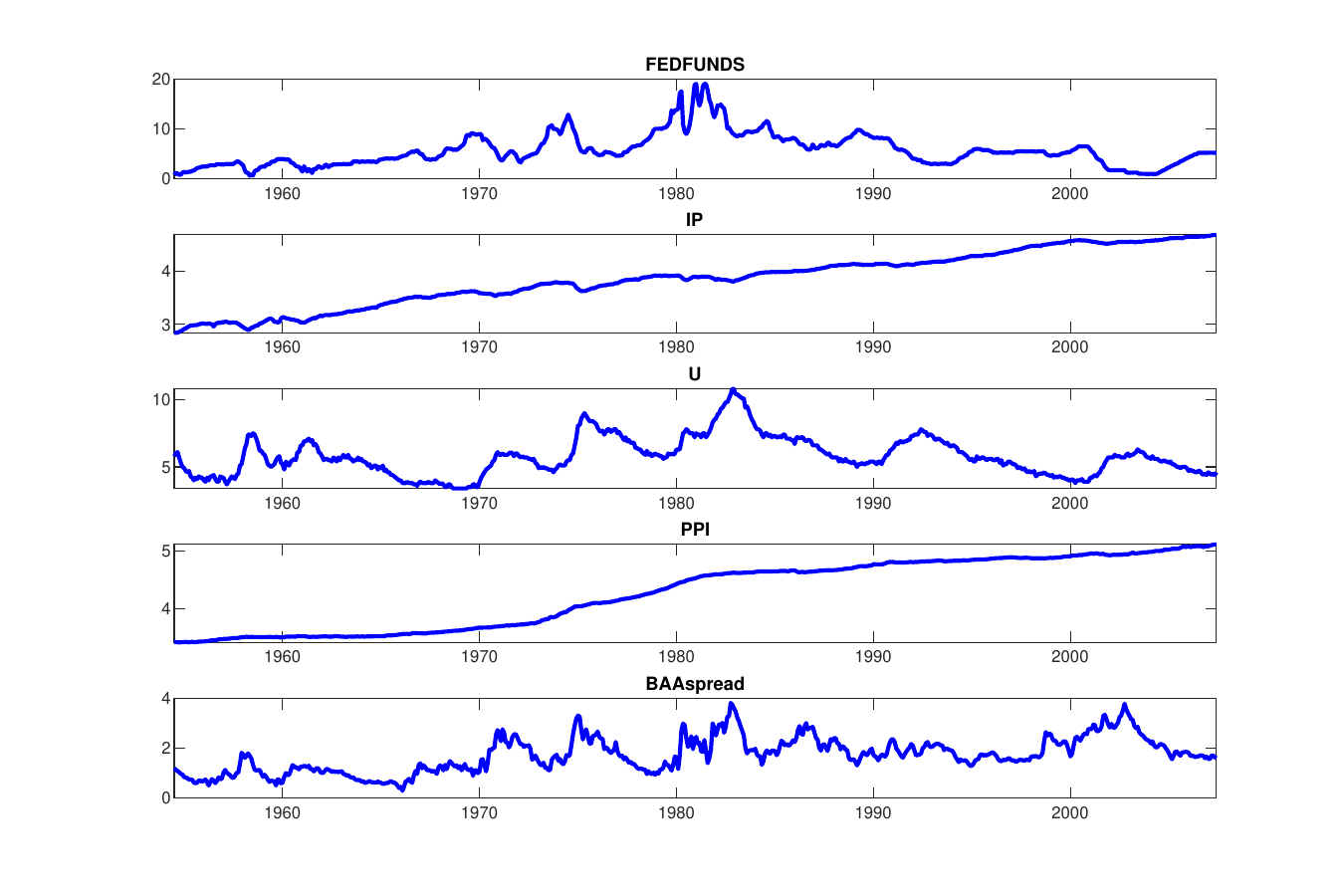}
\end{center}
\label{fig:DataPlot}
\end{figure}

\begin{figure}[ht!]
 \caption{Reduced-form residuals and break date.}
\begin{center}
  \includegraphics[origin=c,scale=0.6]{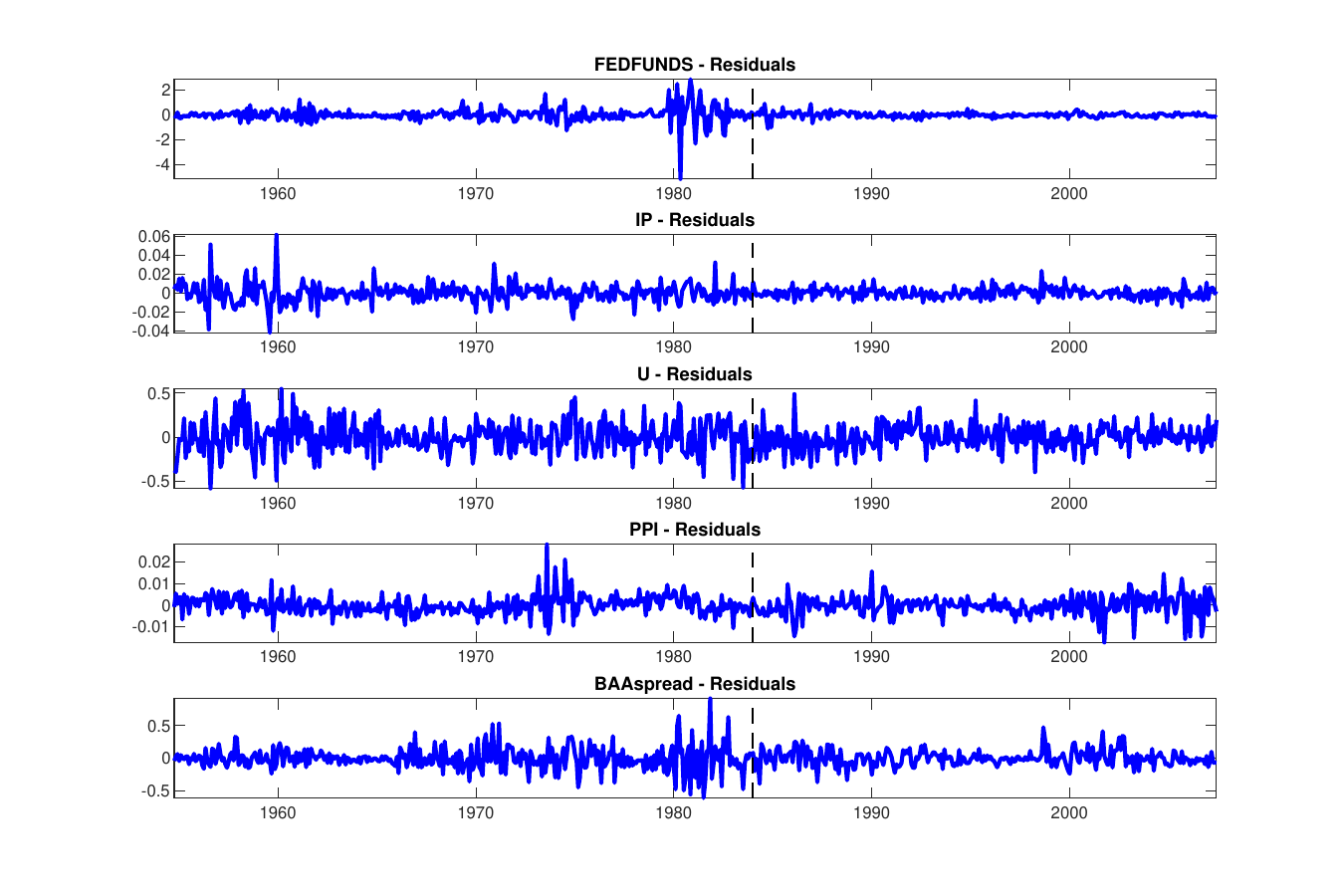}%
\end{center}
\caption*{\scriptsize\textit{Notes}: Reduced form residuals and time of the break (1st January 1984). Monthly data.}
\label{fig:ResPlot}
\end{figure}

The time series included in the VAR and the reduced-form residuals are shown, respectively, in Figures \ref{fig:DataPlot} and \ref{fig:ResPlot}, that also displays a vertical bar in correspondence of the break date, January 1984. 

\clearpage
\section{Estimation and inference algorithms in practice: the non-recursive New-Keynesian SVAR}
\label{sec:EmpAppNK}

We illustrate how our approach works with an empirical application to the non-recursive New-Keynesian SVAR shown in Section \ref{sec:NonRec}. We consider the small scale DSGE model presented in Eq. (\ref{eq:DSGE}), which has the SVAR representation with sign normalizations and the zero restrictions in Eq. (\ref{eq:DSGEsimmat}). The vector of observables is inflation as measured by the GDP deflator ($\pi_t$), real GDP per capita as a deviation from a linear trend ($x_t$) and the federal funds rate ($i_t$).\footnote{The data are used in \cite{AS11} and downloaded from Frank Schorfheide's website: https://web.sas.upenn.edu/schorf/. For details on the construction of the series, see Appendix D from \cite{MSG13} and Footnote 5 of \cite{AS11}.} The data are quarterly from 1965:1 to 2006:1.

As discussed in Section \ref{sec:NonRec}, the imposed restrictions deliver local identification and, given the reduced form parameters, they can yield up to two admissible matrices, $Q_1$ and $Q_2$. To compute them, we apply  Algorithm \ref{algo:Estimation} at every draw of $\phi$ from its posterior, using the Matlab command \texttt{vpasolve} to solve the system of quadratic equations.

We specify non-informative improper priors for the reduced-form parameters. We draw from the posterior of the reduced form 2,000 times and, considering uniquely the zero restrictions in Eq. (\ref{eq:DSGEsimmat}), obtain 2,000 realizations of $\mathcal{Q}_R(\hat{\phi})$, each of which is non-empty and consists of two orthogonal matrices $Q_1$ and $Q_2$. We label them by $Q_1$ and $Q_2$ according to the ordering of the contemporaneous inflation impulse response. 

\begin{figure}[ht]
        \caption{Impulse response functions for the New-Keynesian non-recursive SVAR.}
  \label{fig:IRF_NK_Eq}
\begin{center}
                \subfigure[{Output gap to $\e_t^{mp}$: Single admissible $Q_i$.}]{
                \includegraphics[angle=270,origin=c,scale=0.23]{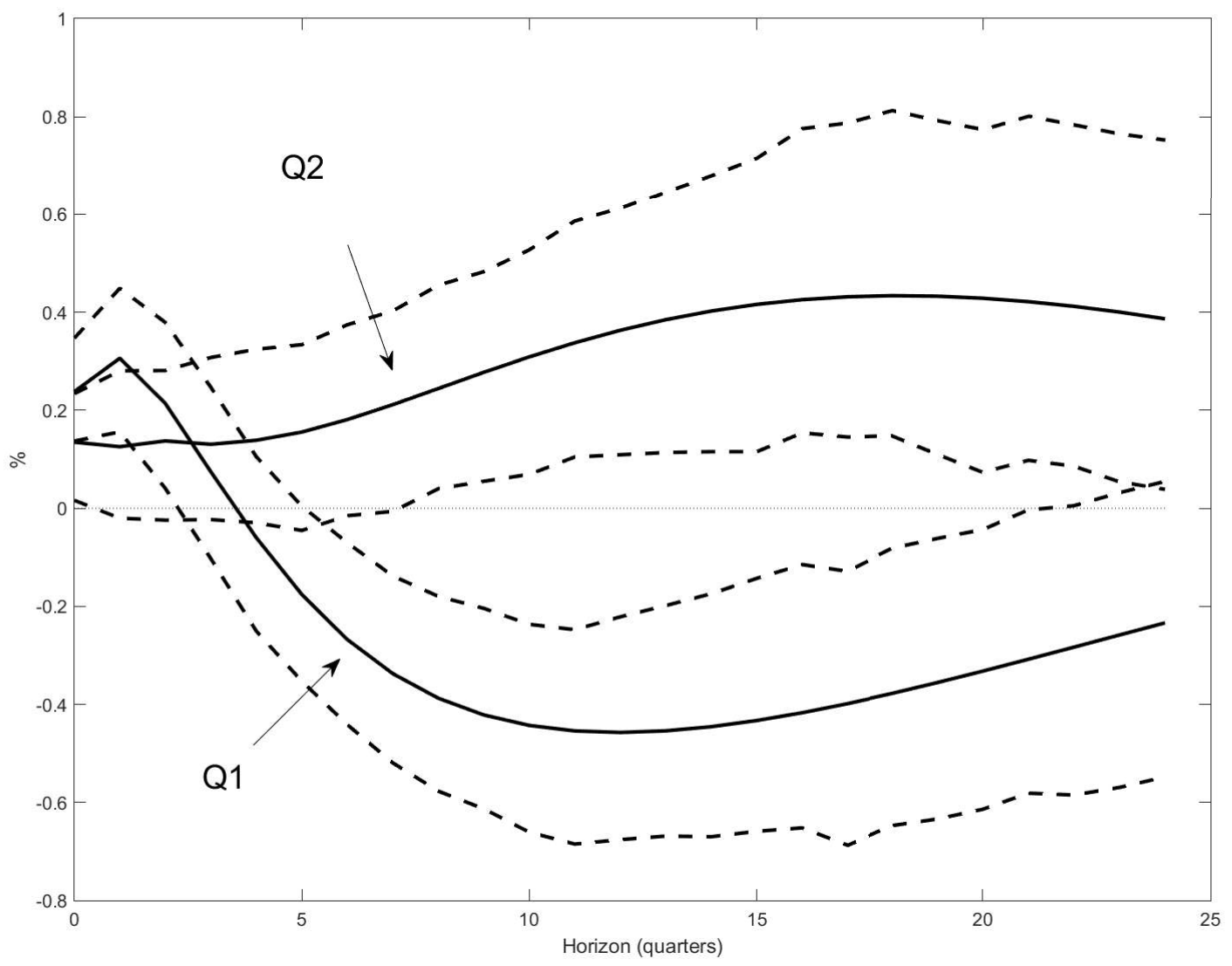}}
                \subfigure[{Inflation to $\e_t^{mp}$: Single admissible $Q_i$.}]{
                \includegraphics[angle=270,origin=c,scale=0.23]{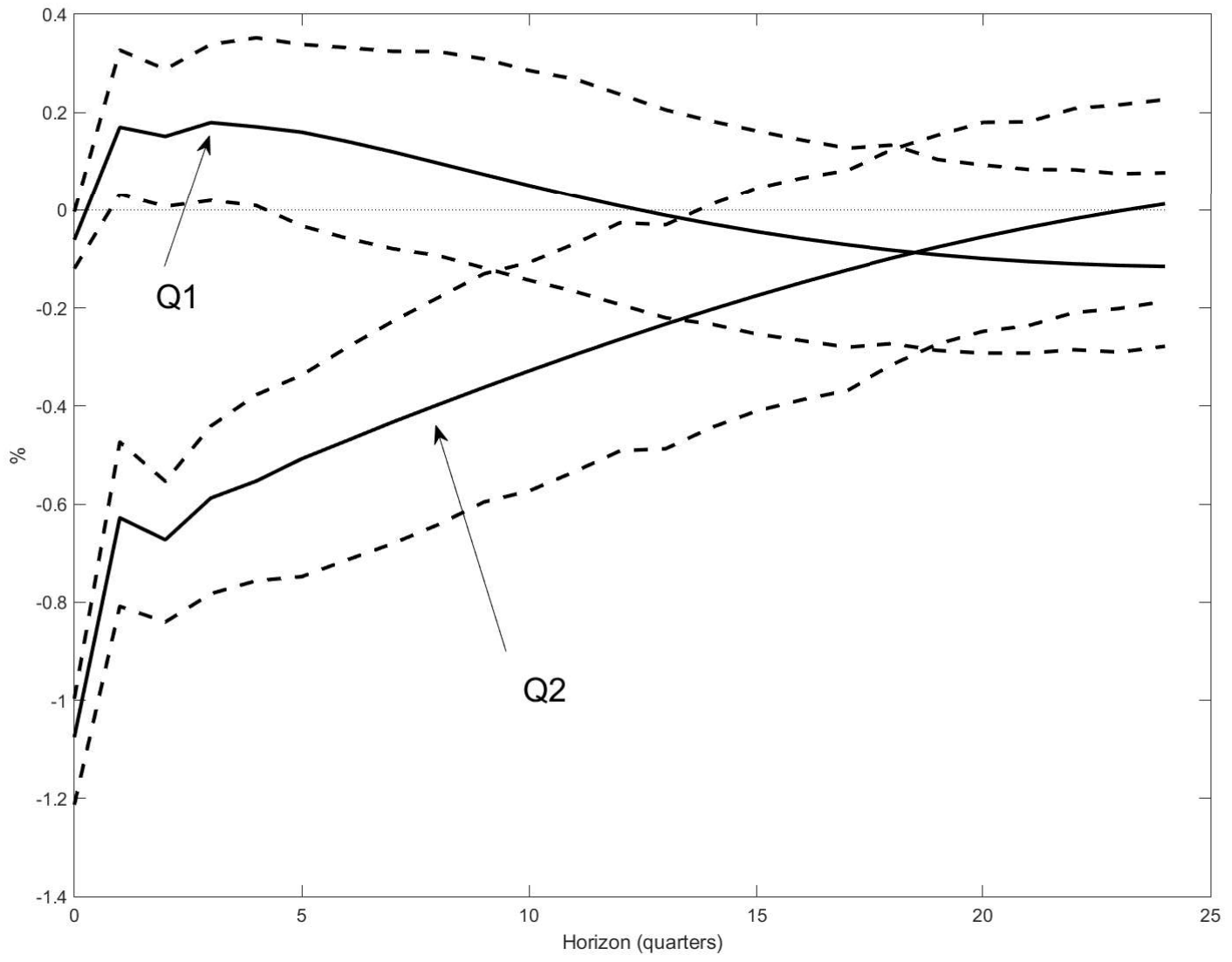}}
\end{center}
\begin{minipage}{\textwidth}%
{\scriptsize{\textit{Notes}: In the left column we report the impulse responses for the output gap 
obtained as the Bayesian posterior means with the upper and lower bounds of the highest posterior density regions with credibility $90\%$
obtained through the admissible $Q_1$ and $Q_2$ matrices, considered separately. Similarly, in the right column we report the impulse responses for 
inflation.\par}}
\end{minipage}
\end{figure}

Figure \ref{fig:IRF_NK_Eq} reports the impulse response to a contractionary monetary policy shock for the output gap (left panel) and inflation (right panel). It shows the posterior means and the highest posterior density regions with credibility $90\%$ that would be obtained if the conditional prior for $Q$ given $\phi$ assigned all probability mass to either $Q_1$ or $Q_2$. That is, reporting one of the inference outputs corresponds to the Bayesian approach that focuses only on one of the posterior modes, ignoring the other.\footnote{Although not reported for saving space, the Bayesian credible intervals of Figure \ref{fig:IRF_NK_Eq} are nearly identical to those obtained by the frequentist bootstrap-after-bootstrap approach of \cite{Kilian98}}

The inference result based on $Q_1$ shows evidence for both price and output puzzles in the short run. In the medium term, on the other hand, a contractionary monetary policy shock triggers a contraction of the output gap, leaving the price dynamics mostly unaffected. Inference based on the other orthogonal matrix $Q_2$, however, leads to a contrasting conclusion. The reaction of prices to the monetary policy shock is significantly negative, while the output gap responds positively and significantly, particularly in the medium-long run. 

This example illustrates that different locally-identified observationally equivalent parameter values can lead to strikingly different conclusions, and ignoring this distorts the information contained in the data. A standard off-the-shelf econometric package could uncover just one of the two results. For instance, Gretl and Eviews both return results in line with those obtained through $Q_1$. These packages rely on algorithms that maximize the likelihood starting from some initial value without checking other local maxima. Thus, we recommend checking for the existence of other local maxima and, if any exist, addressing how the conclusions change among them by applying the methods proposed in this paper.

\begin{figure}
        \caption{New-Keynesian non-recursive SVAR with zero restrictions only}
  \label{fig:IRF_NK_Eq_NoSign}
\begin{center}
                \subfigure[{Output gap to $\e_t^{mp}$: impulse response draws}]{
                \includegraphics[angle=270,origin=c,scale=0.2]{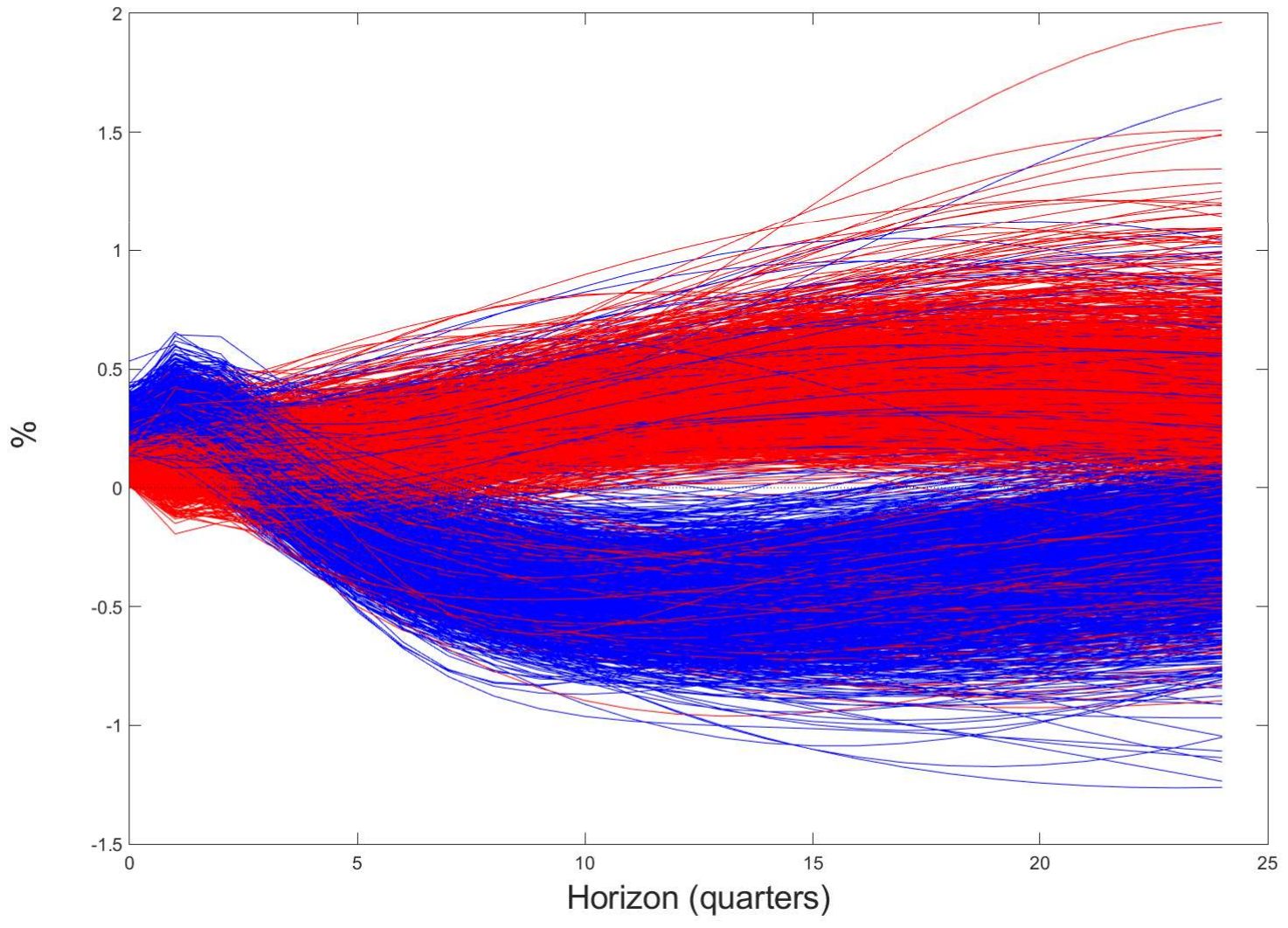}}\hspace{1.5cm}
                \subfigure[{Inflation to $\e_t^{mp}$: impulse response draws }]{
                \includegraphics[angle=270,origin=c,scale=0.2]{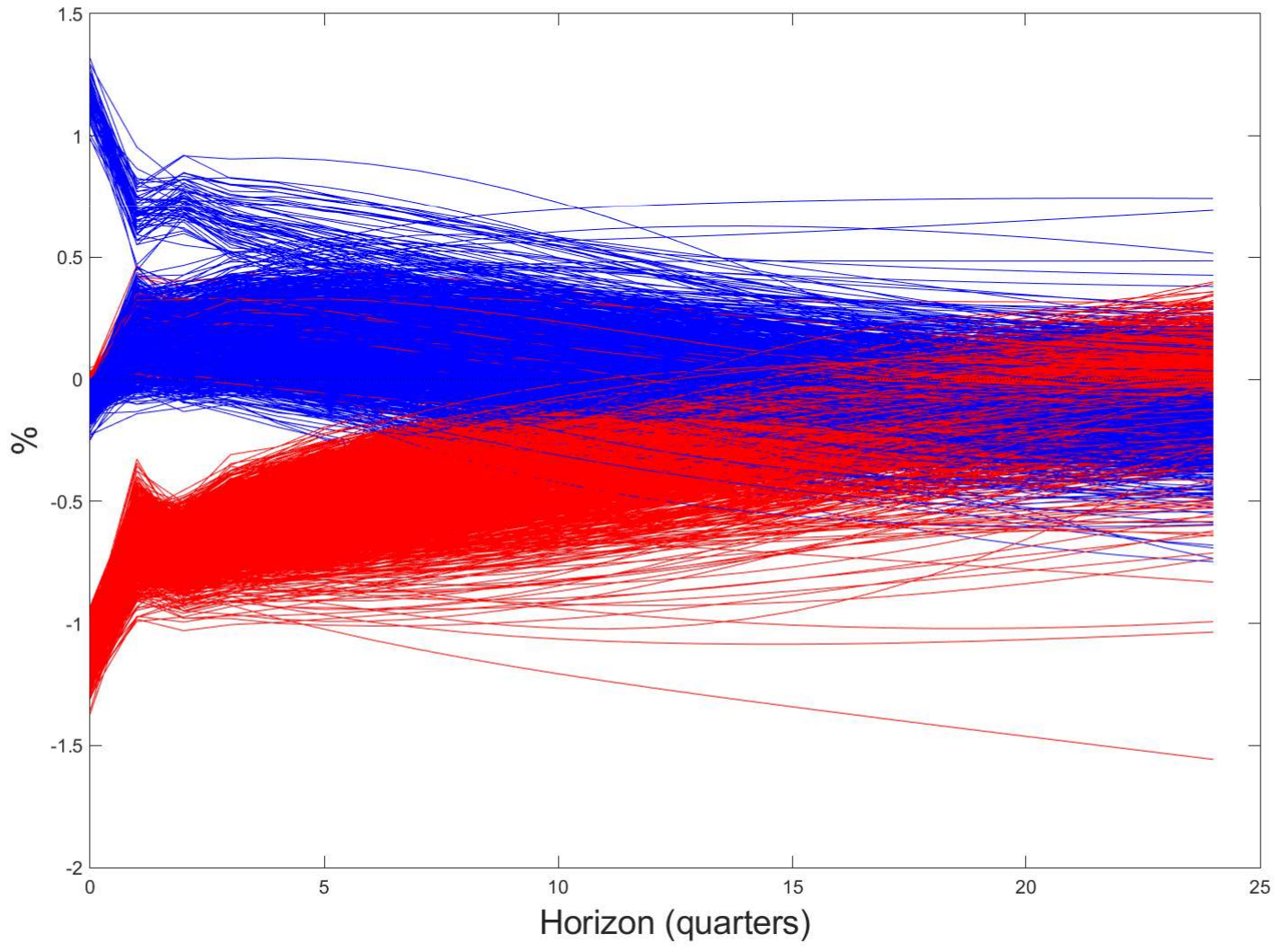}}
                \subfigure[{Credible regions and fixed-label projection confidence sets}]{
                \includegraphics[angle=270,origin=c,scale=0.2]{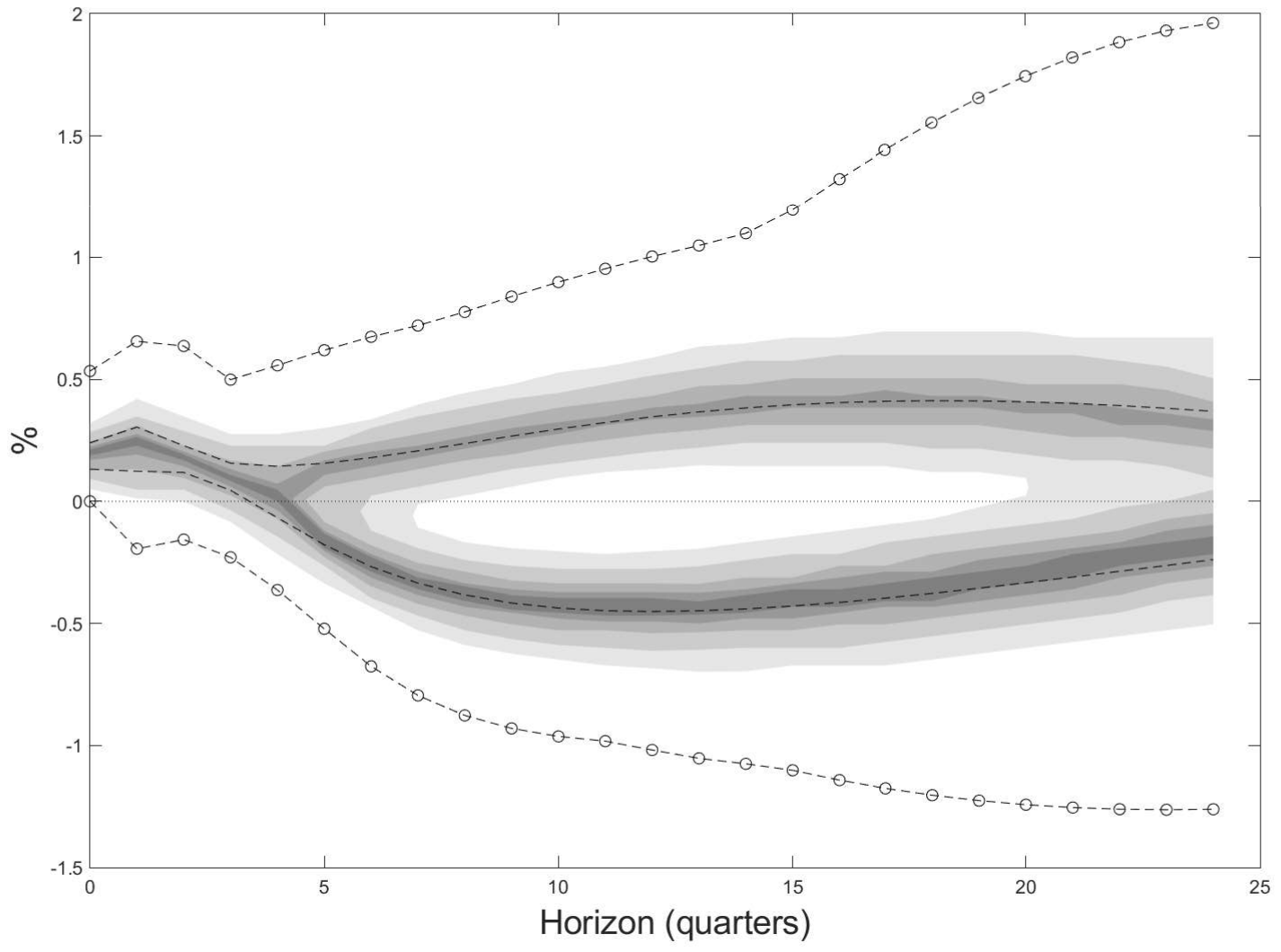}}\hspace{1.5cm}
                \subfigure[{Credible regions and fixed-label projection confidence sets.}]{
                \includegraphics[angle=270,origin=c,scale=0.2]{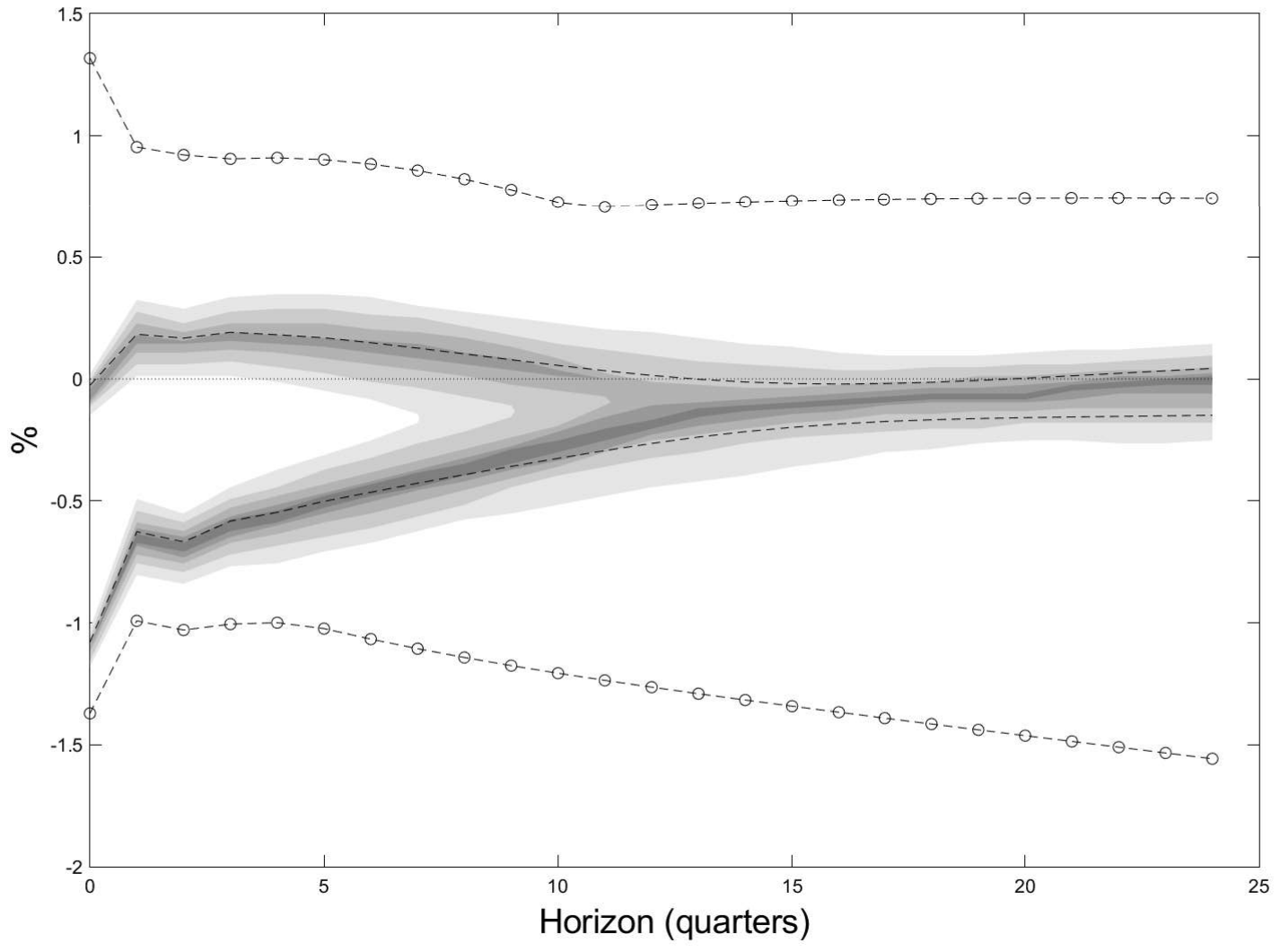}}
                \subfigure[{Credible regions and switching-label projection confidence sets}]{
                \includegraphics[angle=270,origin=c,scale=0.2]{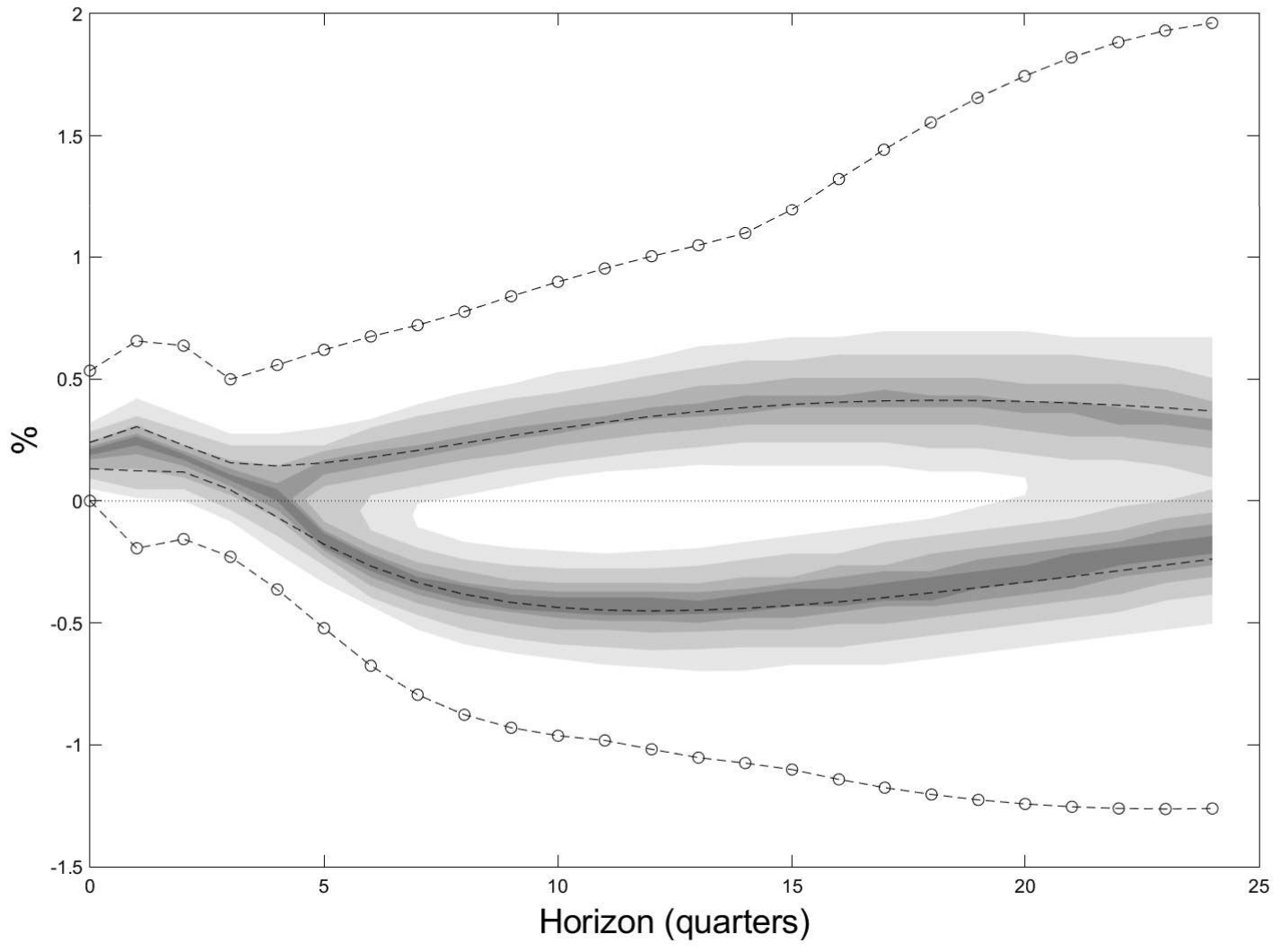}}\hspace{1.5cm}
                \subfigure[{Credible regions and switching-label projection confidence sets}]{
                \includegraphics[angle=270,origin=c,scale=0.2]{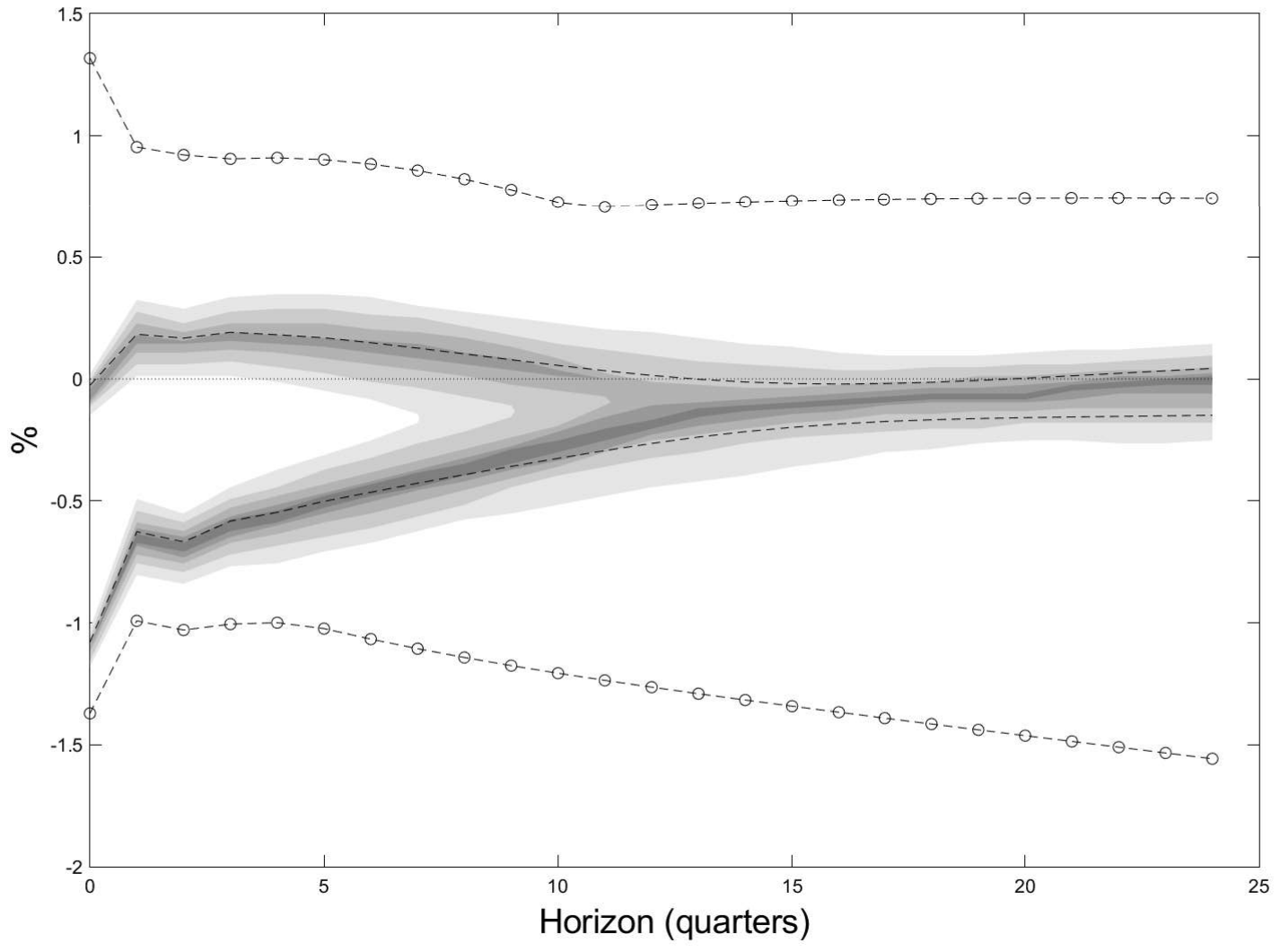}}
\end{center}
\begin{minipage}{\textwidth}%
{\scriptsize{\textit{Notes}: The left column reports the output gap impulse responses and the right column reports the inflation impulse responses, both to a contractionary monetary policy shock. The middle and bottom panels report the posterior highest density regions at $90\%$, $75\%$, $50\%$, $25\%$ and $10\%$ in gray scale. The upper and lower bounds of the frequentist confidence sets are plotted by the dotted-circle lines. The dotted lines in the middle panels plot the set of posterior means.\par}}
\end{minipage}
\end{figure}

The inference approaches proposed in Section \ref{sec:InfLocId} produce the results reported in Figure \ref{fig:IRF_NK_Eq_NoSign}. The left panels plot results for the output gap while the right panels plot those for inflation. The top panels show the draws of the impulse responses obtained based on the draws of $\phi$ from its posterior. 
For each draw, we highlight the two observationally equivalent impulse responses corresponding to admissible $Q_1$ (blue) and $Q_2$ (red) that are coherent with the zero restrictions in Eq. (\ref{eq:DSGEsimmat}). The labels of $Q_1$ and $Q_2$ in the plots of impulse responses for inflation are maintained in the plots for output.

The middle and bottom panels present interval estimates based on the Bayesian and frequentist-valid inference procedures of Sections \ref{sec:BI} and \ref{sec:FVI}. The Bayesian posterior (whose highest density regions are reported both in the middle and bottom panels) is obtained by specifying the uniform conditional prior for $Q$ given $\phi$, i.e., equal weights are assigned to $Q_1$ and $Q_2$ conditional on $\phi$. The highest posterior density regions are plotted with gradation in gray scale, where the credibility levels vary over $90\%$, $75\%$, $50\%$, $25\%$, and $10\%$, from the lightest to darkest.\footnote{The highest posterior density regions are computed by slicing the posterior density approximated through kernel smoothing of the posterior draws of the impulse responses.}

The middle left panel of Figure \ref{fig:IRF_NK_Eq_NoSign} shows the marginal posterior distributions for the output gap impulse response. These are unimodal up to $h=4$, but become bimodal for longer horizons. While there is evidence for output gap puzzle at the shortest horizons, the probability density is tighter and higher for the negative impulse responses in the medium-long run: the darkest gray region (highest $25\%$ and $10\%$ of the distribution) appears mostly for the negative part of the responses. The middle right panel of Figure \ref{fig:IRF_NK_Eq_NoSign} shows the marginal posterior distribution for the inflation impulse response. This is bimodal up to $h=10$ and becomes unimodal at longer horizons. Similarly to the output gap, for the horizons with bimodal distributions, the negative impulse responses have tighter and higher densities than the positive ones.

For both the output gap and inflation, we also present the frequentist-valid confidence intervals (in dotted-circle lines) proposed in Section \ref{sec:FVI}. These are obtained by retaining $90\%$ of the draws of the reduced-form parameters with the highest value of the posterior density function. The middle panels show the fixed-label projection confidence sets of Section \ref{sec:ConstantLabel}, while the bottom panels show the switching-label projection confidence sets of Section \ref{sec:SwitchingLabel}. In addition, for both the output gap and inflation, we show the range of posterior means obtained by the robust Bayesian approach (dotted lines). It is evident that the Bayesian approach gives the narrowest interval estimates, and the highest posterior density regions well visualize the bi-modal nature of the posterior distributions at some horizons. The wider confidence intervals of the frequentist-valid approach reflect a couple of their features. First, they are agnostic over the observationally equivalent parameters in the sense that they do not assign any weights over the observationally equivalent impulse responses. Second, our proposed frequentist-valid procedures rely on projecting the joint confidence intervals for the reduced-form parameters and do not optimize the width of the interval estimates for impulse responses. Concerning the results of the robust Bayesian approach, the bounds of the set of posterior means are in line with the two modes of the posterior distributions.

\begin{figure}[H]
        \caption{New-Keynesian non-recursive SVAR with zero restrictions and one sign restriction}
  \label{fig:IRF_NK_Eq_MildSign}
\begin{center}
                \subfigure[{Output gap to $\e_t^{mp}$: impulse response draws}]{
                \includegraphics[angle=270,origin=c,scale=0.2]{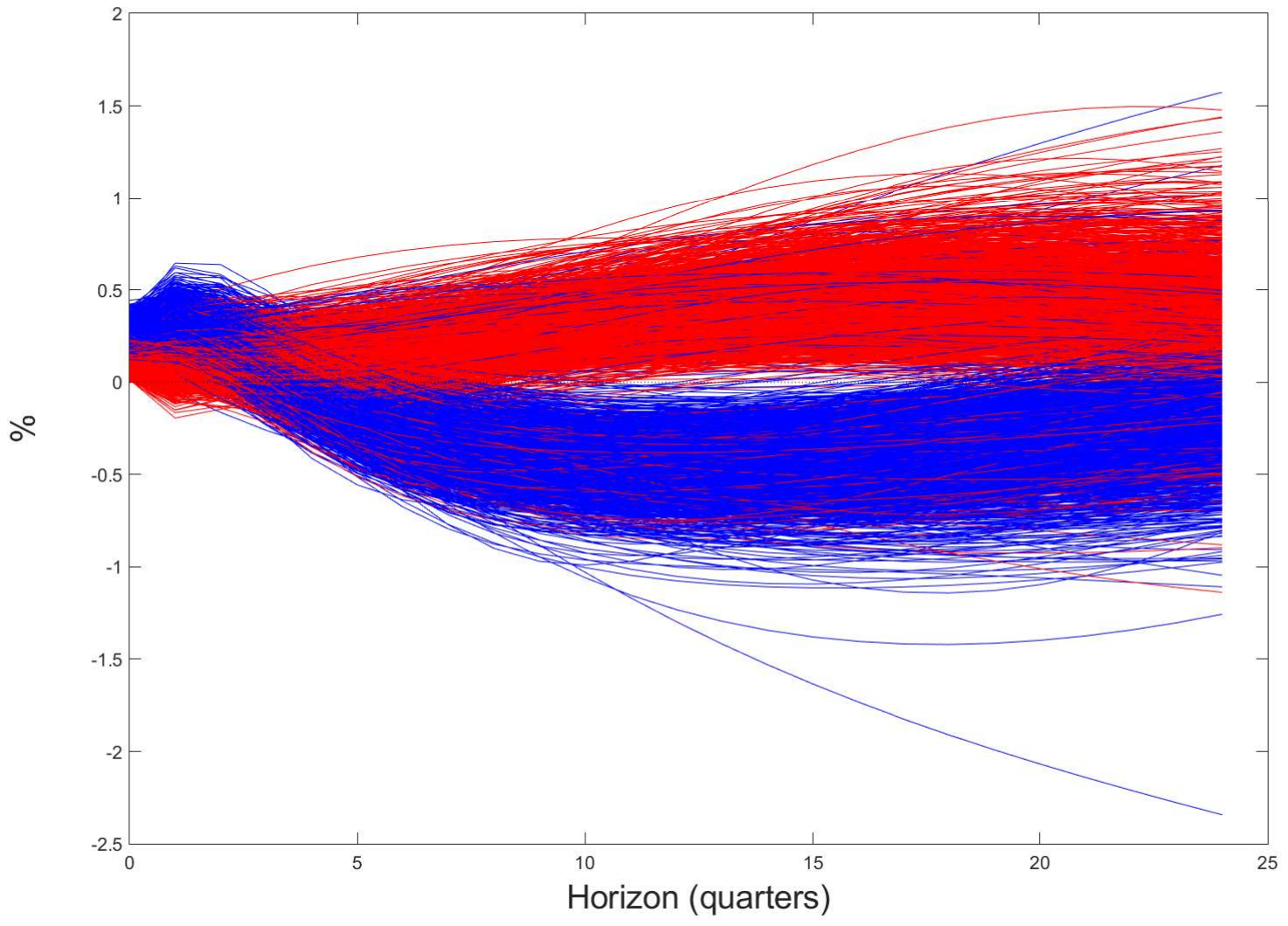}}\hspace{1.5cm}
                \subfigure[{Inflation to $\e_t^{mp}$: impulse response draws }]{
                \includegraphics[angle=270,origin=c,scale=0.2]{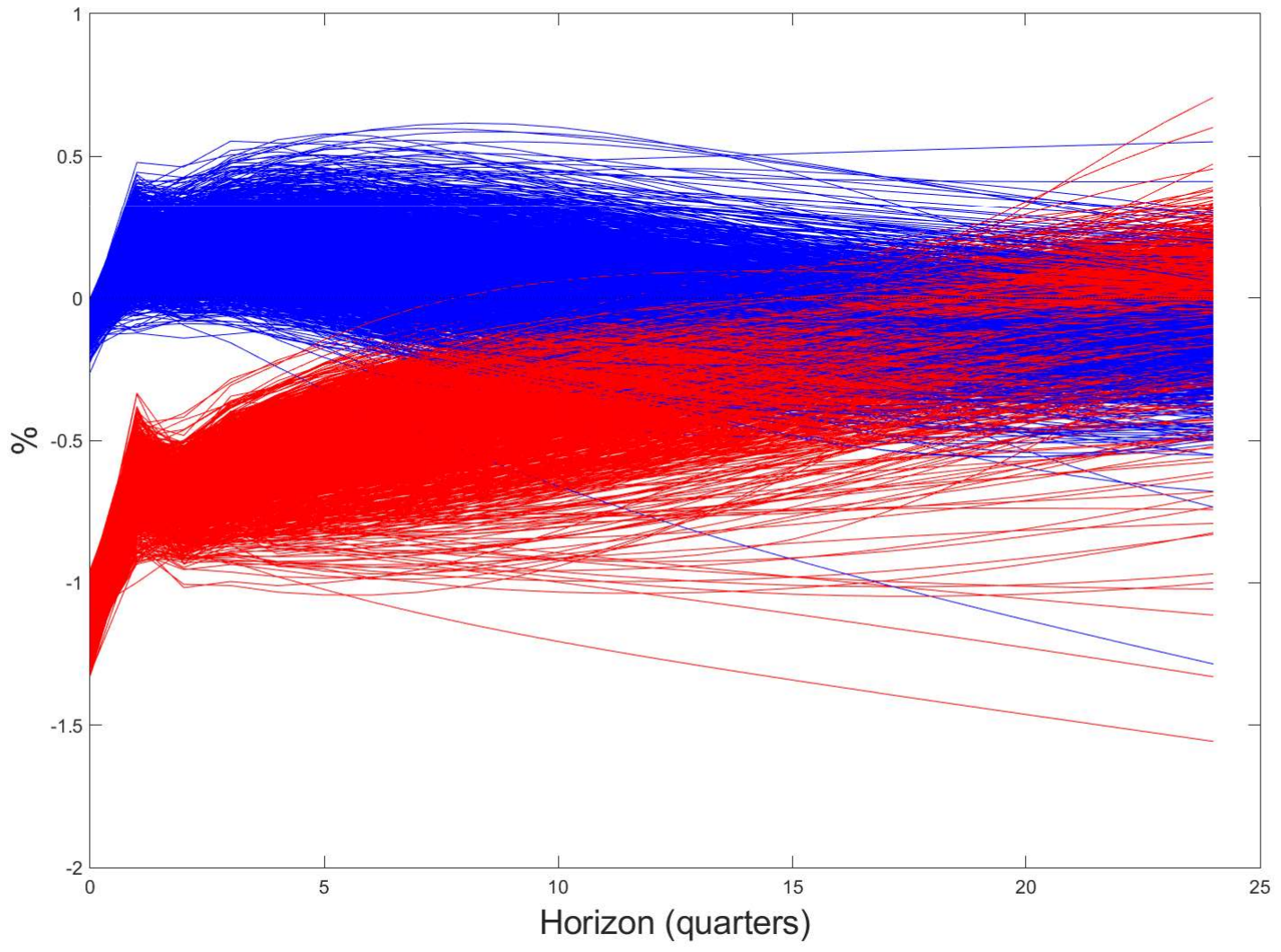}}
                \subfigure[{Credible regions and fixed-label projection confidence sets}]{
                \includegraphics[angle=270,origin=c,scale=0.2]{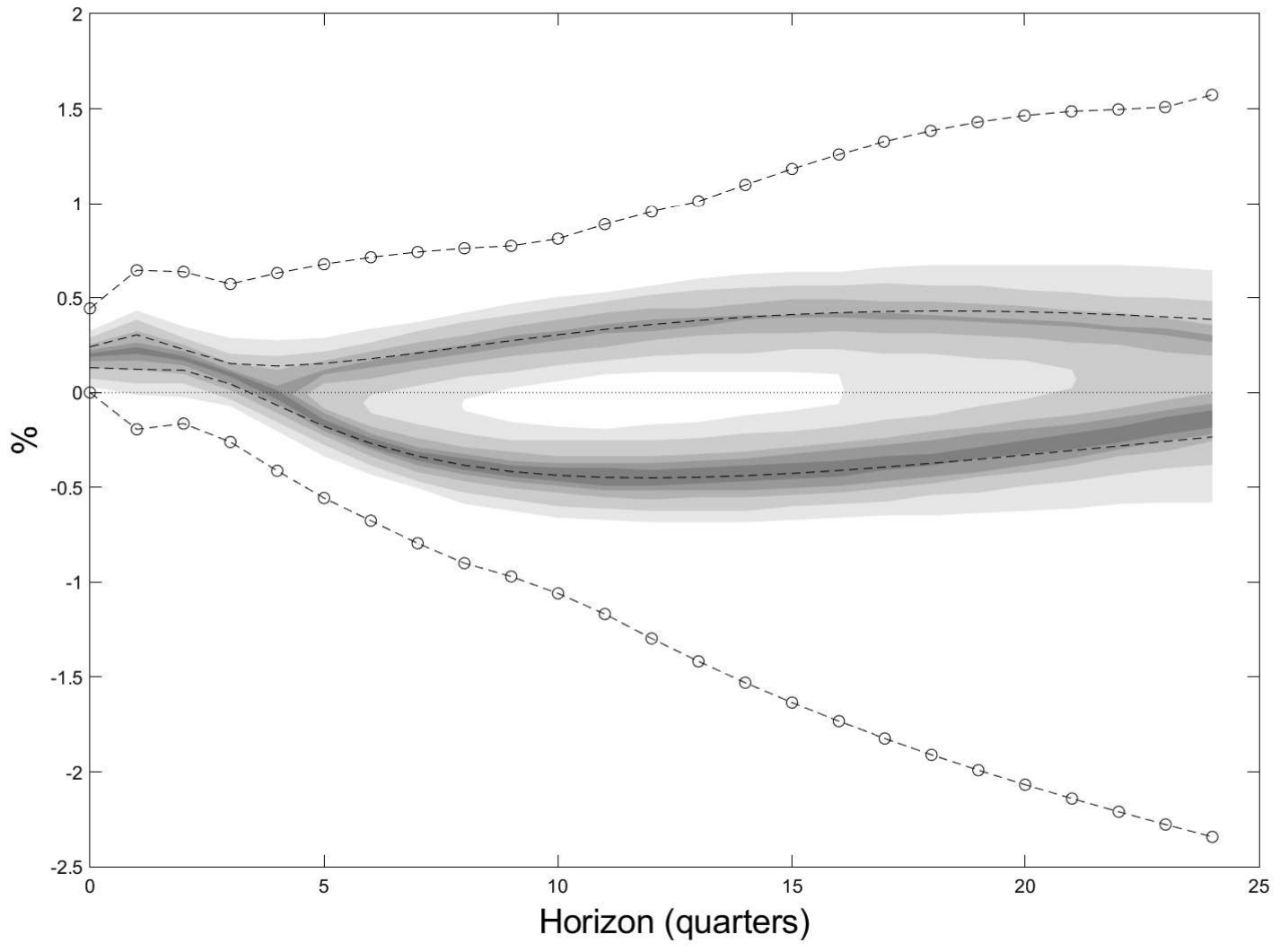}}\hspace{1.5cm}
                \subfigure[{Credible regions and fixed-label projection confidence sets.}]{
                \includegraphics[angle=270,origin=c,scale=0.2]{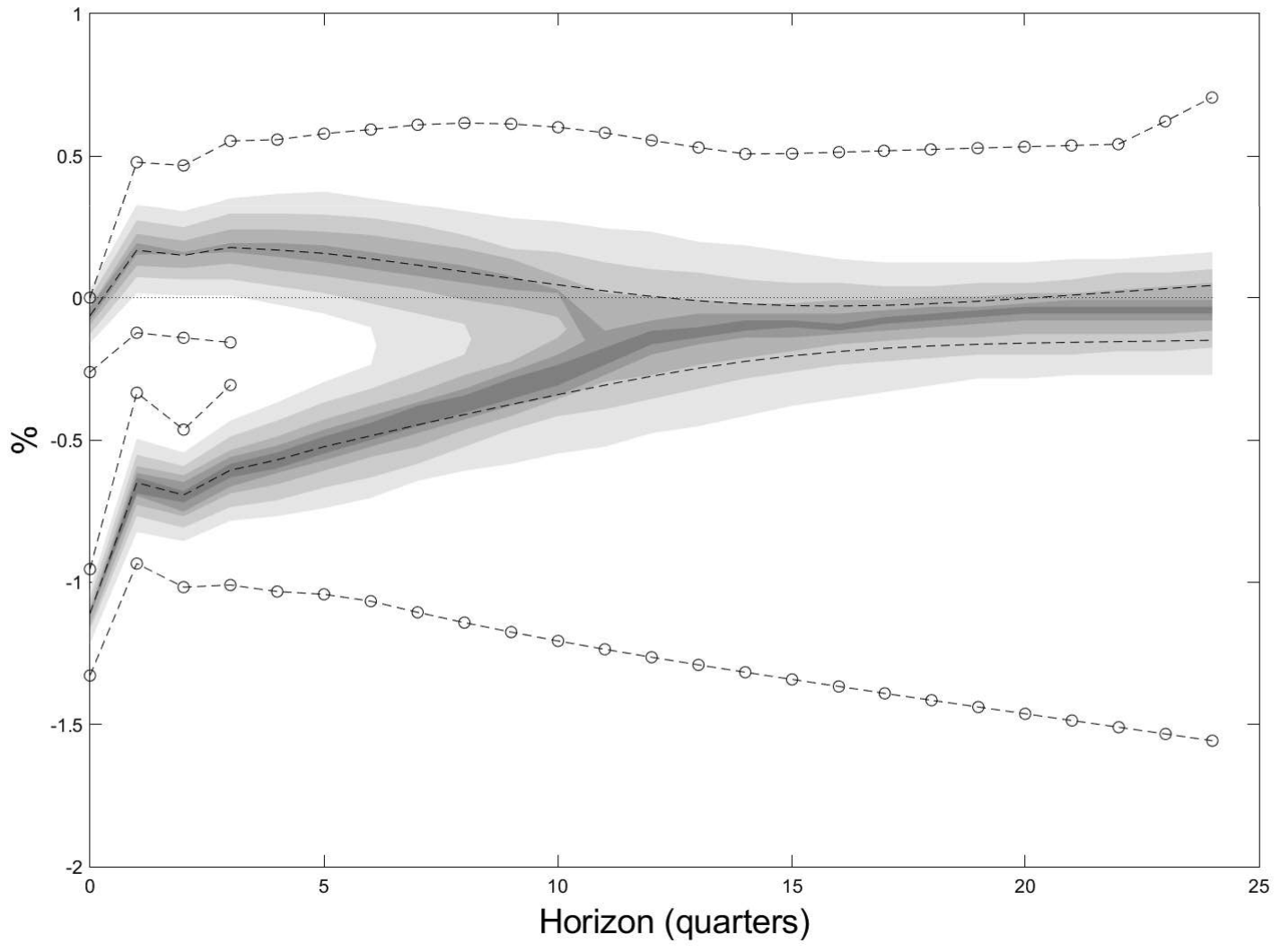}}
                \subfigure[{Credible regions and switching-label projection confidence sets}]{
                \includegraphics[angle=270,origin=c,scale=0.2]{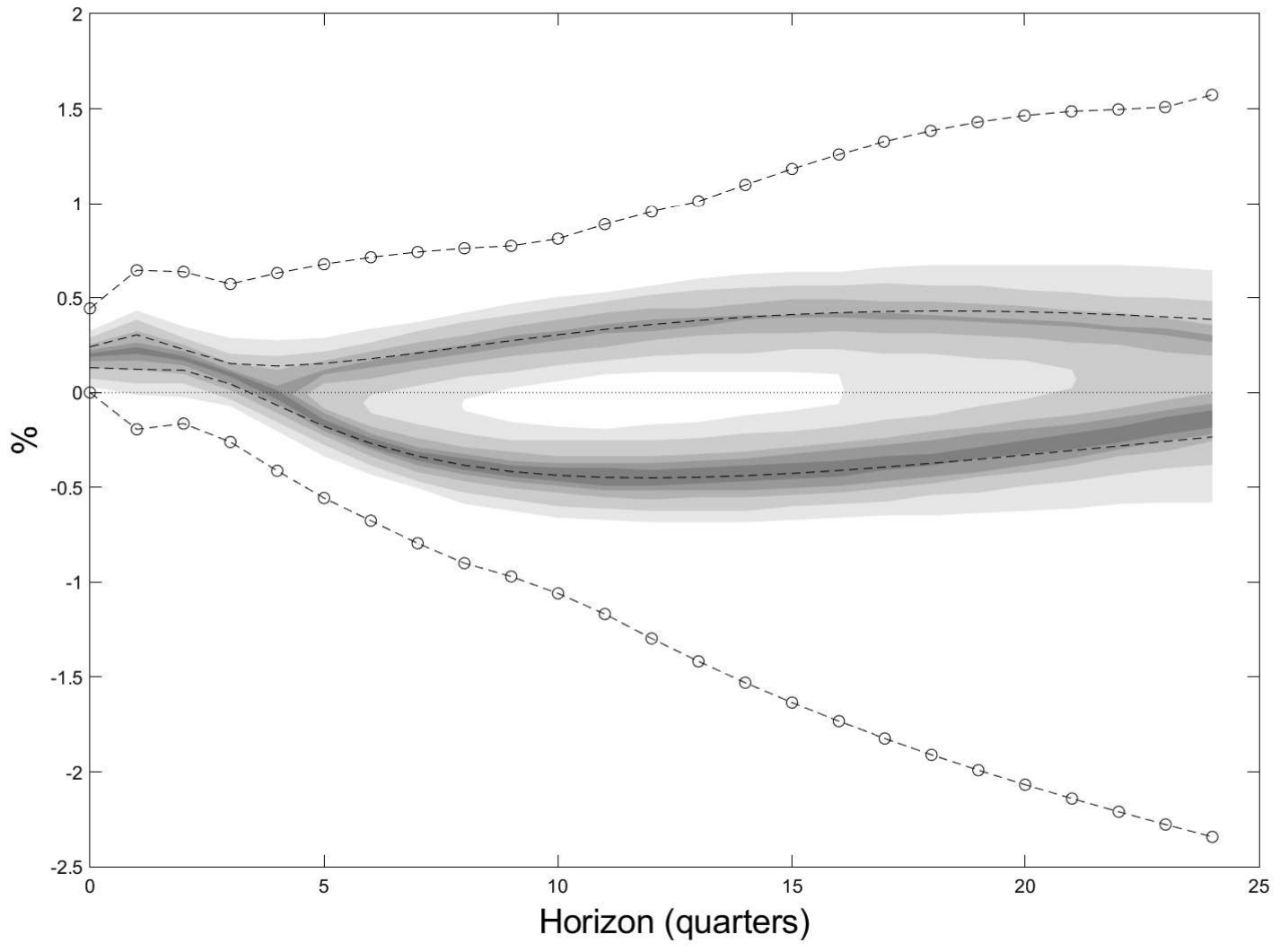}}\hspace{1.5cm}
                \subfigure[{Credible regions and switching-label projection confidence sets}]{
                \includegraphics[angle=270,origin=c,scale=0.2]{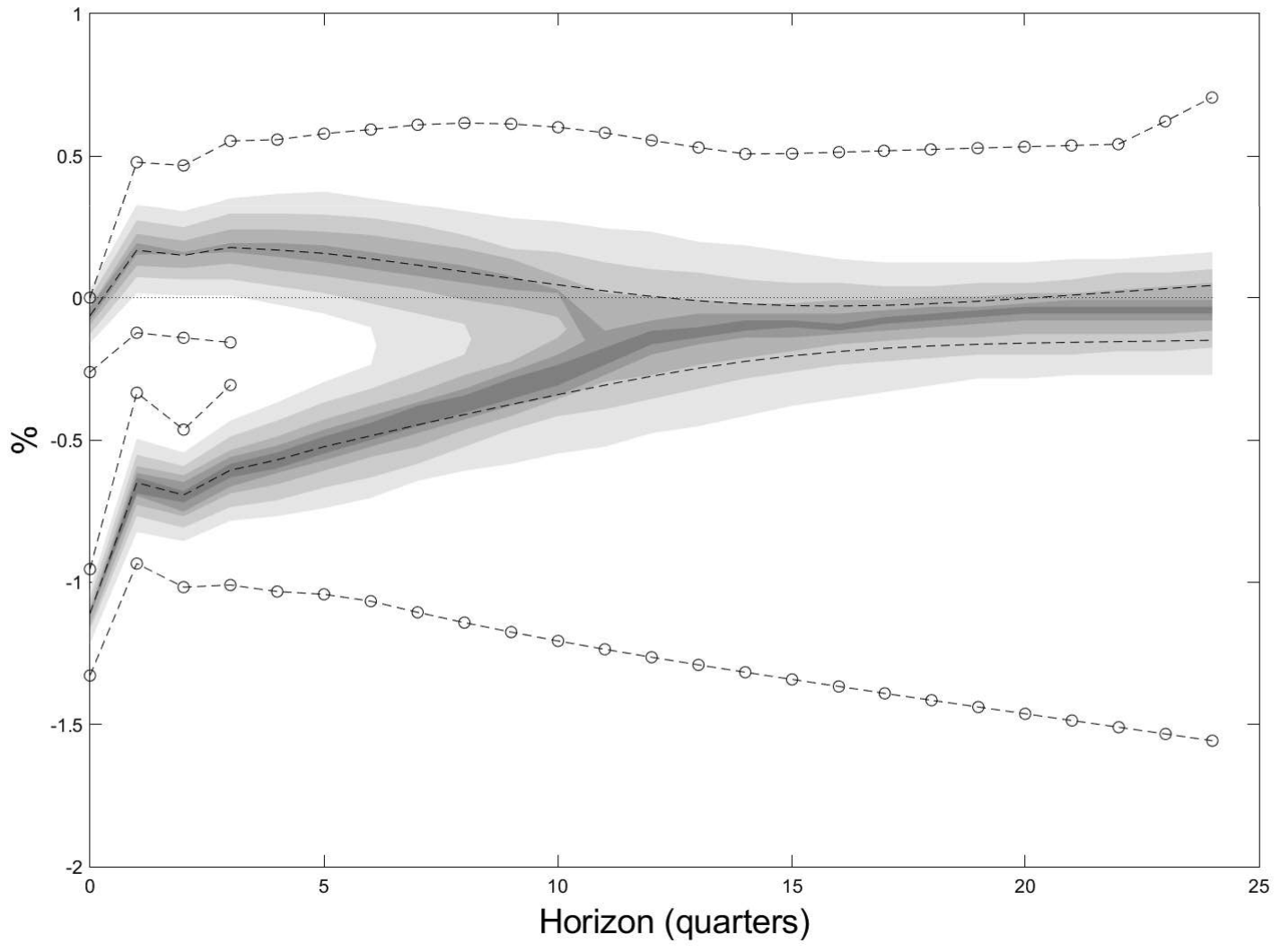}}
\end{center}
\begin{minipage}{\textwidth}%
{\scriptsize{\textit{Notes}: The left column reports the output gap impulse responses and the right column reports the inflation impulse responses, both to a contractionary monetary policy shock. The middle and bottom panels report the posterior highest density regions at $90\%$, $75\%$, $50\%$, $25\%$ and $10\%$ in gray scale. The upper and lower bounds of the frequentist confidence sets are plotted by the dotted-circle lines. The dotted lines in the middle panels plot the set of posterior means.\par}}
\end{minipage}
\end{figure}

\begin{figure}[H]
        \caption{New-Keynesian non-recursive SVAR with zero and additional four sign restrictions}
  \label{fig:IRF_NK_Eq_Sign}
\begin{center}
                \subfigure[{Output gap to $\e_t^{mp}$: impulse response draws}]{
                \includegraphics[angle=270,origin=c,scale=0.2]{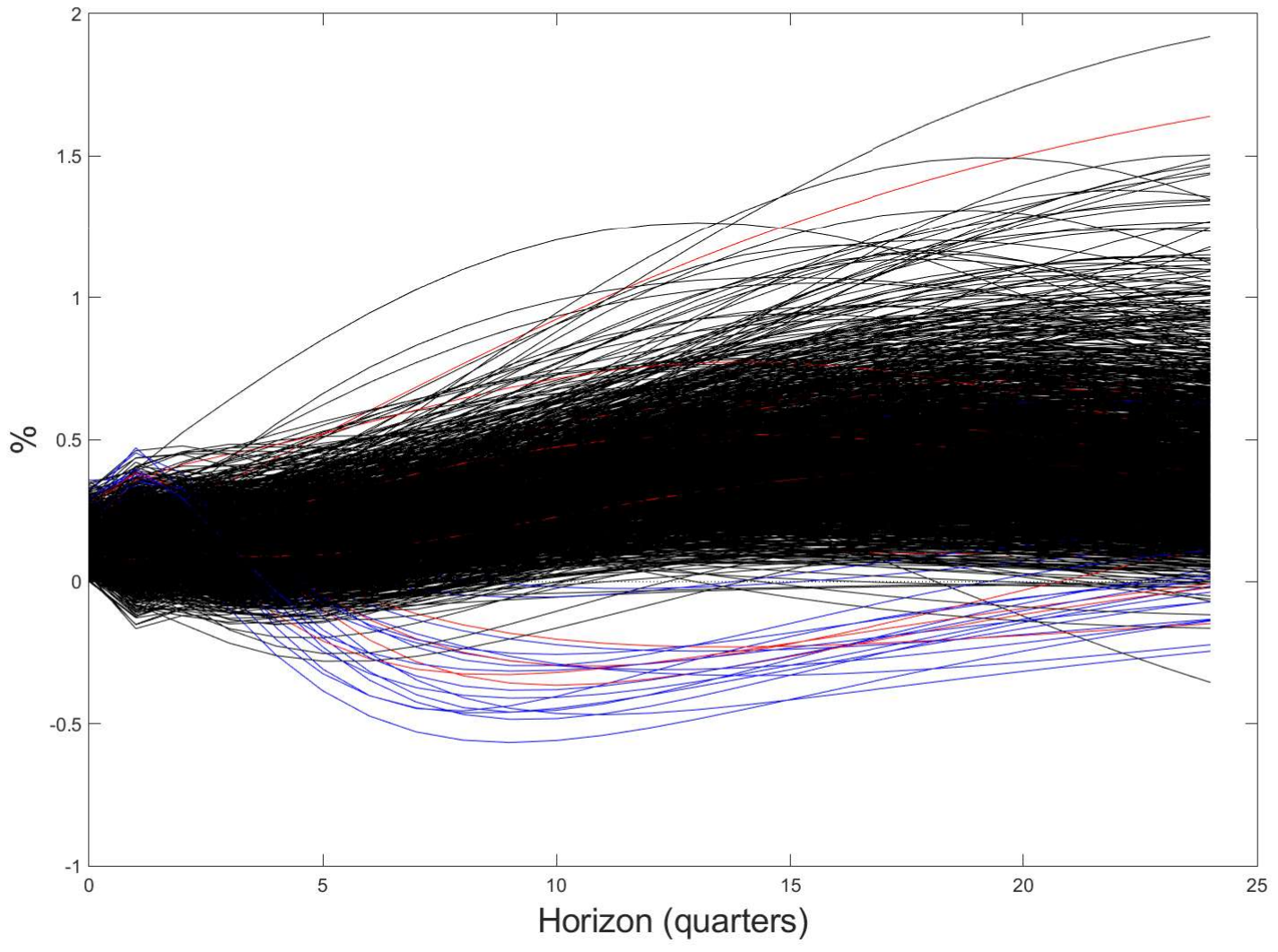}}\hspace{1.5cm}
                \subfigure[{Inflation to $\e_t^{mp}$: impulse response draws }]{
                \includegraphics[angle=270,origin=c,scale=0.2]{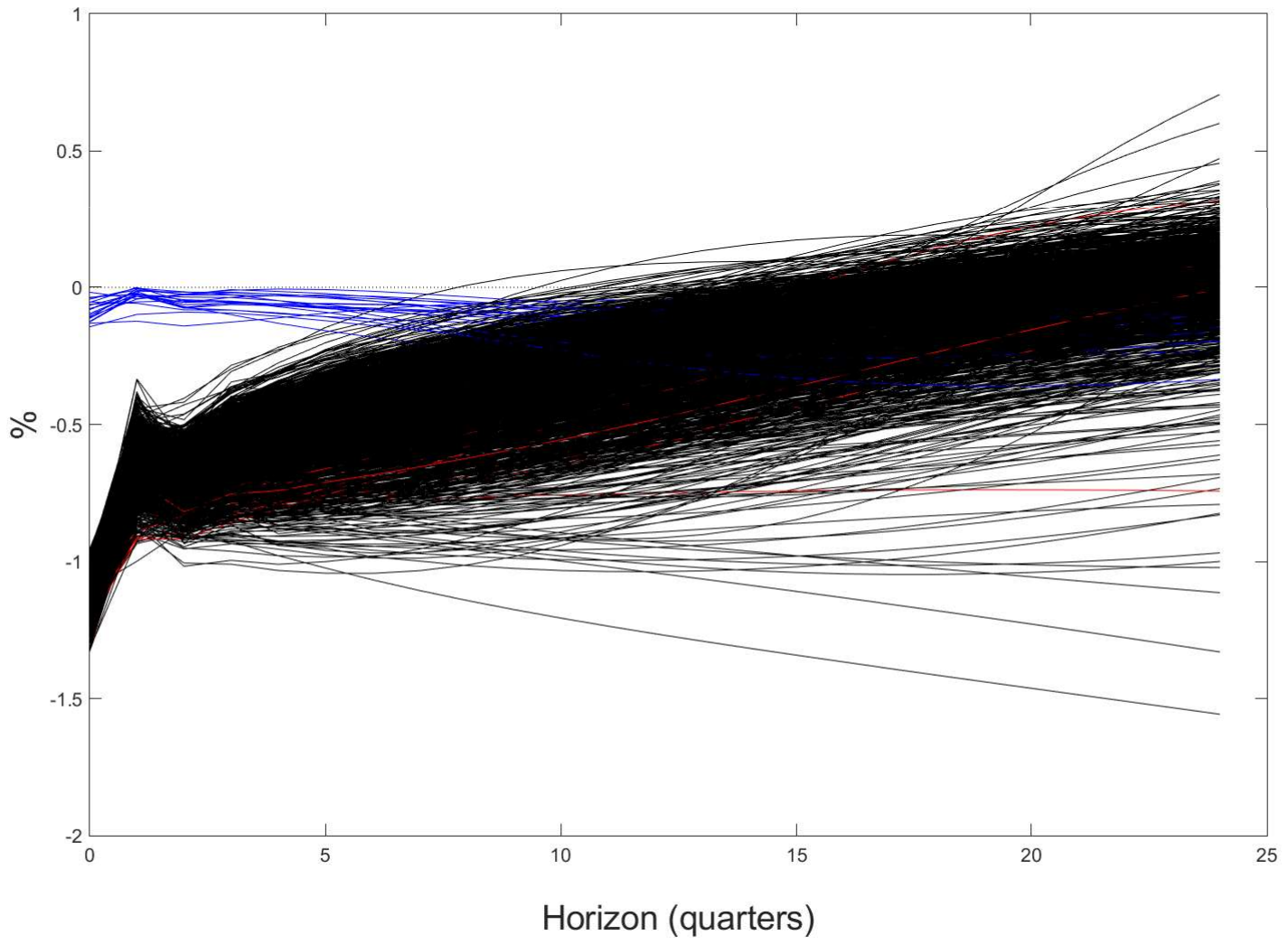}}
                \subfigure[{Credible regions and fixed-label projection confidence sets}]{
                \includegraphics[angle=270,origin=c,scale=0.2]{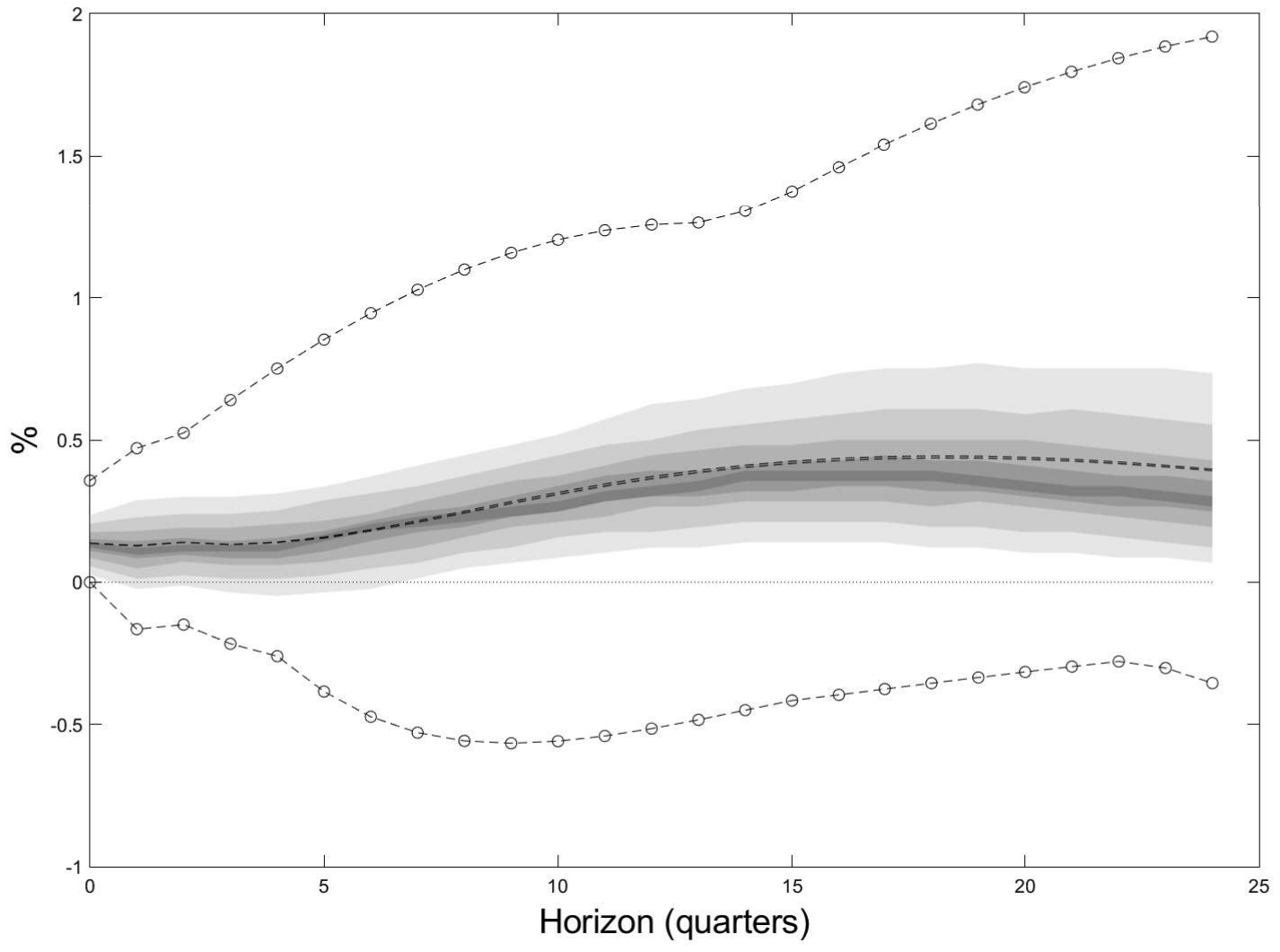}}\hspace{1.5cm}
                \subfigure[{Credible regions and fixed-label projection confidence sets.}]{
                \includegraphics[angle=270,origin=c,scale=0.2]{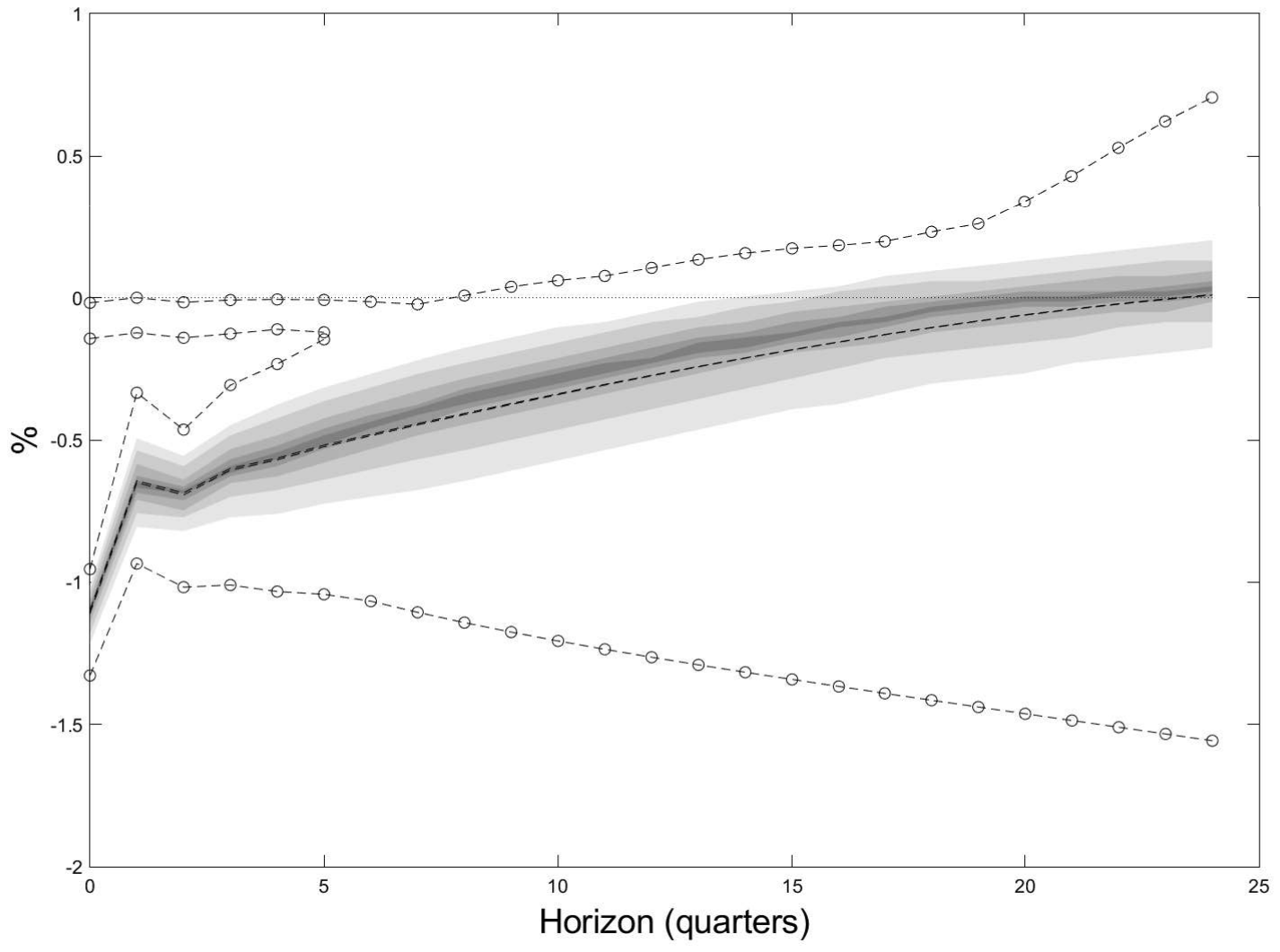}}
                \subfigure[{Credible regions and switching-label projection confidence sets}]{
                \includegraphics[angle=270,origin=c,scale=0.2]{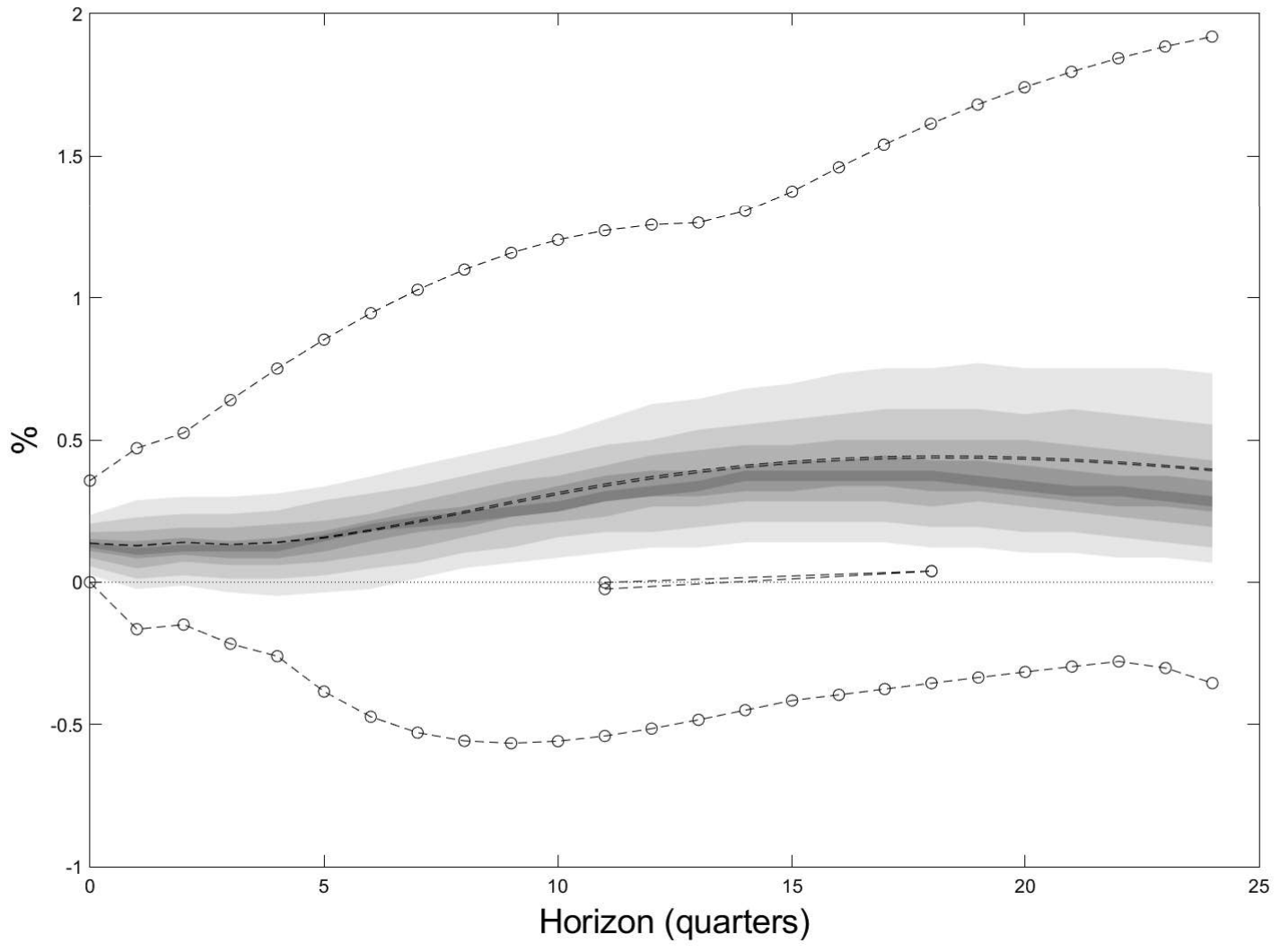}}\hspace{1.5cm}
                \subfigure[{Credible regions and switching-label projection confidence sets}]{
                \includegraphics[angle=270,origin=c,scale=0.2]{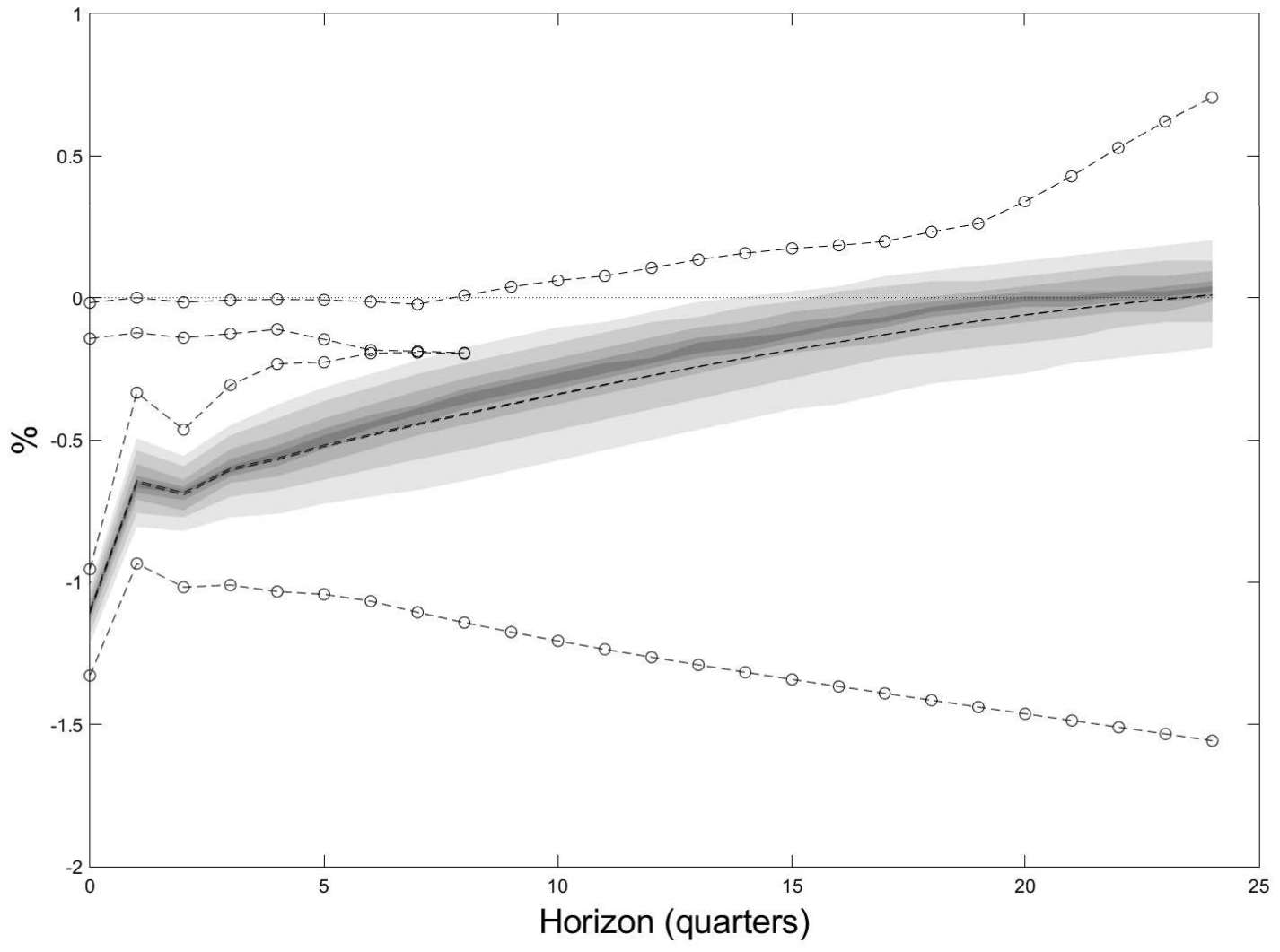}}
\end{center}
\begin{minipage}{\textwidth}%
{\scriptsize{\textit{Notes}: The left column reports the output gap impulse responses and the right column reports the inflation impulse responses, both to a contractionary monetary policy shock. The middle and bottom panels report the posterior highest density regions at $90\%$, $75\%$, $50\%$, $25\%$ and $10\%$ in gray scale. The upper and lower bounds of the frequentist confidence sets are plotted by the dotted-circle lines. The dotted lines in the middle panels plot the set of posterior means.\par}}
\end{minipage}
\end{figure}

A useful strategy for reducing the number of locally-identified admissible solutions is to introduce sign restrictions. To refine the results reported in Figure \ref{fig:IRF_NK_Eq_NoSign}, consider assuming no price puzzle by restricting the inflation responses to be non-positive for (a) the contemporaneous period, or (b) for the four quarters following a contractionary monetary policy shock. The results are reported in Figures \ref{fig:IRF_NK_Eq_MildSign} and \ref{fig:IRF_NK_Eq_Sign}, respectively. 

The results with the contemporaneous non-positivity restriction (Figure \ref{fig:IRF_NK_Eq_MildSign}) appear similar to those in Figure \ref{fig:IRF_NK_Eq_NoSign}. A notable difference is in the upper bound of the frequentist-valid confidence intervals for the inflation response, which now excludes the positive responses shown in Figure \ref{fig:IRF_NK_Eq_NoSign} (top-right panel). For the first few quarters, both switching- and fixed-label projection confidence sets exclude a region between the two modes of the distribution.

Imposing the four sign restrictions (Figure \ref{fig:IRF_NK_Eq_Sign}), allows us to eliminate one of the admissible $Q$ matrices for most of the draws of $\phi$. In the top panels of Figure \ref{fig:IRF_NK_Eq_Sign}, the impulse responses plotted in black have a unique admissible $Q$ under the imposed sign restrictions. Comparing Figure \ref{fig:IRF_NK_Eq_NoSign} and Figure \ref{fig:IRF_NK_Eq_Sign} shows that in this latter the sign restrictions rule out the impulse responses corresponding to  $Q_1$ matrices. The fixed-label and switching-label confidence intervals produce similar results. The only notable difference appears in the response of output gap, where the switching-label confidence intervals in the bottom-left panel have narrow ``gaps'' from $h=11$ to $h=18$, while the fixed-label confidence intervals in the middle-left panel do not. 

\newpage

\newpage

\newpage

\end{document}